\documentclass[10pt]{emulateapj}
\usepackage{graphicx}
\usepackage{amsmath}
\usepackage{natbib}

\begin{document}

\title{Evidence for Cluster to Cluster Variations in Low-Mass Stellar Rotational Evolution}

\author{Carl T.~Coker, Marc Pinsonneault, Donald M.~Terndrup}

\affil{Department of Astronomy, The Ohio State University, Columbus, Ohio 43210, USA\\
coker, pinsono, terndrup@astronomy.ohio-state.edu}

\begin{abstract}

A concordance model for angular momentum evolution has been developed by multiple investigators. This approach postulates that star forming regions and clusters are an evolutionary sequence which can be modeled with assumptions about the coupling between protostars and accretion disks, angular momentum loss from magnetized winds that saturates in a mass-dependent fashion at high rotation rates, and core-envelope decoupling for solar analogs.  We test this approach by combining established data with the large h Per dataset from the MONITOR project and new low-mass Pleiades data.   We confirm prior results that young low-mass stars can be used to test star-disk coupling and angular momentum loss independent of the treatment of internal angular momentum transport.  For slow rotators, we confirm the need for star-disk interactions to evolve the ONC to older systems, using h Per (age 13~Myr) as our natural post-disk case.  Further interactions are not required to evolve slow rotators from h Per to older systems, implying no justification for extremely long-lived disks as an alternative to core-envelope decoupling.  However, our wind models cannot evolve rapid rotators from h Per to older systems consistently; this appears to be a general problem for any wind model that becomes ineffective in low-mass young stars.  We outline two possible solutions:  either there is cosmic variance in the distribution of stellar rotation rates in different clusters or there are substantially enhanced torques in low-mass rapid rotators.  We favor the former explanation and discuss observational tests that could be used to distinguish them.  If the distribution of initial conditions depends on environment, models which test parameters by assuming a universal underlying distribution of initial conditions will need to be re-evaluated.

\end{abstract}


\section{Introduction} \label{sec:intro}

Star and planet formation are complex, and we rely on empirical data to constrain our theories.  The imprint of the formation process is largely erased in older stars by stellar evolution, so we are forced to rely largely on data from young systems.  Stellar rotation provides an intriguing exception: stars arrive on the main sequence with a range of rotation rates, and we can trace the origins of this range back to the star formation process.  It is also difficult to evolve the measured rotation periods of protostars into the rates observed in young clusters without invoking protostar-gaseous disk interactions, which could in turn be impacted by planet formation.  Rotation measurements in older stars therefore may provide a window into their formation and youthful environment.  

However, the angular momentum evolution of stars is also complex. Stars lose mass and angular momentum in magnetized solar-like winds, and there is no current consensus model for the internal transport of angular momentum. These effects are known to be important even early in the lives of solar analogs.  Fortunately, natural structural trends with stellar mass make it possible to separate out different physical phenomena.  Lower mass stars experience both weaker spin down and have deeper convective envelopes, making their surface angular momentum evolution less sensitive to the details of internal processes.  Data for stars below $0.5 M_{\odot}$ has been relatively sparse in young open clusters, however, making this a difficult regime to work in.  In this paper we use new data that fills in the age gap and extends our data into the low mass regime in young open clusters. There has also been a paucity of data between ages typical for star forming regions (0-10 Myr) and the youngest star clusters (40 Myr and more), which has made it difficult to test both the underlying physical picture and the strength of the direct evidence supporting it.  We use these data sets to critically test the hypothesis that gaseous accretion disks are important for angular momentum evolution and to test stellar winds in the low-mass regime independent of the complications arising from internal angular momentum transport.

In the generally accepted picture of pre-main-sequence stellar evolution, there are three basic phases:  first, diffuse gas clouds collapse until they are dense enough to begin fusing deuterium; second, accretion disks form around the newly born protostars which continue to accrete material from these disks until their radiation dissociates the disks; and third, stars contract freely to main sequence and begin fusing hydrogen in their cores.  The hydrostatic phase of evolution is a natural starting point for stellar angular momentum evolution models, and we begin there.

As long as protostars retain a massive gaseous disk, the stars magnetic field can interact with its accretion disk.  The lifetime of this phase was traditionally thought to be of order 6 Myr or less (Haisch et al.~2001), but recent revisions to the ages of star forming regions may imply disk ages of up to 12 Myr (Bell et al.~2013; Somers \& Pinsonneault~2015). Protostar-disk interactions can effectively couple the star's rotation rate to that of the inner disk (Konigl~1991; Edwards et al.~1993; Shu et al.~1994; but see also Matt \& Pudritz 2005).  This “disk-locking” mechanism has become a standard tool for constructing stellar angular momentum evolution models.  Producing definitive tests of this proposed mechanism, however, has proved difficult.  Part of this difficulty has resulted from the aforementioned lack of data for clusters which have just seen all major disks fully evaporate.  This has recently been rectified by the MONITOR project's data release (Moraux et al. 2013) for the cluster h Per, which is $\simeq 13$ Myr old (Bell et al. 2013).  Additionally during this phase, the rotation period distribution becomes strikingly bimodal, split between fast rotators with periods below $\simeq 1$~day and slower stars with periods greater than about 3-4~days.  This bimodality is clearly seen in h Per.  However, it is also present for some mass ranges in much younger clusters such as Orion (Attridge \& Herbst~1992; however, see also Stassun et al.~1999) and NGC 2264 (Lamm et al.~2005), and can persist until the age of the Pleiades (Terndrup et al.~2000).  Some form of protostars-disk interaction is an attractive hypothesis for generating such a bimodality; a range of effective disk coupling timescales, combined with a range of initial rotation rates, can effectively produce a wide range of distributions, including bimodal ones.  We therefore test the impact of disks by comparing h Per with the younger Orion Nebula Cluster, and bypass the question of disks for angular momentum evolution of older systems by using h Per as an initial condition for those tests.

Once the disk surrounding a star largely evaporates, the star's magnetic field can drive a strong stellar wind, stealing angular momentum from the star and acting to spin it down.  These solar-like winds (Weber \& Davis~1967; Kawaler~1988) can efficiently drain angular momentum from young stars, producing a strong $v_{rot} \simeq t^{-1/2}$ spin down at late ages (Skumanich~1972), which can make rotation a potent age indicator at late ages (e.g. Soderblom~2010 and references therein; see Epstein \& Pinsonneault~2014 and van Saders et al.~2016 for age uncertainties for older stars).  However, a comparison of models using simple solar scaled winds with open cluster data revealed two striking phenomena: there was both a “slow rotator problem”, namely a population rotating far too slowly relative to protostars, and a “rapid rotator problem”, arising from the fact that angular momentum loss scaling as the third power of angular velocity would preclude stars from ever becoming rapid rotators.  Solving the rapid rotator problem required a saturation in angular momentum loss (Keppens et al.~1995, Chaboyer et al. 1995) which is mass dependent (Krishnamurthi et al.~1997).  The slow rotator population is typically interpreted as a consequence of core-envelope decoupling (Endal \& Sofia~1979; Pinsonneault et al.~1989; Allain~1998).  The envelope initially spins down in response to the wind torque, and the core responds on an internal transport timescale; there will therefore be a detectable transient phase of anomalously slow surface rotation if the transport timescale is of order tens of Myr or longer.  This core-envelope decoupling can have a large effect on the surface rotation rate as the star ages (Pinsonneault et al.~1990; MacGregor \& Brenner~1991) 

The resulting phenomenological models can successfully reproduce many of the observed time dependence of stellar rotation in open cluster stars (see for example Gallet \& Bouvier~2013, Lanzafame \& Spada~2015).  However, the resulting solutions have numerous free parameters and there could be severe physical degeneracies between them.  Furthermore, it is possible that the assumptions behind this approach may themselves be questionable.  For example, Brown~(2014) has invoked changes in magnetic dynamos, rather than core-envelope decoupling, to explain slow rotators; Matt \& Pudritz~(2005) have questioned the efficacy of protostars-disk coupling; and more physically motivated wind models (Cranmer \& Saar~2011, Matt et al.~2015, van Saders \& Pinsonneault~2013) make different predictions about mass and rotation trends in angular momentum loss than models constructed in the Kawaler~(1988) framework.  There is also a crucial and untested assumption in all empirical angular momentum evolution models: that associations and star clusters form an evolutionary sequence with a universal underlying set of initial conditions.

In this paper, we extend previous investigations into the rotation period distribution of low-mass stars.  We attempt to bridge some of the gaps in the understanding caused by lack of data from the time shortly after all disks have evaporated, as well as the lack of data in the regime below $0.5 M_{\odot}$.  We demonstrate that models constructed within the existing framework naturally predict distinct domains: a high mass regime where initial conditions, core-envelop decoupling, and solar-like winds are all important; an intermediate domain where initial conditions and winds are important; and a low mass, young domain where only the initial conditions are important.  We use these domains to critically test the underlying assumptions, and demonstrate a breakdown in either the evolutionary sequence assumption or the wind behavior in the lowest mass stars.  We also test whether star-disk interactions can produce the bimodality, and confirm their importance for evolving the angular momentum distributions of protostars.

In Section~\ref{sec:data} we present the clusters we use for our datapoints, as well as our reasons for choosing them.  In Section~\ref{sec:models} we present the physics behind our stellar models.  In Section~\ref{sec:results}, we present the results of our models and their implications.  A brief discussion is presented in Section~\ref{sec:discussion}. 

\begin{figure*}
\includegraphics[scale=0.45]{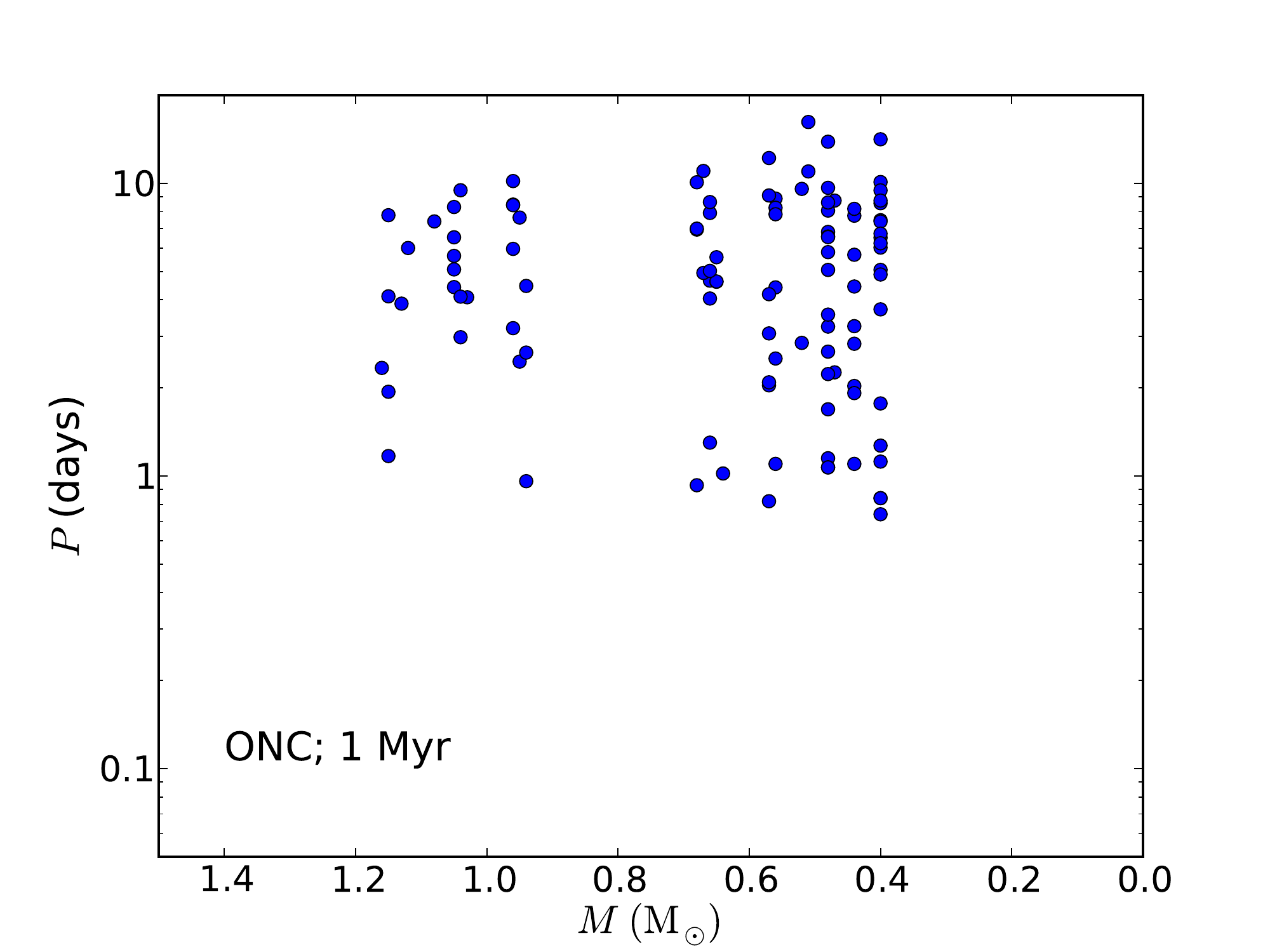}
\includegraphics[scale=0.45]{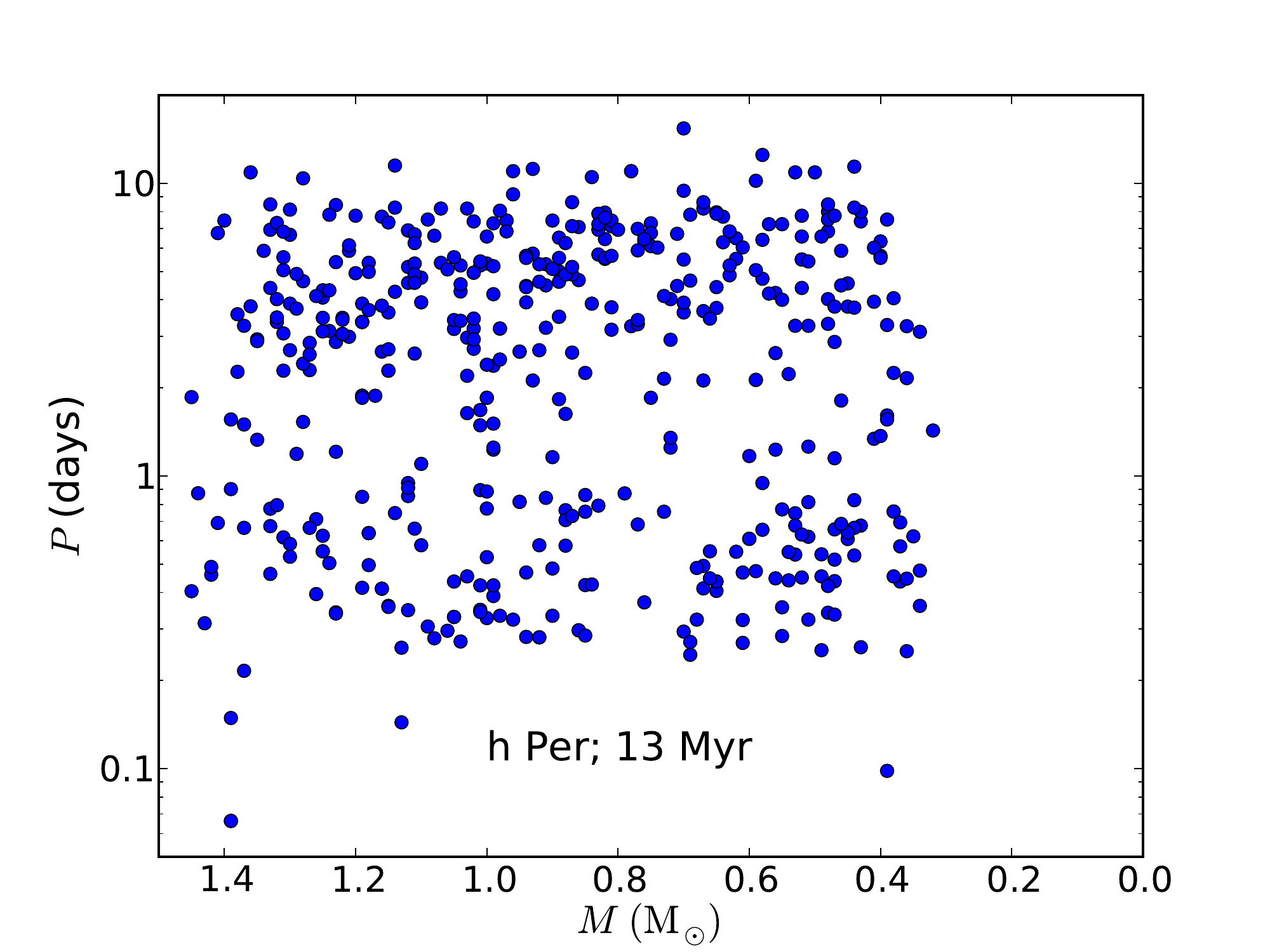}
\includegraphics[scale=0.45]{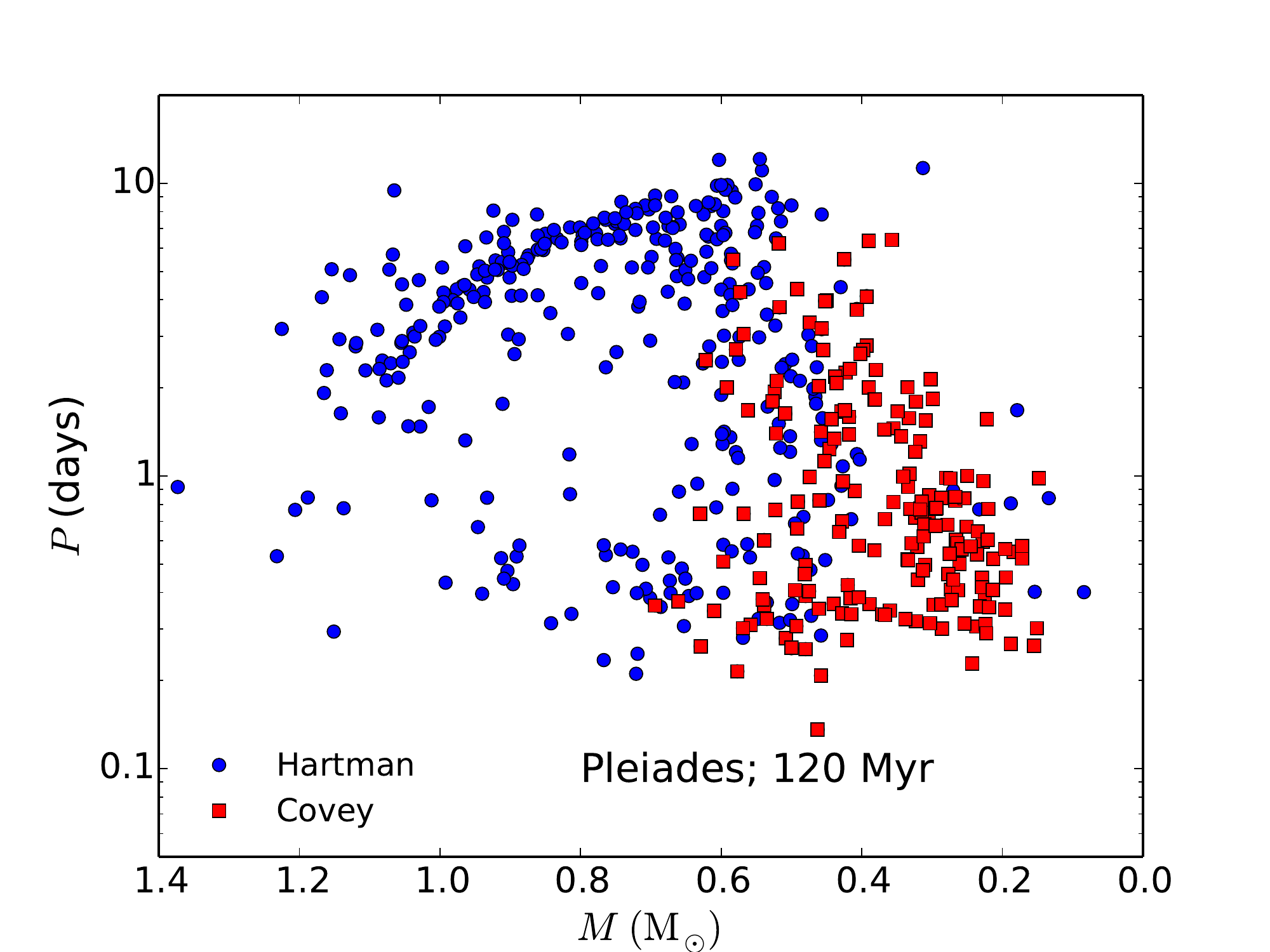}
\includegraphics[scale=0.45]{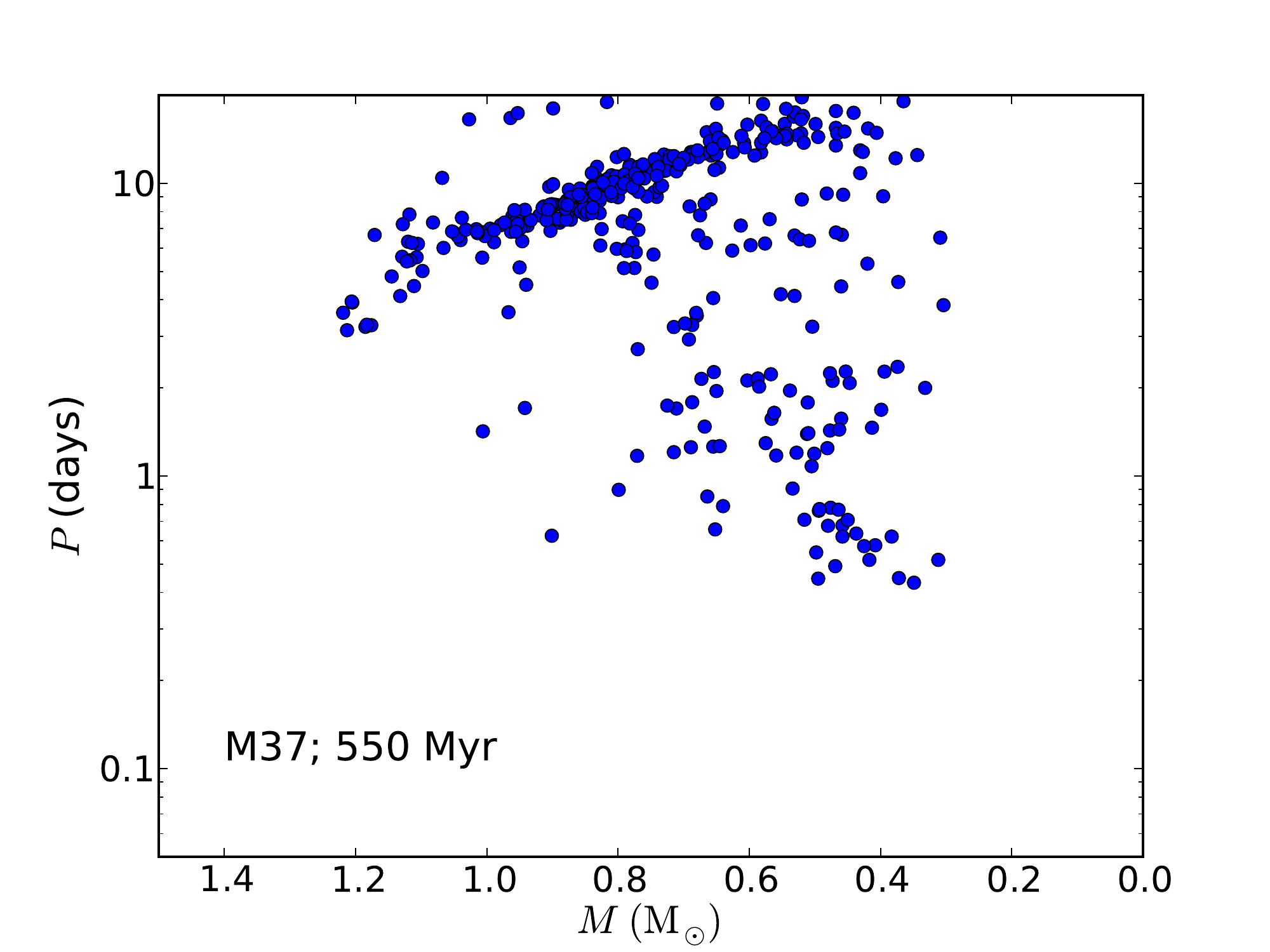}
\caption{Mass-period plots of each of the clusters under discussion, for reference.  The clusters are presented in order of age.  Clockwise from top left:  ONC (data from Rebull et al.~2006), h Per (Moraux et al.~2013), Pleiades (Hartman et al.~2010; Covey et al.~in prep), M37 (Hartman et al.~2009).  For the Pleiades figure, the blue dots represent stars which Hartman et al.~(2010) provided periods for, while the red squares are those from Covey et al.~(2016).}
\label{fig:clusters}
\end{figure*}

\begin{figure}
\includegraphics[scale=0.45]{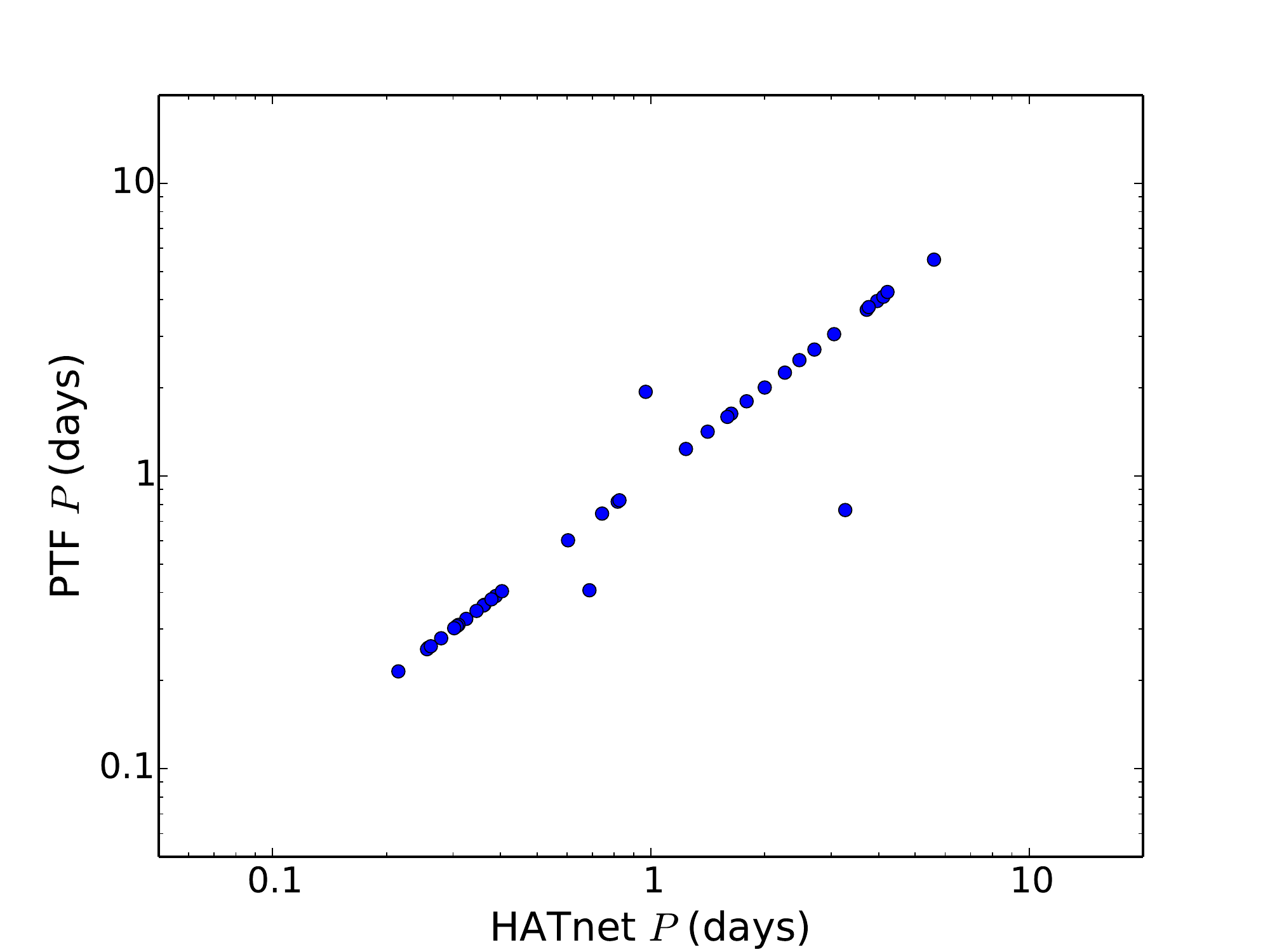}
\caption{A plot of the HATnet Pleiades periods from Hartman et al.~(2010) versus those from Covey et al.~(2016), which were obtained using the PTF.  Except for three outliers, each dataset gives the same answer.  For our models, we use the PTF periods where there is a conflict.  The errors on the periods are not shown because they are far smaller than the dots used.}
\label{fig:PTFvsHAT}
\end{figure}

\section{Data} \label{sec:data}

In order to answer the questions set forth in the introduction, we selected four clusters, each of which provides a relatively large sample of stars for their age ranges:  the ONC (Rebull et al.~2006), h Per (Moraux et al.~2013), the Pleiades (Hartman et al.~2010, Covey et al.~2016), and M37 (Hartman et al.~2009).  The latter three systems have rotation periods for several hundred members.  The basic properties of our chosen clusters are found in Table~\ref{table:clusterprop}.  Figure~\ref{fig:clusters} shows period-mass diagrams for each of our clusters.

We are not attempting to replicate the parametric solutions for angular momentum evolution in previous studies, such as Gallet \& Bouvier~(2013, 2015) or Denissenkov et al.~(2010).  We are focusing instead on low mass stars where the timescale for angular momentum evolution is relatively long and where, as we demonstrate in Section 4, the assumptions about core-envelope coupling do not significantly impact our models.

Our first goal is to compare rotation distributions between young systems where there is a large fraction of protostars with gaseous accretion disks and ones where there are few.  For this comparison we chose the ONC data of Rebull et al.~(2006) and the h Per data of Moraux et al.~(2013) respectively.  We chose the Rebull et al.~(2006) data set because it combines a statistically significant dataset with comprehensive disk diagnostics, and we are interested in comparing angular momentum evolution in stars both with and without disks.

Our second goal is to test angular momentum evolution in the low mass domain.  For sufficiently low mass and young stars, little angular momentum loss is expected; a comparison of the distributions of rotation rates in that circumstance will be a test of the universality of the initial conditions and the Rossby number-scaled treatment of angular momentum loss that predicts mimimal loss in this mass domain.  For this purpose the Pleiades data sets from Hartman et al.~(2010) and Covey et al.~(2016) provide an ideal young stellar template to compare with evolution from H Per and the ONC.  These samples are comprehensive and well-vetted for membership.  To calibrate angular momentum loss in these models we used the M37 data of Hartman et al.~(2009).  As demonstrated in Epstein \& Pinsonneault~(2014), an M37-based calibration is consistent with data in clusters of similar age such as the Hyades, Praesepe, and NGC 6811.

We note that there are other stars clusters with data.  However, inclusion of these additional systems would not impact our core science goals, but would add additional model-dependent uncertainties arising from sample selection, background contamination (which can be very high in systems without membership constraints), and the relative metallicities and ages of the systems.

\subsection{Cluster Age and Metallicity} \label{sec:clusterprop}

For the ONC, D'Orazi et al.~(2009) used FLAMES/UVES and Giraffe spectra to get an average metallicity of the low-mass members, while Hillenbrand~(1997) estimated the age by fitting to the isochrones presented in Swenson et al.~(1994) and D'antona \& Mazzitelli~(1994).  For h Per, Mayne \& Naylor~(2008), present an age of 13~Myr, which they derive from isochrone fitting of the main-sequence turnoff.  We take the metallicity of h Per to be solar.  There is some evidence that this is the case; substantially subsolar metallicity isochrones are unable to reproduce the observed M supergiant $V-J$ colors (Currie et al.~2010), and disk objects that are around the solar distance from the galactic center are usually around solar metallicity (Friel et al.~2002; Cunha et al.~2016).  We discuss the potential impact of a non-solar abundance for h Per in Section~\ref{sec:metallicity}.  For the Pleiades, the age was estimated by Stauffer et al.~(1998) using the lithium depletion of the boundary method, which is robust for this cluster (Burke et al.~2004).  The metallicity was estimated by Soderblom et al.~(2009) using abundance measurements of several elements in Pleiades stars with $T_{\rm eff}$ of roughly solar and low $v\sin i$.  For M37, the age and metallicity come from Hartman, et al.~(2008).  They used measurements of several transition metal lines to estimate the metallicity, and then fit to a set of YREC isochrones to get the age.  We used the age estimate from their models which included convective overshoot.

\subsection{Stellar Mass Estimates} \label{sec:massest}

For the Pleiades, Hartman et al.~(2010) used the K-band magntiude and the YY2 isochrones to estimate the mass of each star, while Covey et al.~(2016) used the K-band magnitude and the Dartmouth isochrones.  For h Per, the MONITOR group used the $i'_{CFHT}$ magnitude and models from Siess et al.~(2000).  The ONC mass estimates were provided to us in a private communication with Rebull, and are derived from fitting photometry to the models from D'Antona \& Mazzitelli~(1994).  For M37, mass estimates were not provided in Hartman et al.~(2009), so we used the values provided by Van Saders \& Pinsonneault~(2013).  They used a combination of $V$ and $B-V$ photometry and fit the photometric data to the An et al.~(2007) isochrones.  

\subsection{Cluster Period Distribution Features} \label{sec:clusterdiscussion}

The period-mass diagrams for each of our clusters are shown in Figure~\ref{fig:clusters}.  As can be seen from following the sequence, stars are generally born rotating fairly slowly, separate into slowly and rapidly rotating populations by the time they are $\sim$10~Myr old, then spin up until they hit the main sequence.  Once they are either on or close to the main sequence, magnetically-driven stellar winds start spinning them down.  By the age of the Pleiades, the bimodality seen in h Per is largely eliminated - the hotter stars converge to a tight sequence, and this progresses to cooler stars as time goes on.  By the age of M37, the bimodality is effectively gone, and stars are busy spinning down, all ending up at roughly the same period.

Also of note are the hard edges to the rotation period distribution seen in the Pleiades and M37.  In the Pleiades, there are no/very few stars with both very low mass and very slow rotation.  The same thing occurs in M37, although in a different region of the diagram.  There, there are very few fast rotators beyond about 0.6-0.7$M_{\odot}$. This is because stellar winds have spun down all of the more massive stars (see Sections~\ref{sec:models} and \ref{sec:domains} for more details).  It should be noted that there are also possible background sources in the M37 data:  potential nonmembers at the slow end, and synchronized binaries at the fast end of the distribution.  Hartman et al.~(2009) estimate a 20\% contamination rate overall, and the detection limit rises to nearly 0.1 magnitudes in both Hartman et al. studies at the faint end.

\begin{deluxetable} {llllll}
\tabletypesize{\scriptsize}
\tablecolumns{3}
\tablewidth{0pt}
\tablecaption{Cluster Data}
\tablehead{
                        \colhead{Cluster} &
                        \colhead{Age (Myr)} &
		        \colhead{[Fe/H]} & 
			\colhead{No. of Stars} &}
\startdata
ONC  &  1  &  $-0.01 \pm 0.04$  &  105  \\ 
h Per  &  13  &  $0.0$  &  586  \\ 
Pleiades  &  120  &  $+0.042 \pm 0.021$  &  526 \\ 
M37  &  550  &  $+0.045\pm0.044$  &  575\\ 
\enddata
\label{table:clusterprop}
\end{deluxetable}

\section{Models} \label{sec:models}

We constructed all of our stellar models using the Yale Rotation Evolutionary Code (YREC; Van Saders \& Pinsonneault~2012).  This code allows us to specify a set of initial conditions for a star and evolve it forward in time.  It includes the structural effects of rotation, stellar winds, angular momentum transport, and disk locking.  We investigate different angular momentum evolution scenarios in this paper, as described below.  For our models, we use the atmosphere and boundary conditions of Kurucz~(1997), nuclear reaction rates of Adelberger et al.~(2011) with weak screening (Salpeter~1954), and we employ the mixing length theory of convection (Cox~1968; Vitense~1953) with no convective overshoot. Opacities are from the Opacity Project (Mendoza et al.~2007) for a Grevesse \& Sauval~(1998) solar mixture, supplemented with the low temperature opacities of Ferguson et al.~(2005). We utilize the 2006 OPAL equation of state (Rogers \& Nayfonov~2002).  For the angular momentum transport, we only model hydrodynamic mechanisms (see Pinsonneault~1997 for a review).  When we tested the effect of core-envelope decoupling (see Section~\ref{sec:domains}), we used a simple two-zone model where stars were rotationally divided into a core and an envelope, each of which rotated as if it were a solid-body; after the core-envelope decoupling timescale, $\tau_{\rm cpl}$, had passed, the star resumed rotating as one solid body with the new value for the total angular momentum.

For our disk models, we adopted the limiting case of strong star-disk coupling.  Therefore, if a star is tagged with a disk in our simulations, its period is locked to its initial value for the lifetime of the disk, at which point it is released and allowed to change freely.

YREC natively has support for the Kawaler~(1988) formulation of the wind law, as modified by Sills et al.~(2000). Our prescription takes the form

\begin{equation}
 \displaystyle \frac{dJ}{dt} = \left\{
	\begin{array}{l l}
	 \displaystyle f_{\rm K} K_{\rm w} \omega_{\rm crit}^2 \omega \left(\frac{R}{R_{\odot}}\right)^{\frac{1}{2}} \left(\frac{M}{M_{\odot}}\right)^{- \frac{1}{2}} \quad \omega > \omega_{crit} \\
	 \displaystyle f_{\rm K} K_{\rm w} \omega^3  \left(\frac{R}{R_{\odot}}\right)^{\frac{1}{2}} \left(\frac{M}{M_{\odot}}\right)^{-\frac{1}{2}} \quad \omega \leq \omega_{\rm crit} \\
	\end{array} \right. 
	\label{eqn:Kawaler}
\end{equation}
where $f_{\rm K}$ is a constant factor used to scale the loss law to reproduce the solar rotation at the solar age, and $\omega_{\rm crit}$ is the saturation threshold.

\subsection{Wind Law Calibration} \label{sec:windcalib}

With our wind models, we wanted to be able to generate self-consistent evolutionary tracks throughout our age range.  In particular, we wanted to be able to reproduce the evolution of stars when massive gaseuous accretion disks become rare.  For such stars, stellar winds are the major angular momentum loss agent.  We calibrated the torque by requiring that a successful model reproduce the observed spindown in young open cluster stars.  This approach has been extensively employed in the literature (see, e.g., Krishnamurthi et al.~1997; MacGregor \& Brenner~1991; Gallet \& Bouvier~2013).  These goals are part of why we selected the age reference points we did: the age of M37 is much longer than the inferred core-envelope decoupling timescales for stars (Denissenkov et al.~2010), and stars are either on or close to the main sequence at the age of the Pleiades.

Angular momentum loss is a strong function of both mass and spin rate.  Therefore, since we want a wind law that is consistent with the empirical data, we adopt a procedure that is commonly used in the literature to calibrate a median fit to a solar calibrator, and an upper envelope/fast rotator fit for the saturation threshold.

To meet the requirements stated above, we attempted to fit the modified Kawaler wind law described in Section~\ref{sec:models} to our data.  In order to get the fitting constant $f_{\rm K}$, we first ran a solar-mass calibrator model which, starting with a 6.4 day rotation period at 13 Myr, was required to match the solar luminosity and rotation period at the solar age; the assumed properties of this model are listed in Table~\ref{table:solarcalib}.  The initial period was chosen to match the mean period of the slow rotator peak in h Per.  For solid-body models, we found a best-fit $f_{\rm K}$ = 2.562, while for models including angular momentum transport, the best fit $f_{\rm K}$ was 20.418.  The fitting constant is larger for the differentially rotating case because of core-envelope decoupling:  while the surface is able to spin down very quickly initially, the core represents a large reservoir of angular momentum that comes into play once the envelope and core recouple (see Pinsonneault et al.~1989).  Other physical processes, such as wave-driven transport or magnetic fields, could be operating and would produce behavior between these limiting cases.  We then took the 90th percentile period in four evenly spaced, 0.1~$M_{\odot}$ wide mass bins from 0.3 to 0.7 solar masses and created a grid of tracks in mass-$\omega_{\rm crit}$ space.  We then fit these tracks separately to the Pleiades and M37 to find $\omega_{\rm crit}$.

We then took the median absolute deviation (MAD) of the period of the 100th-80th percentile periods in the Pleiades and M37 with respect to the 90th percentile periods.  We calculated our figure of merit for each $\omega_{\rm crit}$ model track in each mass bin and each cluster:

\begin{equation}
\frac{1}{MAD^2} \frac{1}{\sqrt{N_*}} \ln \left(\frac{P_{\rm med}}{P_{\rm model}} \right)^2 ,
\label{eq:FOM}
\end{equation}

where $P_{\rm med}$ is the median period in the mass bin, $P_{\rm model}$ is the period for the model track corresponding to the particular mass bin, and $N_*$ is the number of stars in the mass bin.  The minimum in this figure of merit corresponded to our best fit $\omega_{\rm crit}$ for that cluster and mass bin.  

From this process, we constructed a single Rossby-scaled wind law fit to the shoulder of the fast rotator population seen around $0.6-0.7 M_{\odot}$.  We fit the wind law to the shoulder because that is where the fast rotators start to disappear in M37, so that area is the most sensitive to the saturation threshold.  We found a best-fit $\omega_{\rm crit}$ of $11.03_{-0.58}^{+0.51} \omega_{\odot}$ for the solid-body case and one of $3.82_{-0.32}^{+0.25}\omega_{\odot}$ for the angular momentum transport/differentially rotating case when fitting to M37, and $22.3_{-1.8}^{+1.9}\omega_{\odot}$ and $6.57_{-0.59}^{+0.80}\omega_{\odot}$, respectively, when fitting to the Pleiades.  The error bars were calculated by running tracks one MAD above and below the fast rotator median period and finding their best-fit $\omega_{\rm crit}$s as well.  What is clear is that using one Rossby-scaled wind law to simultaneously fit the Pleiades and M37 is not possible; although this result is obtained only at the $3\sigma$ level or better for the transport case, it is obtained at better than the $5\sigma$ level in the solid-body case.  We use our global wind law, fit to M37, as our wind law for all of the bulk cluster simulations presented in Section~\ref{sec:toPleiades}.  We further discuss the discrepancy between the wind laws fit to the Pleiades and M37, as well as the assumption of Rossby scaling in general, in Section~\ref{sec:toM37}.

\begin{deluxetable} {lll}
\tabletypesize{\scriptsize}
\tablecolumns{2}
\tablewidth{0pt}
\tablecaption{Solar Calibrator Properties}
\startdata
$t_{\rm end}$  &  4.568~Gyr  \\ 
$P_{\rm final}$  &  25.4~days  \\ 
$Y$  &  0.271  \\ 
$Z$  &  0.0182  \\
Mixing Length & 1.933 \\
$f_K$ & 2.562 (solid-body)\\
\enddata
\label{table:solarcalib}
\end{deluxetable}

\section{Results} \label{sec:results}

With our models, we wish to do two things:  first, we want to test the initial conditions of star formation in clusters, and second, we want to test the general assumption of the literature that clusters form a single evolutionary sequence.  Additionally, testing the initial conditions as cleanly as possible means two things:  first, it means we must begin modeling in the youngest systems possible; and second, that we must establish a regime where our models are least sensitive to the input physics, so that the initial conditions have the greatest leverage on the final result.  We find in Section~\ref{sec:domains} that looking at the lowest-mass stars (below $0.5M_{\odot}$) offers the cleanest test of the initial conditions.

We test whether clusters form a single evolutionary sequence by starting at a single cluster, the ONC, and seeing whether a fit of stellar tracks to one cluster at a given age also fits another cluster when evolved to a different age.  We first look at the case of evolving the ONC to h Per, and then reset, evolving h Per to the Pleiades age (Section~\ref{sec:toPleiades}), and subsequently to the age of M37 (Section~\ref{sec:toM37}).  We find that the slowly rotating population can indeed be fit as a single sequence, unbroken from the ONC to M37, but that the rapidly h Per rotating stars cannot be fit simultaneously to the Pleiades and M37 using a single Kawaler wind law prescription.  We argue in Section~\ref{sec:rapidrot} that this is potentially the result of cosmic variance in the rapid rotator fraction in different environments.

\subsection{Protostar-disk interactions:  ONC to h Per} \label{sec:bimodality}

\begin{figure}
\includegraphics[scale=0.45]{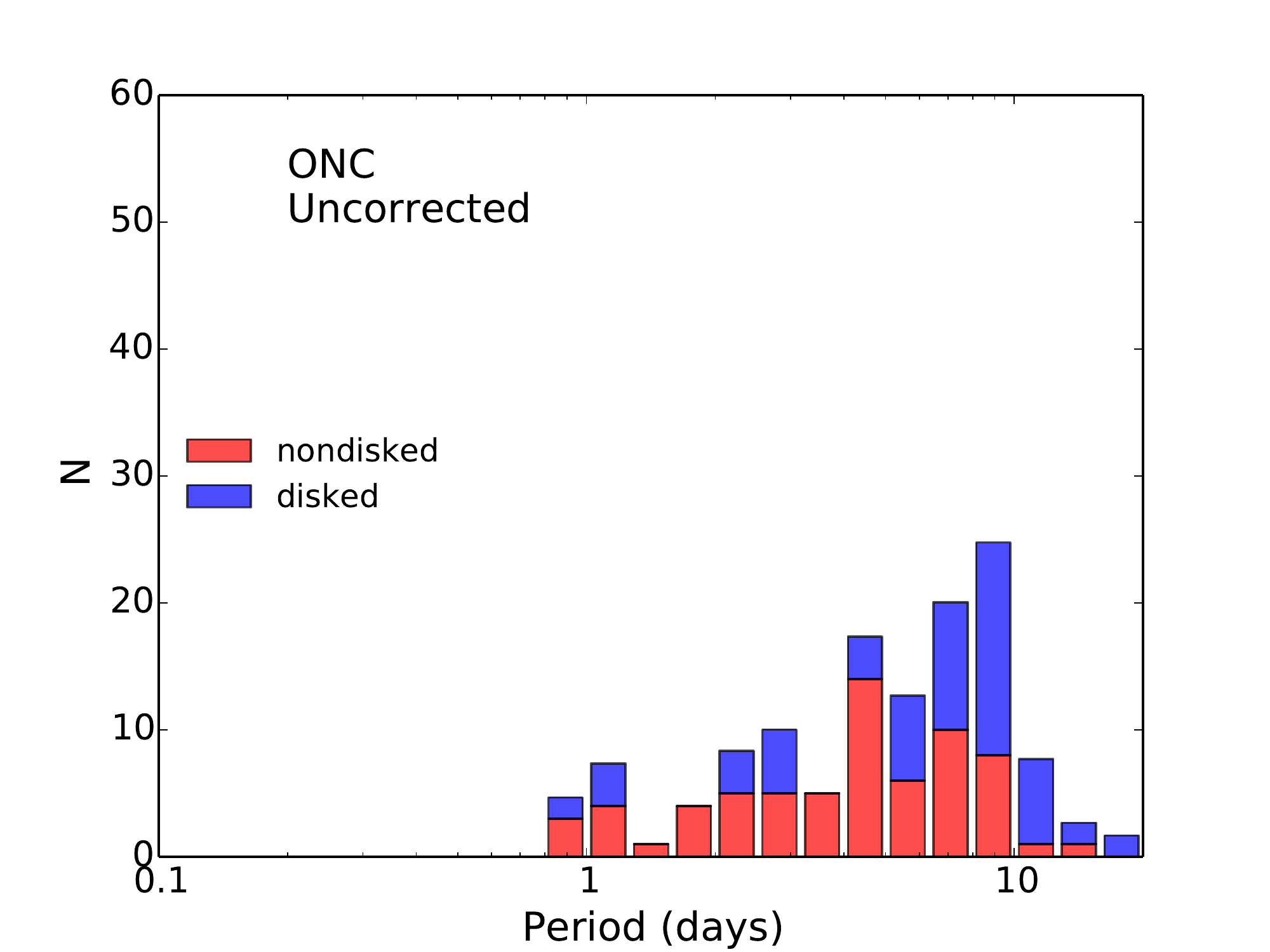}
\includegraphics[scale=0.45]{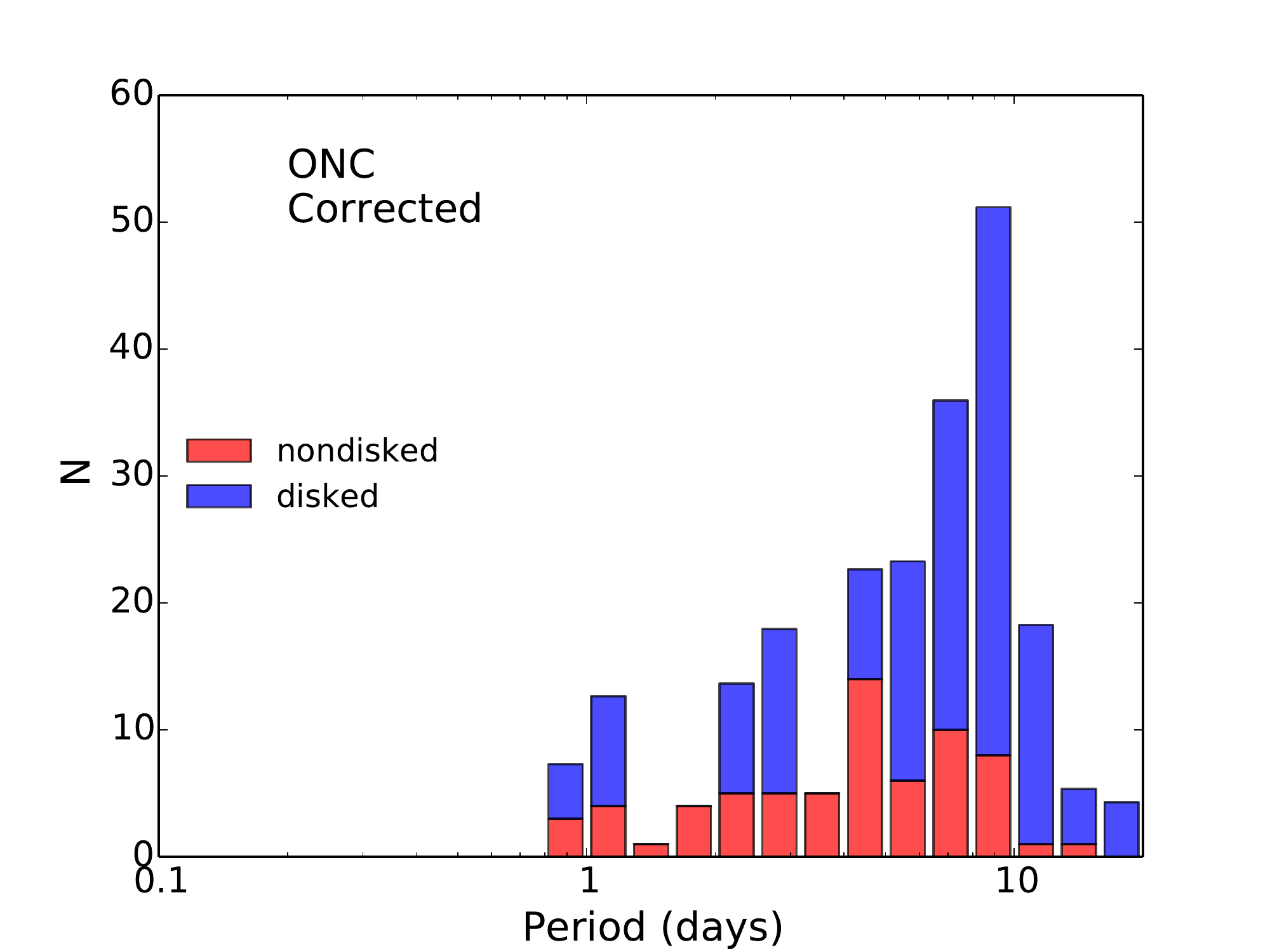}
\caption{Histograms of the unevolved Rebull et al.~(2006) sample.  As described in Section~\ref{sec:data}, we only used the subset with mass estimates.  This subsample has a disk fraction of 34\%, while the full sample has a disk fraction of 57\%.  To remove this bias, we weighted the disked stars more heavily when creating our histograms.  The top panel shows the uncorrected distribution, while the bottom panel shows the corrected one.  The red represents non-disked stars, while the blue represents disked stars.}
\label{fig:Rebull_hist}
\end{figure}

It is interesting that the h Per period distribution is bimodal.  At masses greater than $\sim0.25M_{\odot}$, the ONC data is not (Rebull et al.~2006; but, see, e.g., Herbst et al.~2002), so the mechanism which creates this bimodality must act or start to act sometime between 1 and 10 Myr.  Figure~\ref{fig:Rebull_hist} shows a histogram of those stars in the Rebull sample with mass estimates; no bimodality is visible.  The h Per distribution is strongly bimodal, with more slow rotators than fast.  To generate such a distribution, there must be some differentiating mechanism by which one population of stars loses large amounts of angular momentum, while another loses very little.

\begin{figure}
\includegraphics[scale=0.45]{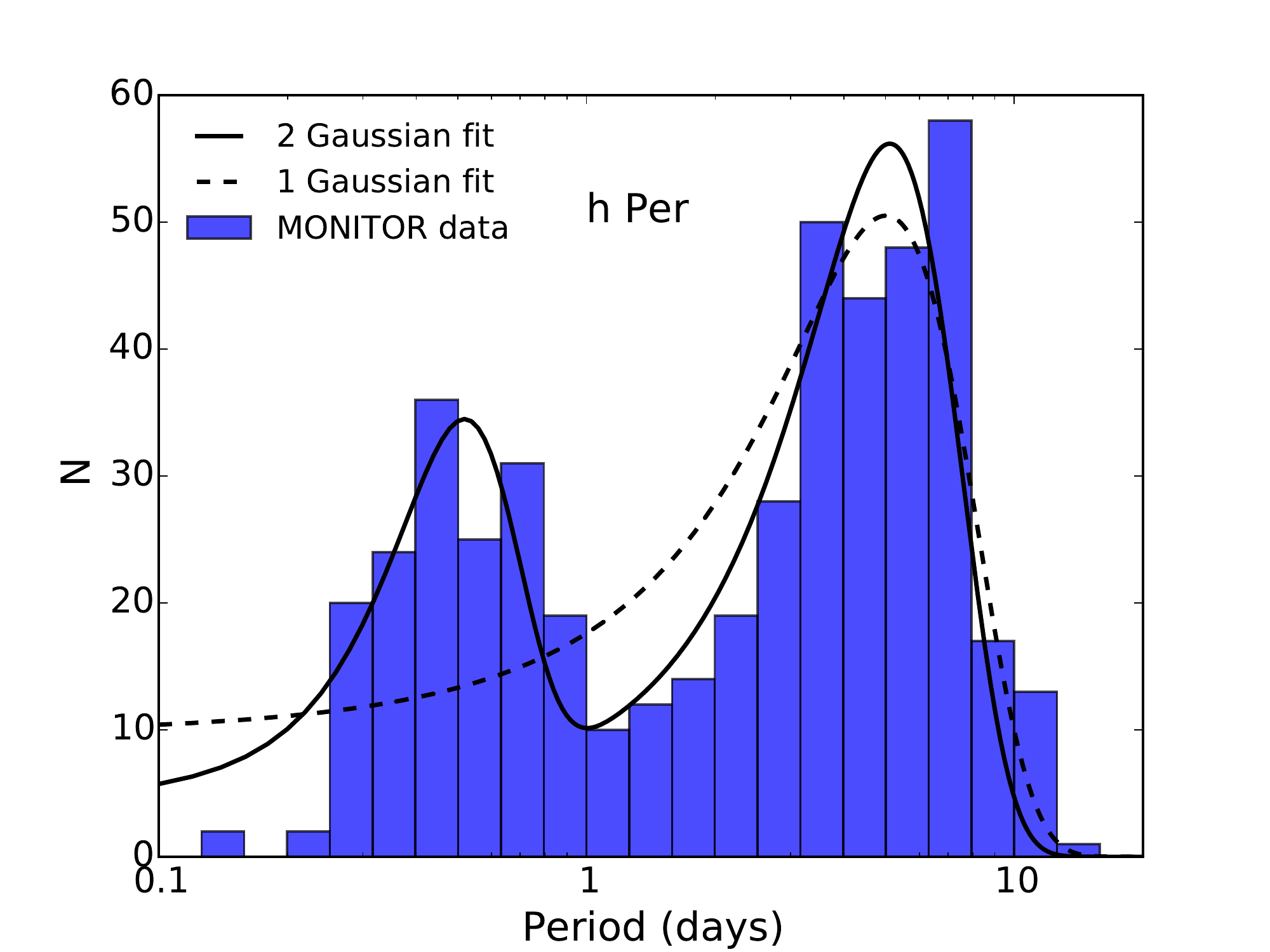}
\includegraphics[scale=0.45]{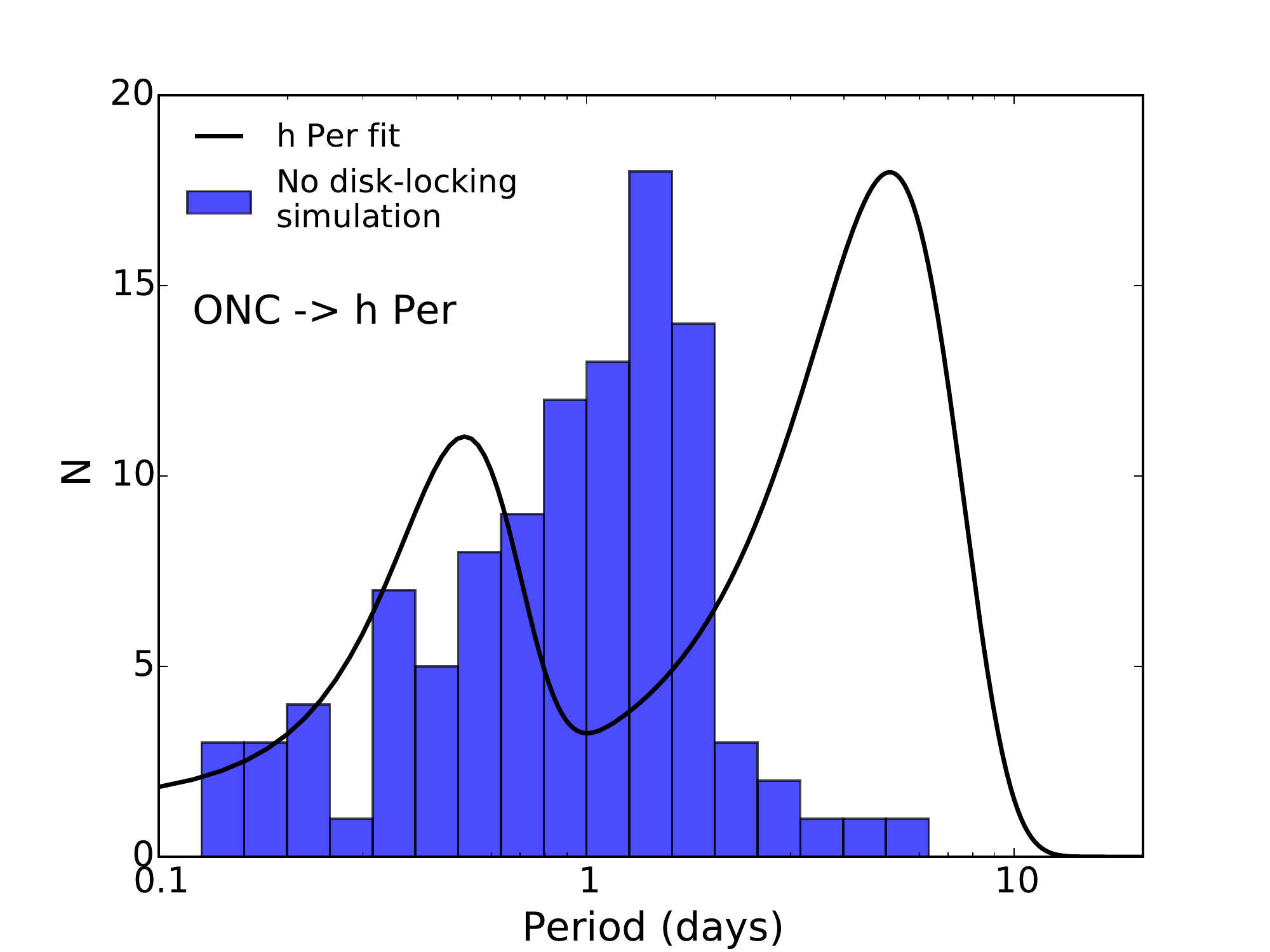}
\caption{Histogram of h Per in the top panel, using the periods from Moraux et al.~(2013).  The period distribution is evidently strongly bimodal; to be judged a success, any attempt to evolve the ONC to this age should reproduce a distribution which looks at least something like this one.  The bottom panel shows the results of models evolving the ONC stars forward to the age of h Per using no disks or other loss mechanisms.  These models generate stars that on average spin much faster than the observed period distribution.  Unlike the other figures in this section, the distribution has not been scaled in any way.}
\label{fig:hPer_hist}
\end{figure}

Disks are a promising candidate for this mechanism, as their presence is relatively easily detected, and even by the age of the ONC, there appears to be a sizeable portion of stars which have lost their disks.  The disk fraction in our full ONC sample is 57\%, while the disk fraction in the sample with mass estimates is just 34\%.  We gave additional weight to the disked stars in making our histograms to eliminate this bias; the effects of this can be seen in the second panel of Figure~\ref{fig:Rebull_hist} and all subsequent histograms in this section.  As described in Rebull et al.~(2006), the periods of the disked stars are almost all longer than about two days, while the periods of the non-disked stars extend further below one day (see Figure~3 in their paper).  For comparison, a histogram of the h Per period distribution is shown in the top panel of Figure~\ref{fig:hPer_hist}.

Disk-locking has previously been proposed as a mechanism for angular momentum loss in ONC-age stars (Edwards et al.~1993; Shu et al.~1994; Rebull et al.~2006; Cieza \& Baliber~2006), although other authors have disputed its importance (e.g., Stassun et al.~1999). Because disk-locking can easily create a population of slow rotators, it is a promising candidate for causing the bimodality.  Given that, we wanted to test the current disk-locking hypotheses with respect to how long disks last and the distribution of disk lifetimes.  Therefore, as a control case, we ran models starting from the ONC with no disk-locking or other angular momentum loss mechanism.  The results of these models are shown in the bottom panel of Figure~\ref{fig:hPer_hist}.  It is easy to see that the stars end up with periods that are far too fast compared to observations, even at an age of 10-13~Myr.  For example, the slowest star in our models has a period of 6~days at 13~Myr and only 1.8~days at 120~Myr, compared to the upper envelope of around 10~days seen in both h Per and the Pleiades.  Given this, some form of disk-locking or other loss is necessary to explain the evolution of young pre-main-sequence stars.

We then added disk-locking back into our models.  We used a uniform distribution of disk-locking lifetimes.  Starting from the disked stars in the Rebull dataset, we generated a grid of models using their initial periods and a range of disk lifetimes from 1-12~Myr.  For each star, we then sampled the grid 1000 times and scaled the resulting distribution down to the correct number of stars and disk fraction from the Rebull sample.  We sampled three disk lifetime ranges using this process:  1-6~Myr, 1.5-9~Myr, and 2-12~Myr.  We chose these ranges to sample a variety of claimed ages for the ONC and disk-locking timescales from other authors; e.g., Tinker, et al.~(2002) claims a disk-locking timescale of no more than 6~Myr, while results from Bell, et al.~(2013) and Somers \& Pinsonneault~(2015) suggest that the ages of young open clusters given in the literature may be too young by as much as a factor of two or more, albeit through different mechanisms.  This implies that disks may be older than previously claimed as well.

\begin{figure}
\includegraphics[scale=0.45]{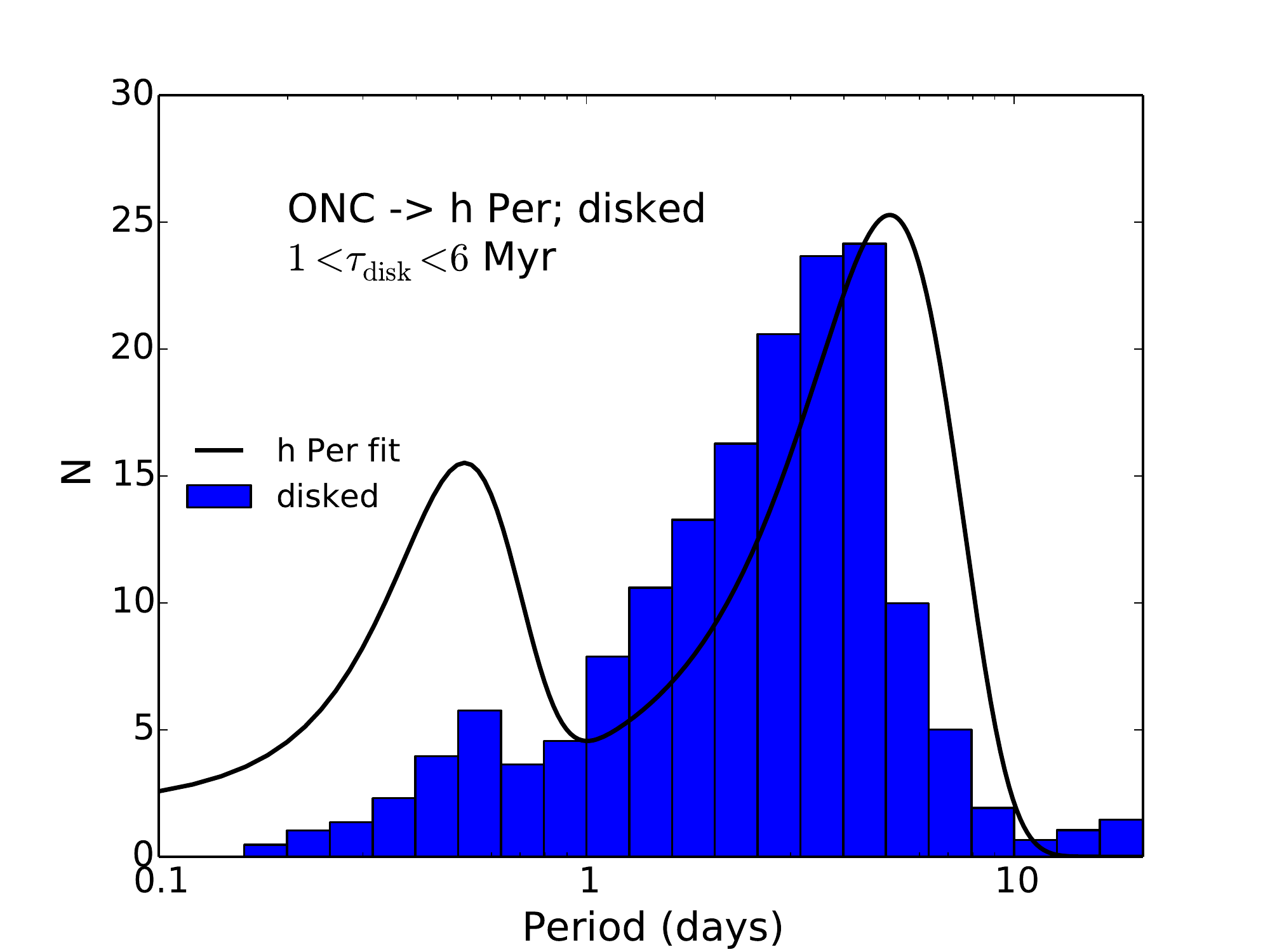}
\includegraphics[scale=0.45]{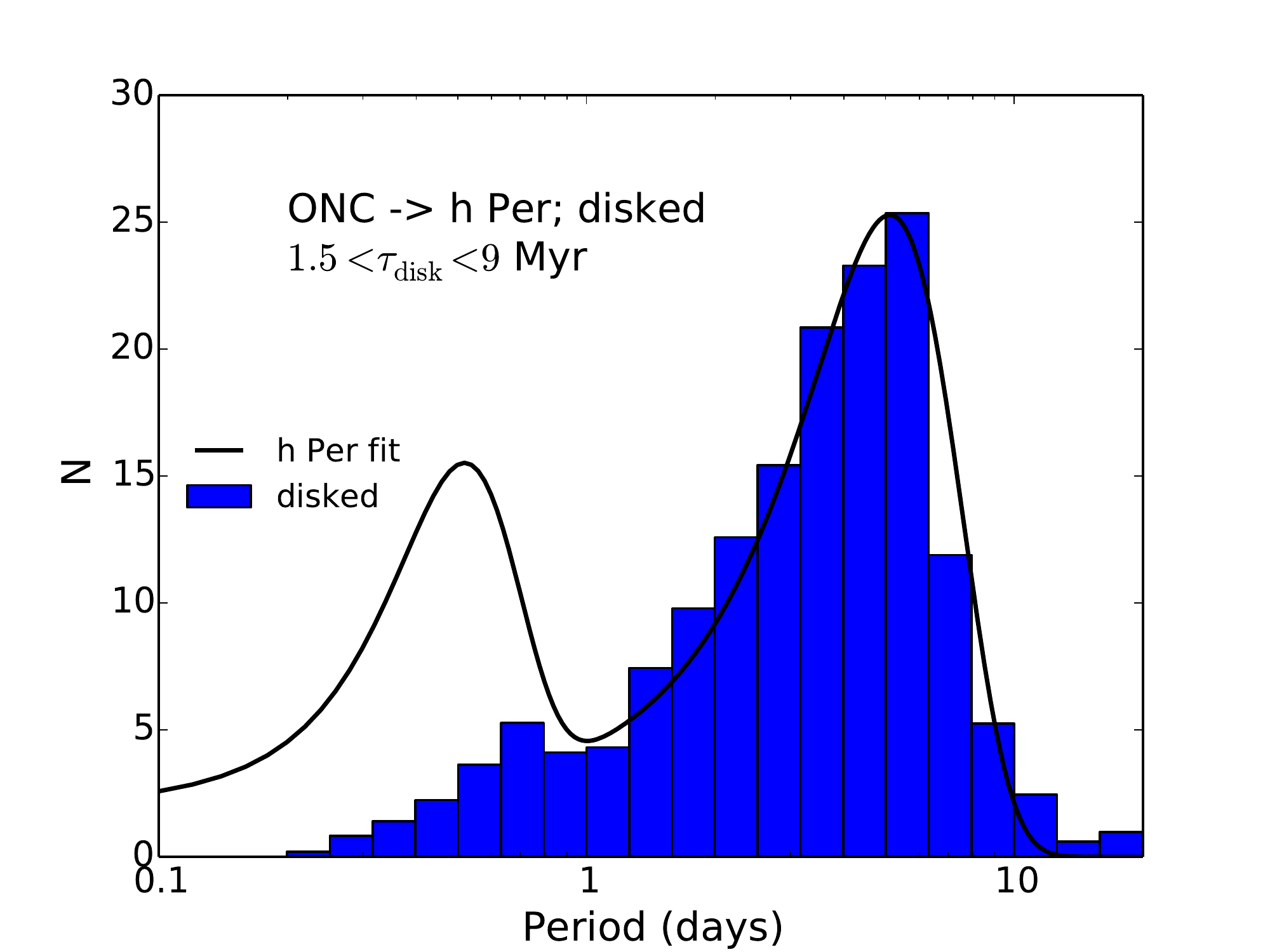}
\includegraphics[scale=0.45]{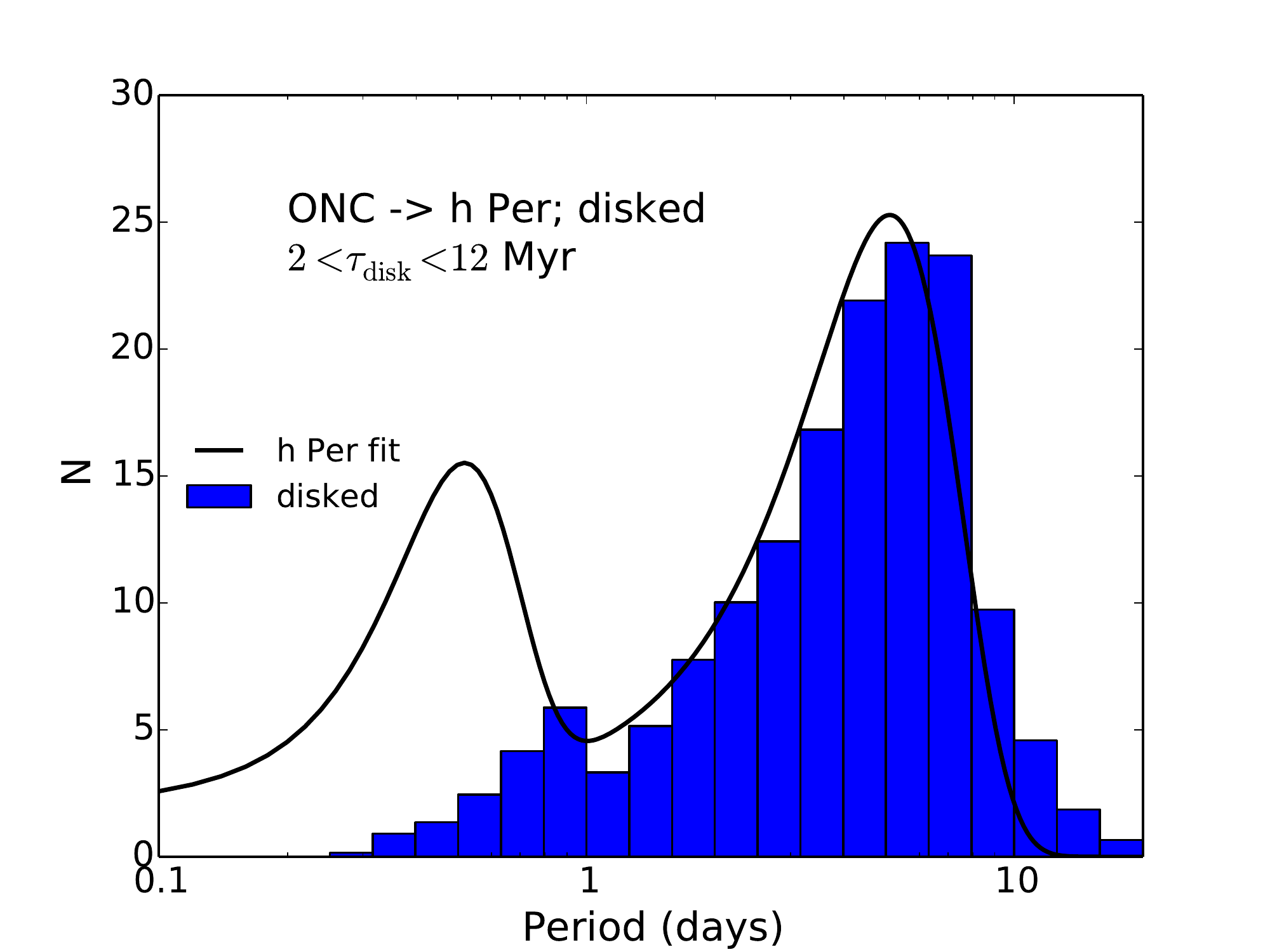}
\caption{Histograms of the scaled disked star distribution for our ONC models, shown at the age of h Per.  The black line, as before, is the Double Gaussian fir to the h Per period distribution from Figure~\ref{fig:hPer_hist}, scaled to the size of the ONC models peak, so as to provide a good comparison.  The simulation results match the slow rotator peak from h Per quite well, and provide a good test of the disk-locking timescale.  Anything longer than a maximum disk lifetime of 12~Myr does not match the observed distribution.}
\label{fig:diskfrac}
\end{figure}

The results of these models are shown in Figure~\ref{fig:diskfrac}, which shows only the disked stars.  We find that disk-locking provides a very natural way to reproduce the slowly rotating peak in h Per.  As seen in the figure, simply by changing the disk-locking timescale, the peak of the period distribution can be moved around to match the cluster, and the distributions appear to have roughly the same shape as well.  This is only weak evidence in favor of the disk-locking model, but it does show that our models have no problem reproducing the slow rotators in h Per starting from the ONC.

\begin{figure}
\includegraphics[scale=0.45]{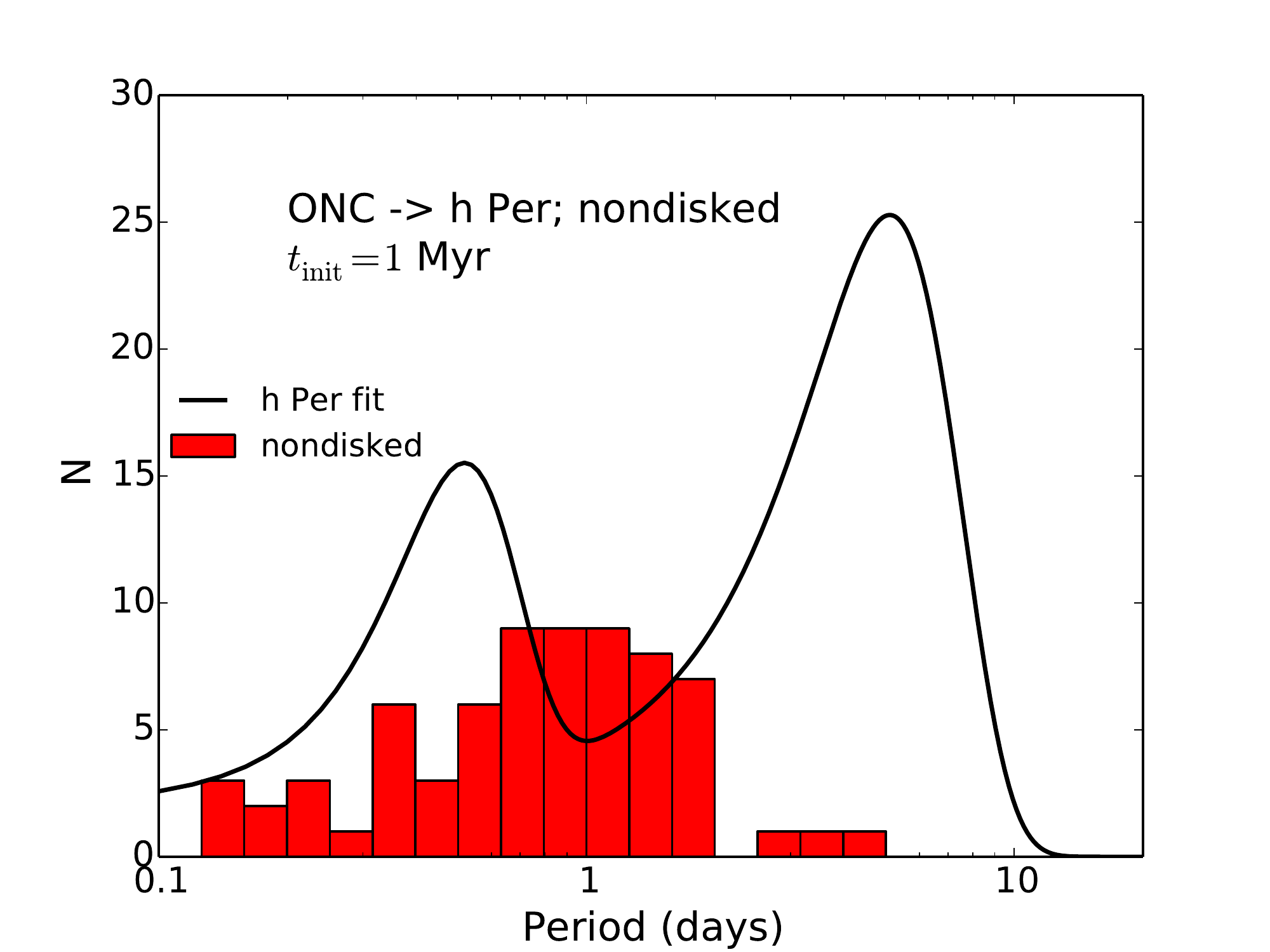}
\includegraphics[scale=0.45]{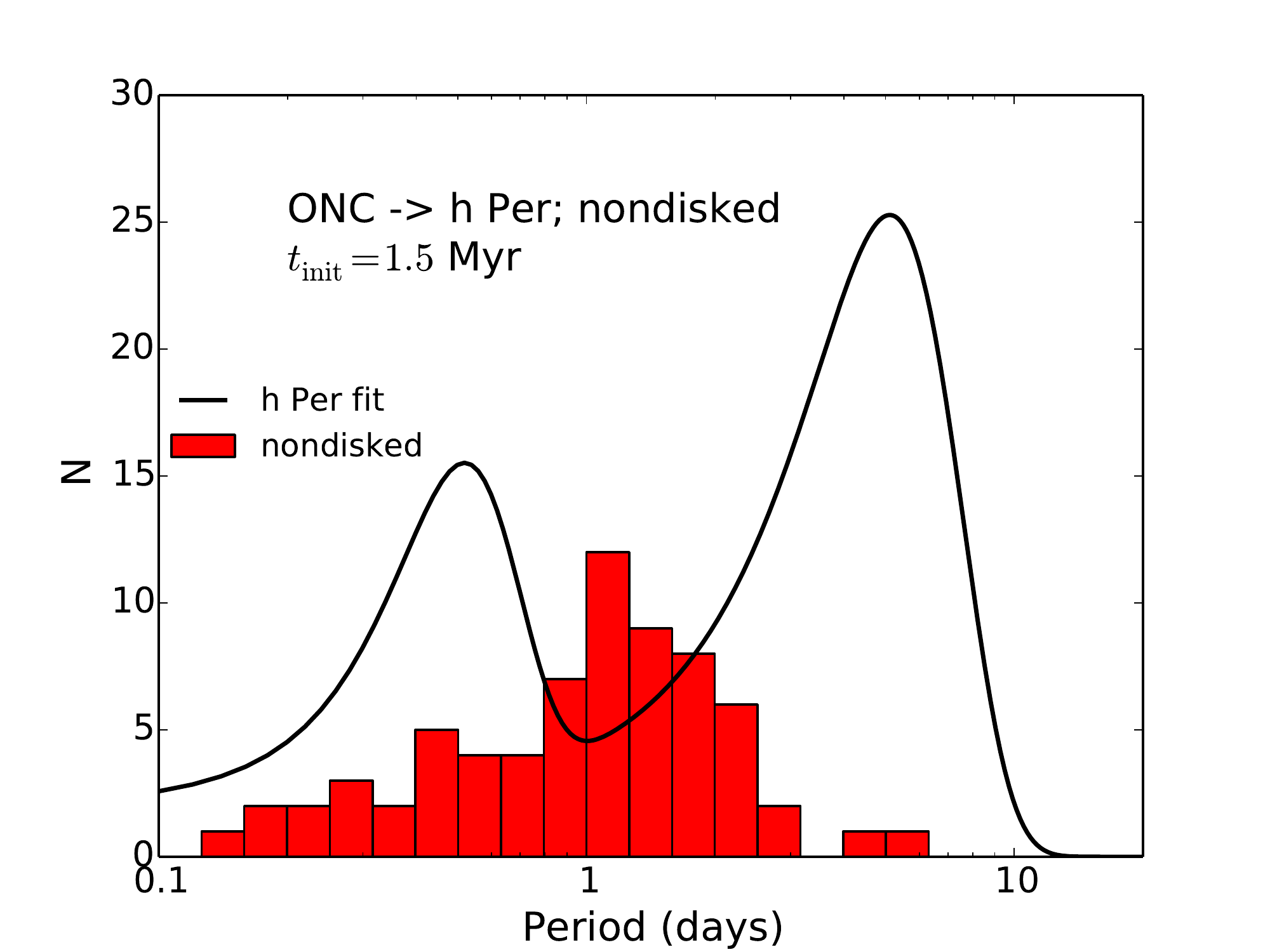}
\includegraphics[scale=0.45]{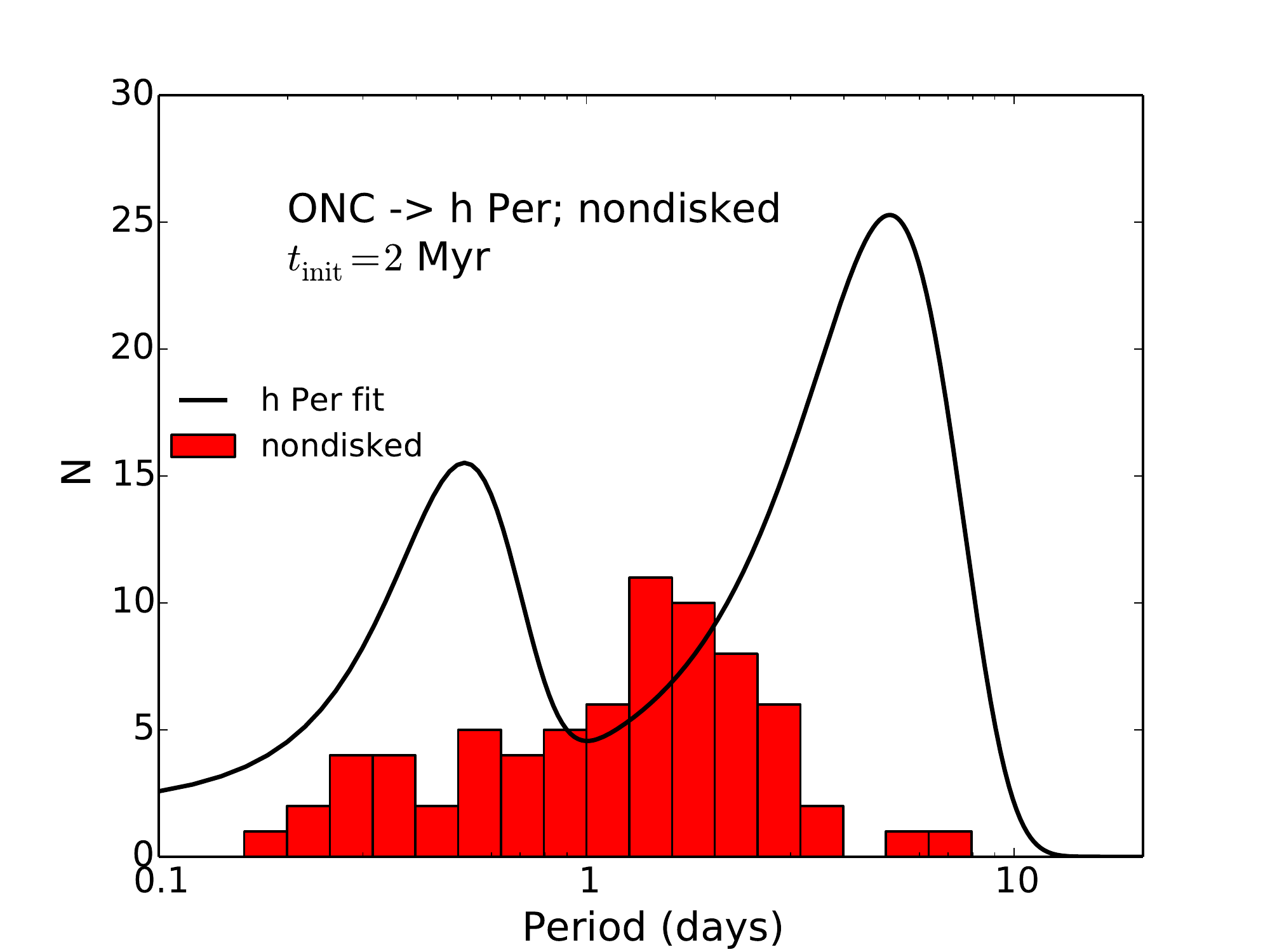}
\caption{Histograms of the nondisked star distribution for our ONC models, shown at the age of h Per.  The black line is the same fit shown in previous figures.  The top panel shows a 1~Myr age for the ONC, the middle panel a 1.5~Myr age, and bottom panel a 2~Myr age.  None of those simulations provide a match for the h Per fast rotator peak.}
\label{fig:nondisked}
\end{figure}

\begin{figure}
\includegraphics[scale=0.45]{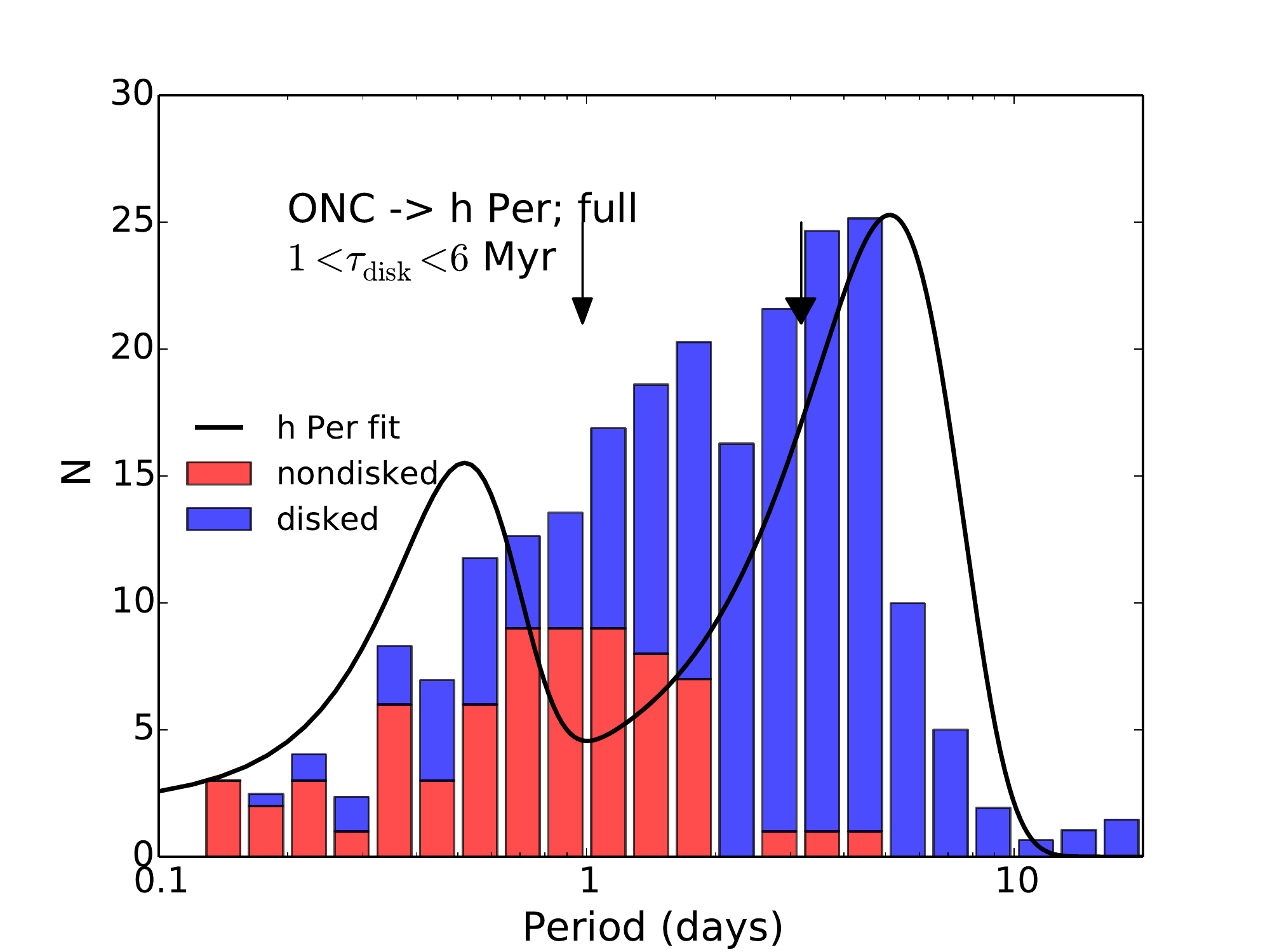}
\includegraphics[scale=0.45]{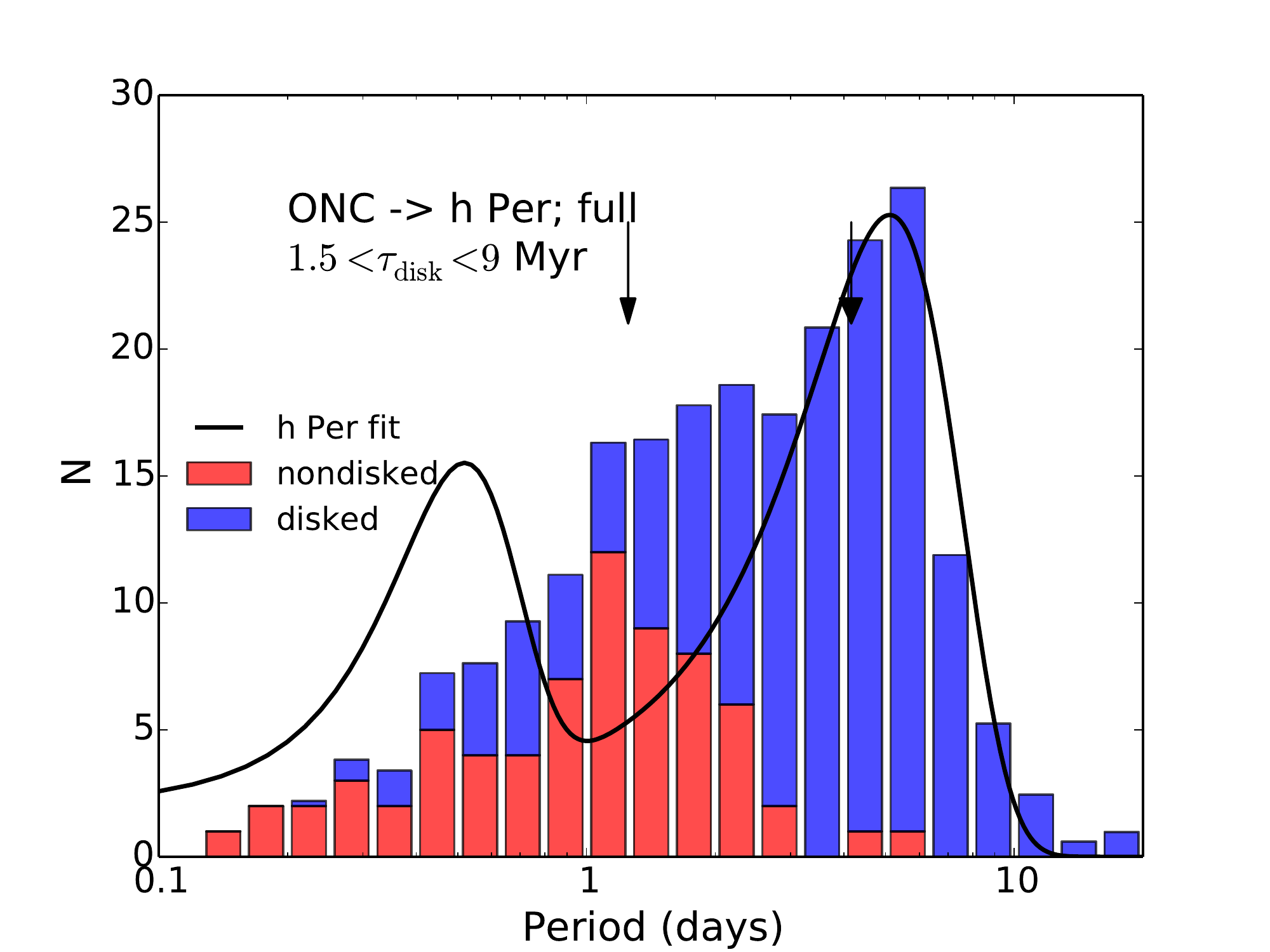}
\includegraphics[scale=0.45]{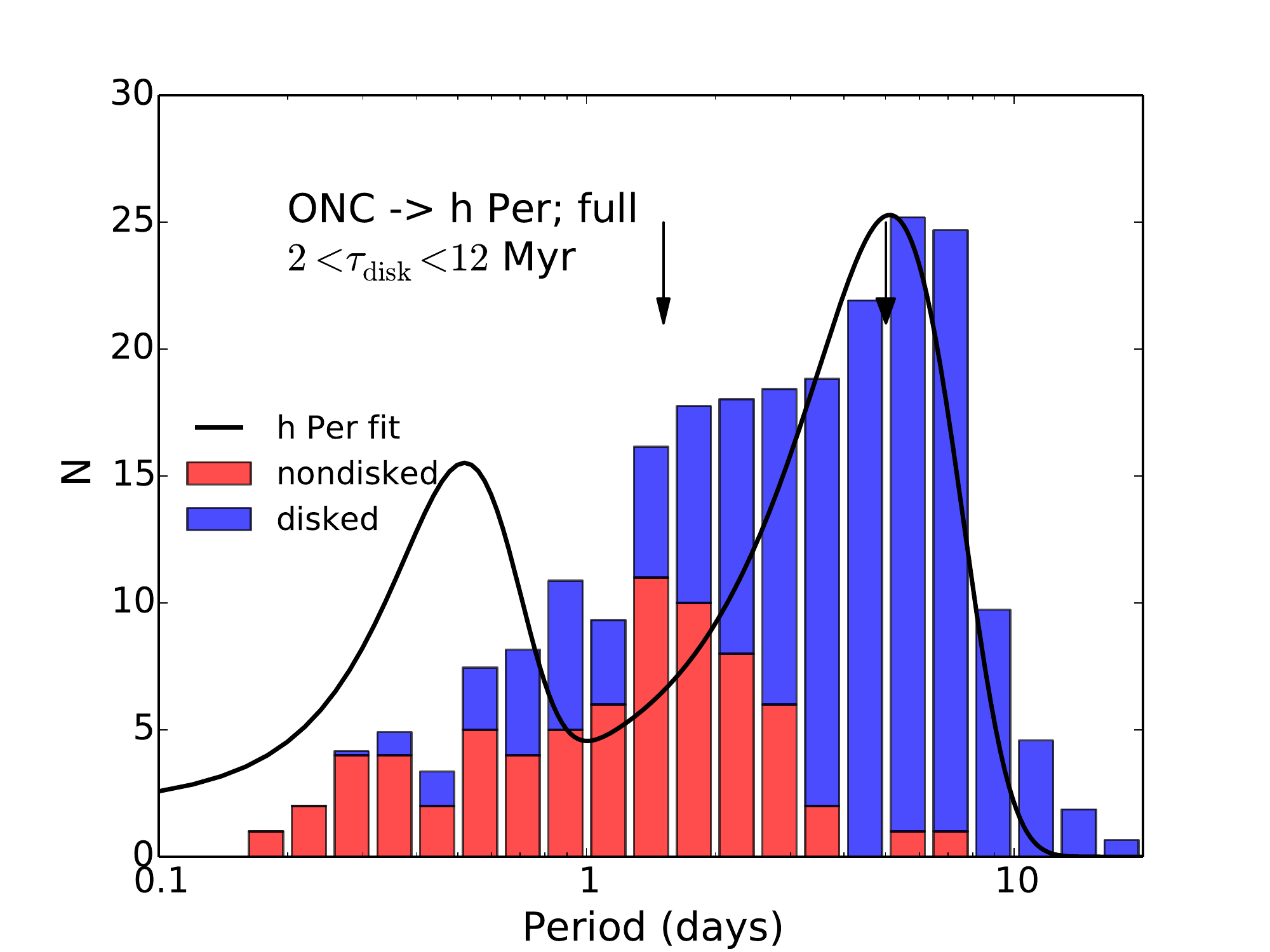}
\caption{Histograms of the evolution of ONC stars from Rebull et al.~(2006) to the age of h Per.  The red bars show the non-disked stars, while the blue bars show the disked stars.  The black arrows show the mean period of the disked and non-disked populations, while the thick black line shows our double Gaussian fit to the h Per period distribution from Figure~\ref{fig:hPer_hist}; it has been scaled to provide a useful comparison.  The top panel uses a uniform distribution of disk-locking timescales from 1-6~Myr and a 1~Myr starting age, the middle panel uses a 1.5-9~Myr distribution and 1.5~Myr starting age, and the bottom panel uses a 2~Myr starting age and 2-12~Myr disk lifetime distribution.  No bimodality is seen in the overall evolved distribution, and the non-disked stars do not appear to correspond to the h Per fast rotator peak, as they do not spin up enough.  A maximum disk-locking timescale of 12~Myr is favored for the slow rotators.}
\label{fig:Rebullflat}
\end{figure}

We find that the fast rotators are not matched nearly so simply.  In fact, we did not reproduce the fast rotator peak of h Per starting from any population of stars from the ONC.  If disk-locking is correct and the bimodality seen in h Per is generic, the fast rotators should come from stars which lose their disks very early on.  We therefore used the non-disked ONC stars to model them. We varied their age exactly as we did for the disked stars, assuming that they were 1~Myr, 1.5~Myr, or 2~Myr old.  The results of these models are shown in Figure~\ref{fig:nondisked}.  It is clear that in no case do the non-disked stars in the ONC spin up enough to match the h Per fast rotator peak.  Instead, the peak always resides in the ``no-man's land'' between the two peaks, with a long tail towards short periods.  We combine the results of the disk and non-disk simulations in Figure~\ref{fig:Rebullflat}.  In no case do we recover an obviously bimodal distribution.  If our models are good depictions of the stellar physics at work, this result argues that cosmic variance plays a larger role in the rotational periods of low-mass stars than previously thought, even very early in their evolution.  This is particularly striking becuase these models are fully convective, so core envelope decoupling is not a factor, and they are also very young, so angular momentum loss is also not a factor for these uncoupled stars.

\subsection{From h Per to the Pleiades} \label{sec:toPleiades}

In order to test the initial conditions as cleanly as possible, we must figure out where the input physics provide the least leverage on the final result.  In Section~\ref{sec:domains}, we do this by turning on and off the various switches in our models and comparing the results through KS tests.  We find that the lowest-mass stars are ideal test-beds for analyzing the initial conditions of star formation, for they are insensitive to changes in the input physics.  The initial conditions also include whether circumstellar disks are important past the commonly assumed maximum lifetime of $\sim10$~Myr.  We find in Section~\ref{sec:slowrotators} that such long-lived disks are not necessary to explain the observed rotation period distributions.

\subsubsection{Domains} \label{sec:domains}

We know from observations and a long history of models that main sequence stars, even in relatively restricted mass ranges, can behave very differently from each other.  We also know that the fraction of the star's angular momentum associated with the convective zone is a function of mass.  Therefore, we expect core-envelope decoupling to be more important in higher-mass stars, as the envelope carries less and less of the angular momentum of the star.  This makes the predicted surface rotation period evolution of such stars more model-dependent.  We also know that spindown timescale for rapid rotators increases as mass decreases (Figure~\ref{fig:clusters}).  General domains exist, in order of decreasing mass:  one where angular momentum transport and loss are required to explain the observed rotation periods; one where angular momentum loss alone is sufficient; and for young enough systems with small enough predicted torques, there could also be one where we need neither loss nor non-solid-body rotation to explain the data.  We expect these domains to exist because the depth of the convection zone changes with mass.  The lower the mass of the star, the deeper the convection zone gets, and thus the more efficient angular momentum transport becomes.  Solid-body rotation represents the limiting case of infinitely efficient angular momentum transport.  As stellar mass decreases, winds also become less efficient because the convection zone represents a larger and larger reservoir of angular momentum, making it harder for them to slow the star down.

\begin{figure*}
\centerline{
\includegraphics[scale=0.45]{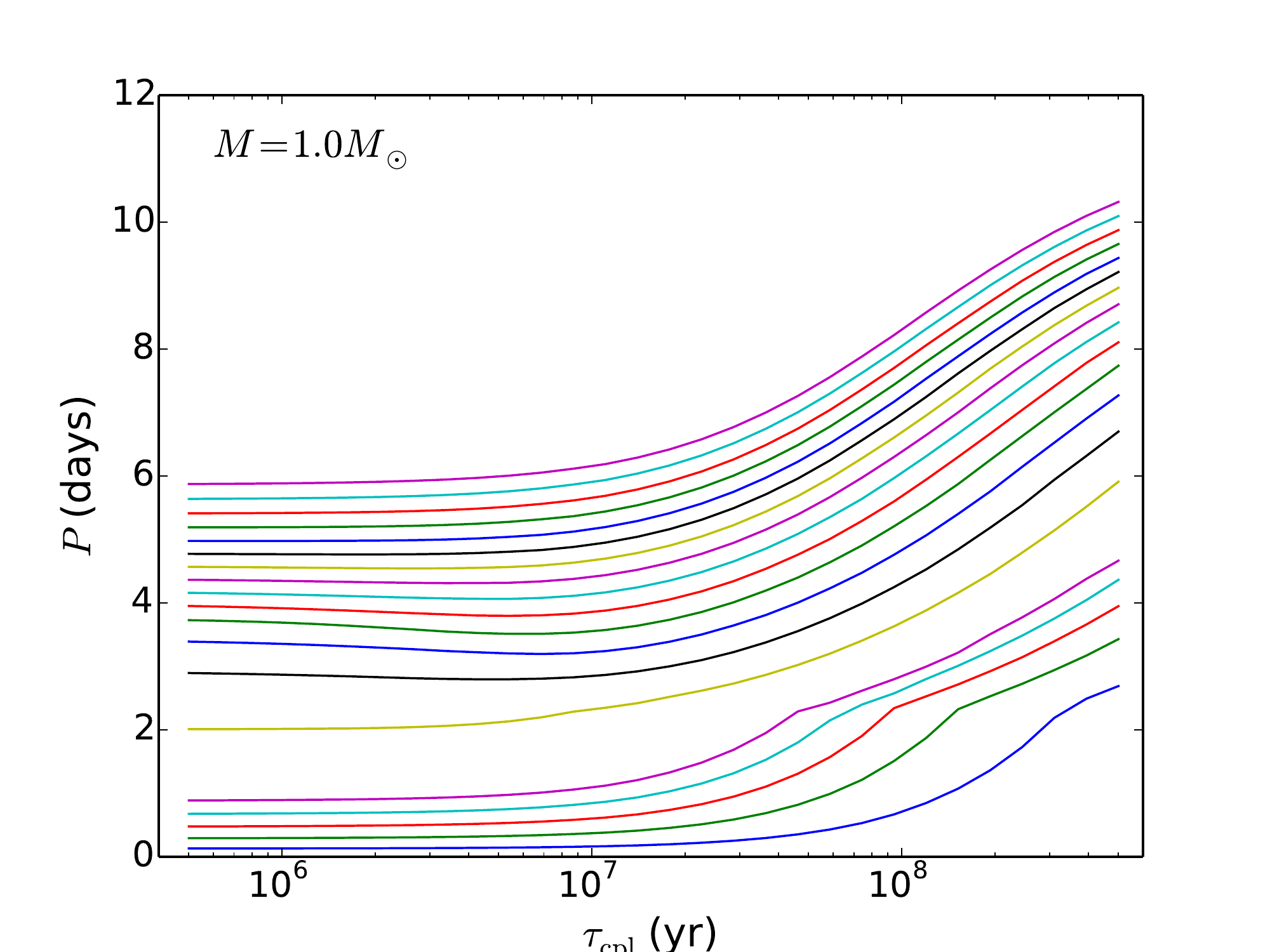}
\includegraphics[scale=0.45]{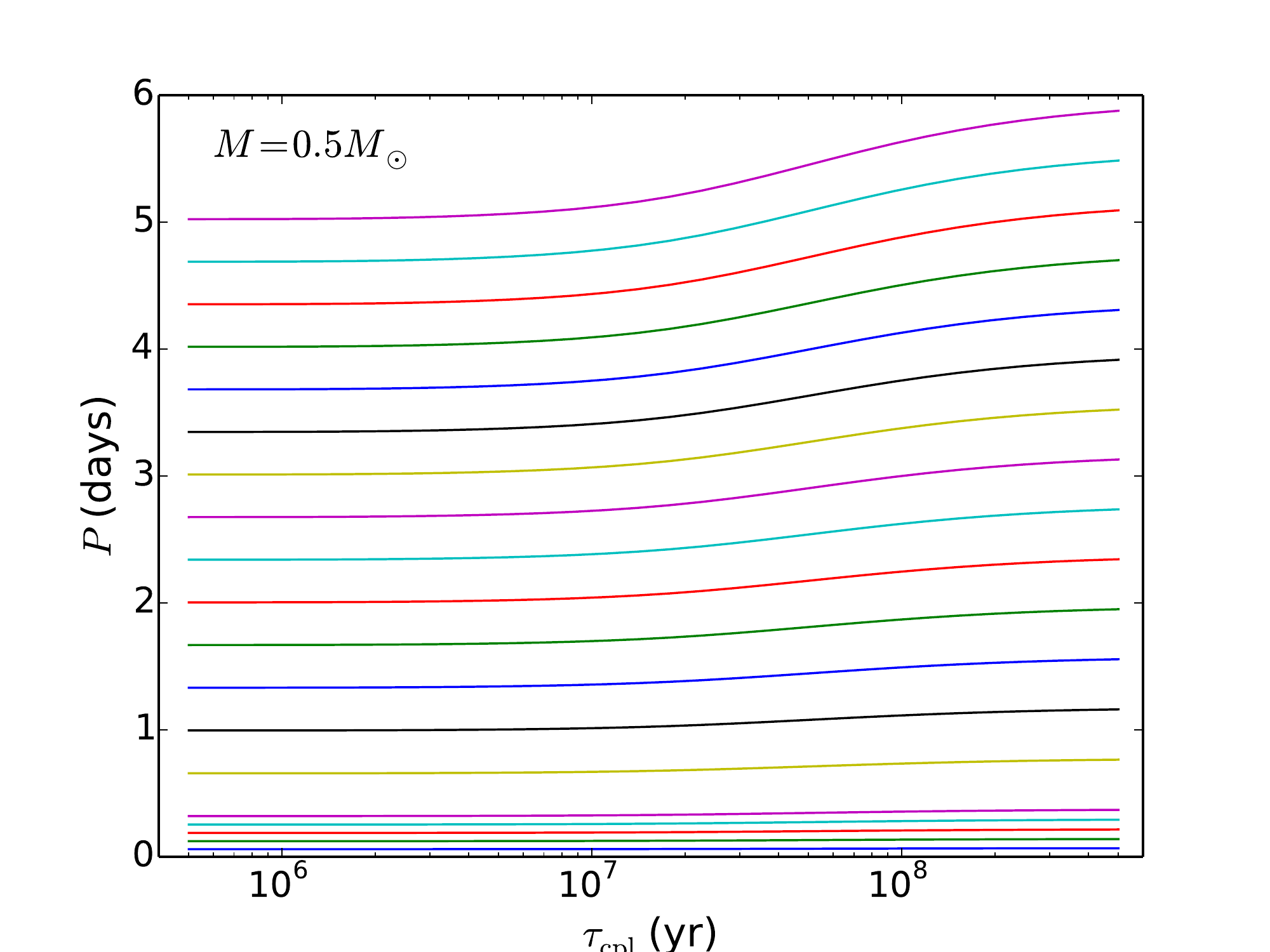}
}
\caption{Slices of our core-envelope decoupling timescale / initial period grid.  Stars were assigned initial periods of 0.2, 0.4, 0.6, 0.8, 1, 2, 3, ... , 15 days, and evolved from an age of 12 Myr to the age of the Pleiades (120 Myr).  Given the very low effect of changing the core-envelope decoupling timescale on the final period of the low-mass stars, we ignored it for that mass range in our subsequent models.}
\label{fig:core-envgrid}
\end{figure*}

To find what relevance core-envelope decoupling had to our models, we made mass-period grids of model stars that were 12~Myr old and evolved them to 120~Myr old, the age of the Pleiades.  We did this over the entire mass range we were concerned with; two sample slices are shown in Figure~\ref{fig:core-envgrid}.  We studied stars over a period range from 0.2 to 15~days and a core-envelope decoupling timescale ($\tau_{\rm cpl}$) ranging from 5~Myr to 500~Myr in 30 logarithmically-spaced increments.  Figure~\ref{fig:core-envgrid} shows the end results of those simulations; each line corresponds to a star with a given initial period.  As the mass of the star increases, the importance of core-envelope decoupling increases as well.  This can be seen in the figure as how far away from flat each line is; if core-envelope decoupling doesn't matter at all, then the final period of the star should not change if the core-envelope decoupling timescale does.  In Figure~\ref{fig:core-envgrid}, it is therefore clear the core-envelope decoupling can have a large effect on the rotational evolution of the solar-mass stars in the left panel, while the $0.5M_{\odot}$ stars in the right panel, especially the faster rotators, are not affected to anywhere near the same degree by modifying $\tau_{\rm cpl}$.  Gallet \& Bouvier~(2015) found that the rotation periods of low-mass stars could be matched to the data if very long core-envelope decoupling timescales were employed.  However, while they found best-fit numbers, they did not conduct sensitivity testing of the kind we perform here.  We therefore anticipate, based on these results, that their models are relatively insensitive to the adopted core-envelope decoupling timescale in this domain.

\begin{figure}
\includegraphics[scale=0.45]{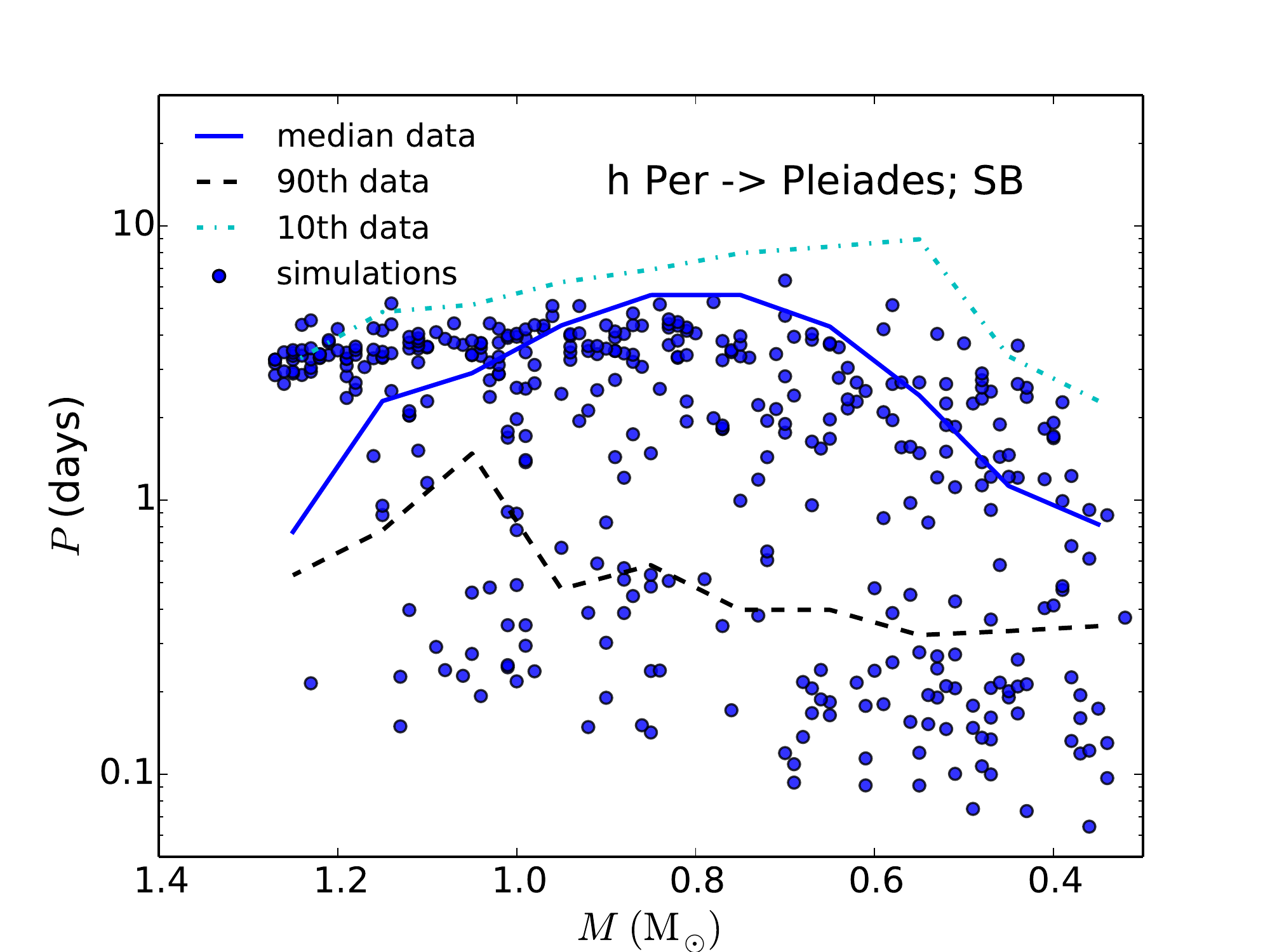}
\includegraphics[scale=0.45]{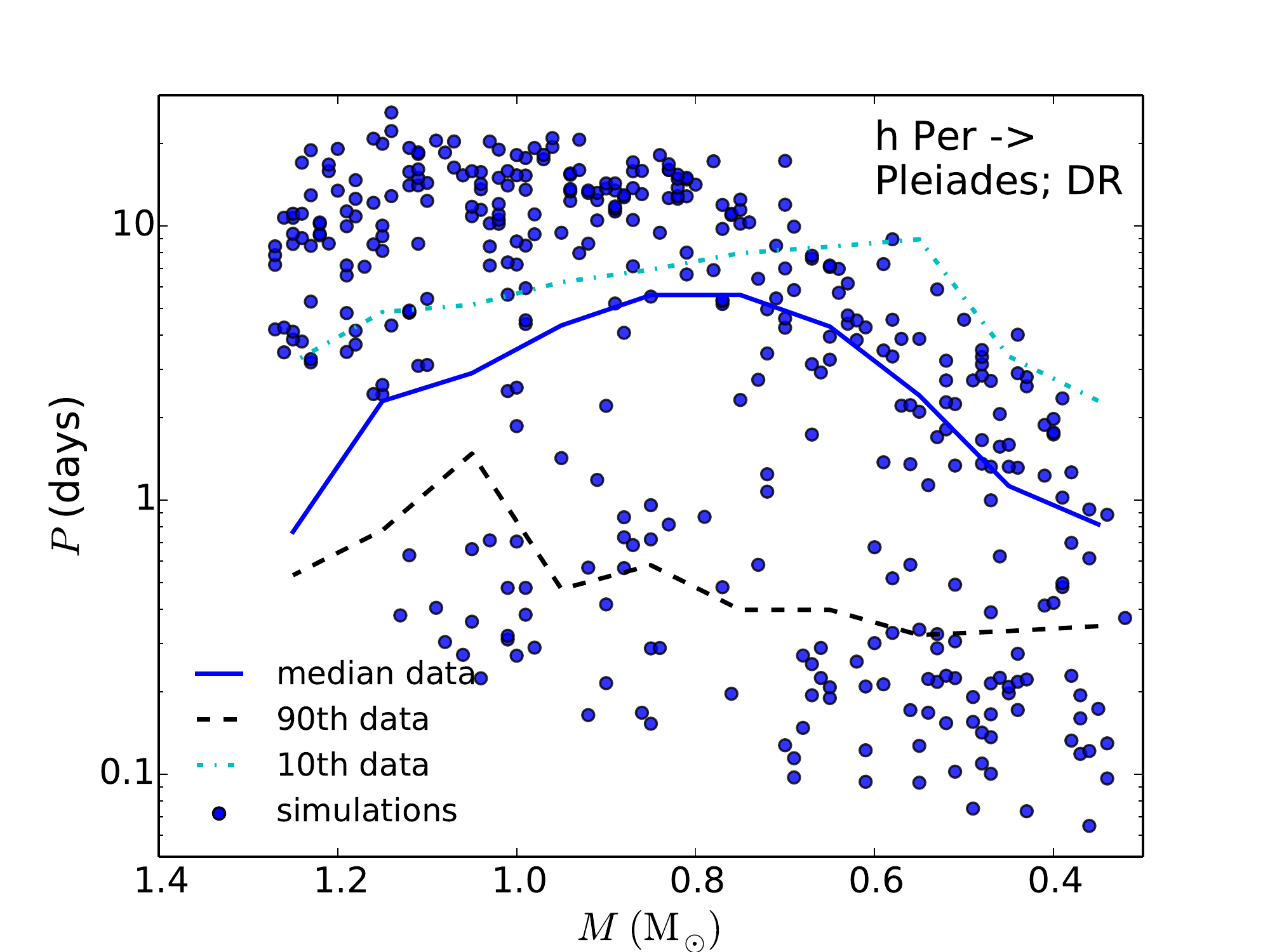}
\includegraphics[scale=0.45]{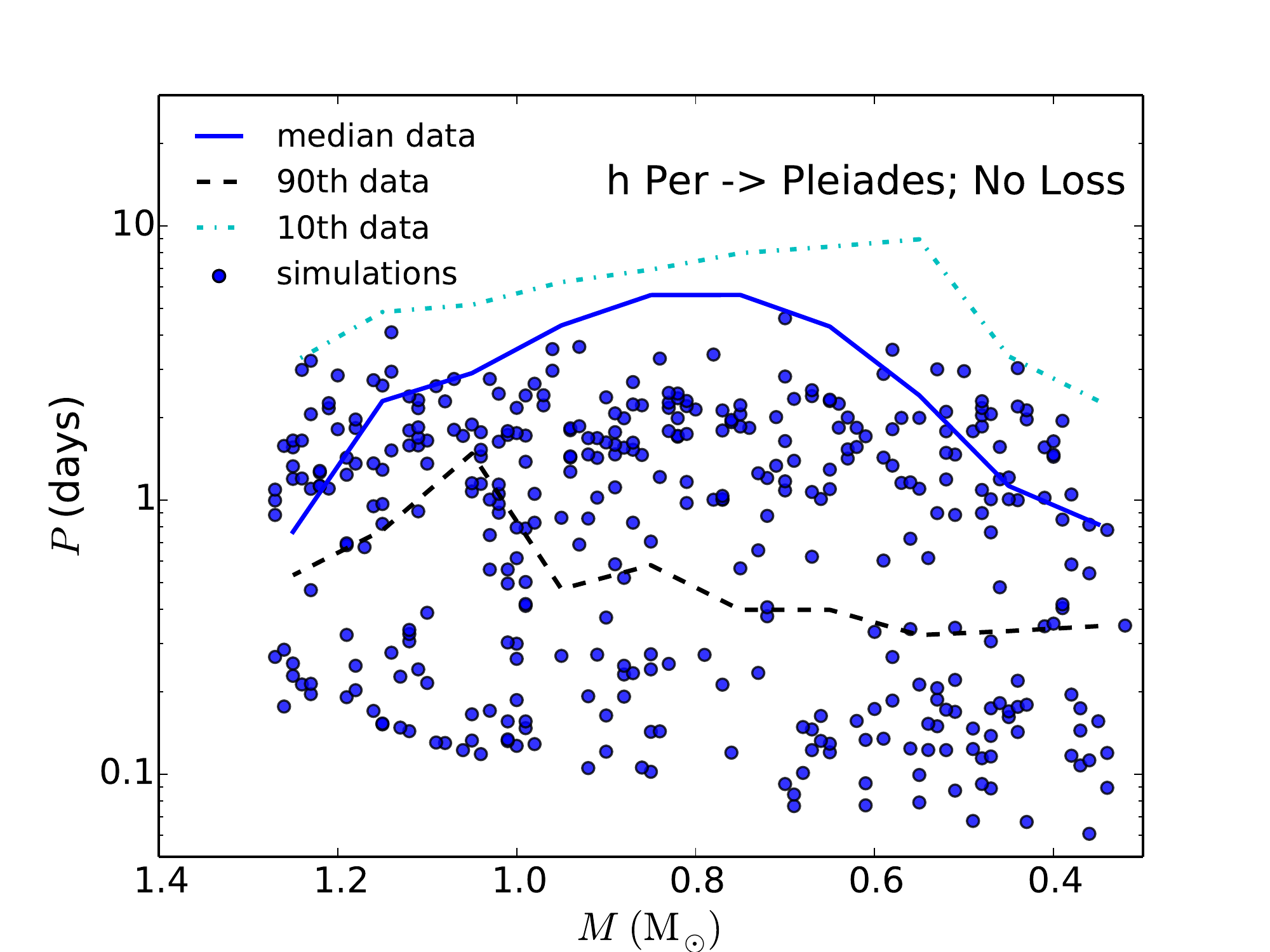}
\caption{Results of simulations running the MONITOR data for h Per forward to the age of the Pleiades.  These models use a Kawaler type wind law.  The top panel shows solid-body models, while the middle panel shows models with full angular momentum transport.  The bottom panel shows the results of running simulations with no angular momentum loss mechanisms.  The cyan dot-dashed, blue solid, and black dashed lines shown in each panel are the 10th-percentile, median, and 90th-percentile periods for each mass bin in the Pleiades dataset; the bins are each 0.1$M_{\odot}$ wide.}
\label{fig:calibratedwinds}
\end{figure}

To compare the solid-body and differentially rotating/angular momentum transport cases, we used the MONITOR data for the h Per stars and again evolved them to the age of the Pleiades.  These models excluded explicit core-envelope decoupling of the type used in the two-zone model described in Section~\ref{sec:models}, but included the calibrated Kawaler wind law described in Section~\ref{sec:windcalib}.  The results are shown in Figure~\ref{fig:calibratedwinds}.  We ran KS tests on the two distributions to find out where they matched each other, although it is clear from simply looking at the figure that the rotational evolution of stars of $0.8M_{\odot}$ and above is heavily affected by whether they are modeled as solid bodies or not.  The KS test returned a $p$-value of $>0.05$ below $\sim0.67M_{\odot}$, meaning that models with and without core-envelope decoupling are statistically consistent below that mass.  We compare these two limiting cases in the remainder of our work, and restrict our analysis to the low-mass regime where the assumed transport properties do not strongly impact our work.

We then compared these models to otherwise identical ones with no angular momentum loss at all.  The results of those simulations are shown in the third panel of Figure~\ref{fig:calibratedwinds}.  We ran KS tests for the lower mass end of the distribution, and found that a KS test returned a $p$-value of 0.05 or above below $\sim0.59M_{\odot}$.  This result means that stellar winds have only a modest impact below 0.6~$M_{\odot}$, at least until the age of the Pleiades.  However, they are important by the age of M37, so we include angular momentum loss in our models.

\begin{figure}
\includegraphics[scale=0.45]{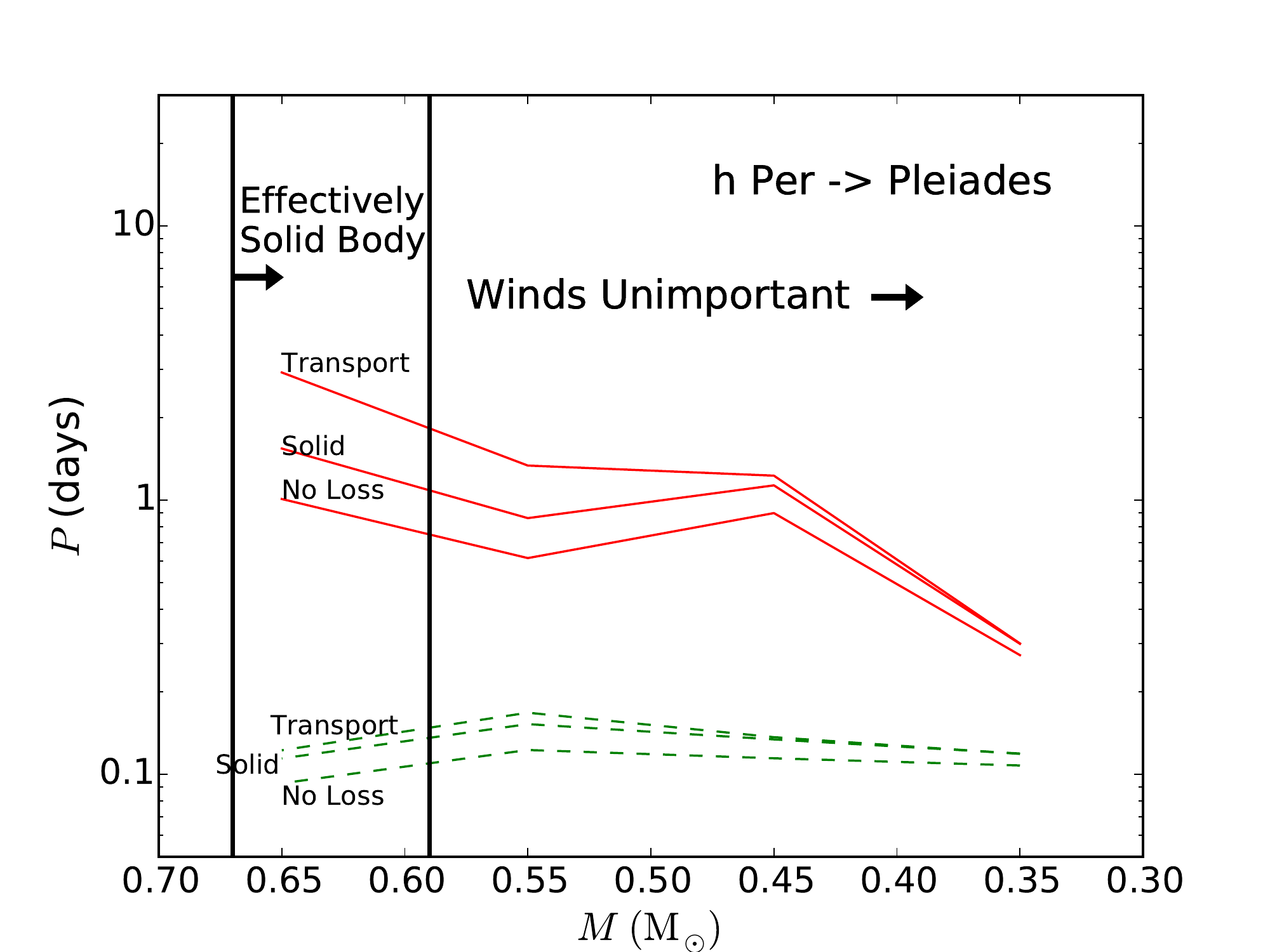}
\includegraphics[scale=0.45]{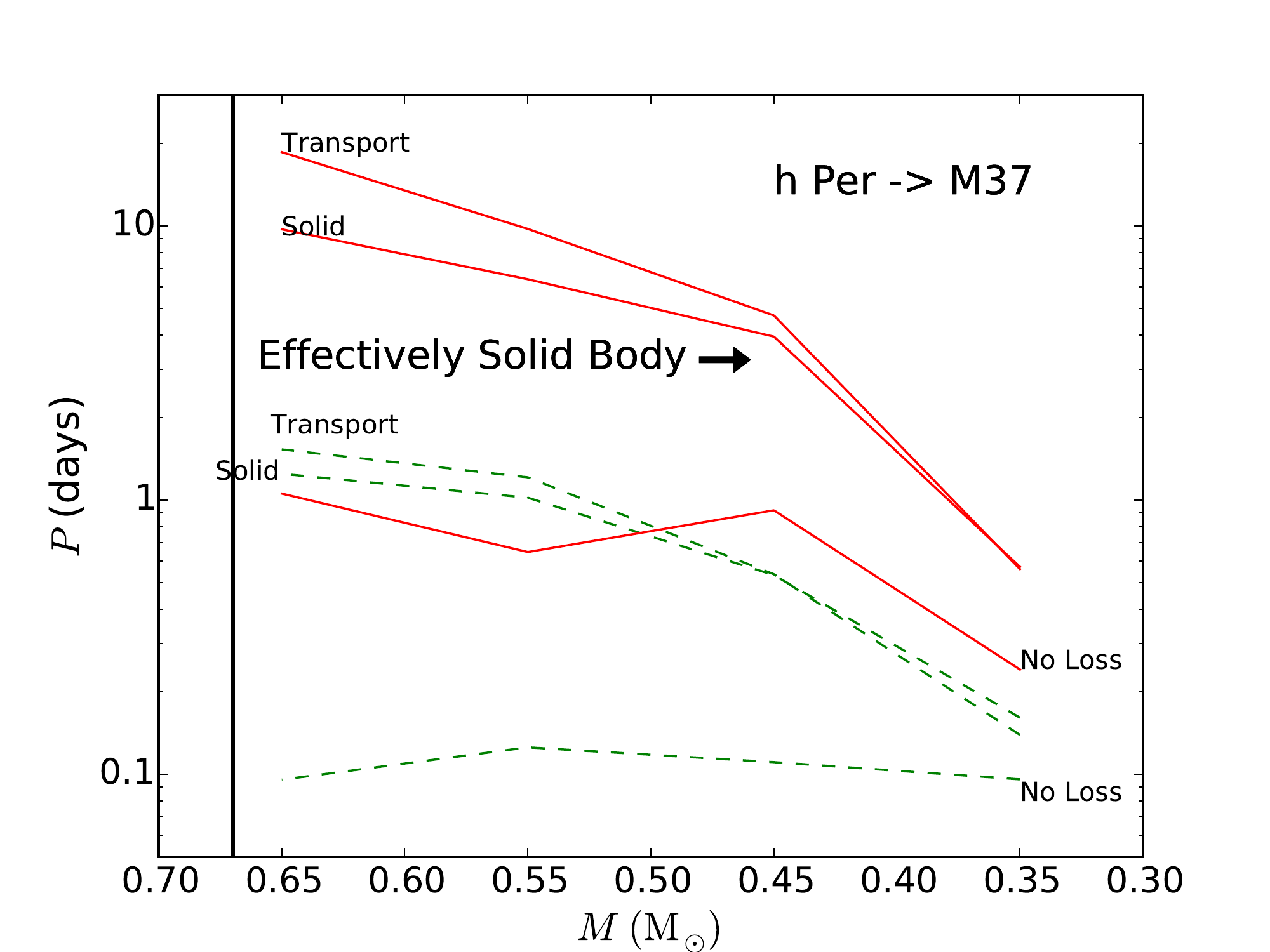}
\caption{Comparison of predictions of models with different assumptions about angular momentum loss and transport.  The red solid and green dashed lines represent the median and 90th percentile periods respectively, in our simulations.  The black vertical lines divide the parameter space into the domains discussed in Section~\ref{sec:domains}; the differential rotation/solid body division is at 0.67$M_{\odot}$ in our models, while winds are unimportant to the age of the Pleiades below 0.59$M_{\odot}$.  We determined these divisions by comparing models with the different physics using a KS test; a $p$-value of 0.05 or less was used as the breakpoint.  The top panel shows our simulations at the age of the Pleiades, while the bottom one shows them at the age of M37.}
\label{fig:losslawcomp}
\end{figure}

Figure~\ref{fig:losslawcomp} shows a further visualization of the points in this section.  At very low masses, the three different angular momentum evolution cases effectively merge at young ages.  All three are consistent with the data on the slowly rotating end, while the fast rotators present a problem.  As we move to higher and higher masses, the no loss and solid-body cases diverge further and further from the differentially rotating case, as expected.  By the age of M37, it is clear that some form of angular momentum loss is entirely necessary to match the data; the no loss case is wildly different from the other two.  Because of this behavior, the results we present below are insensitive to the exact physical model used.

\begin{figure*}
\includegraphics[scale=0.45]{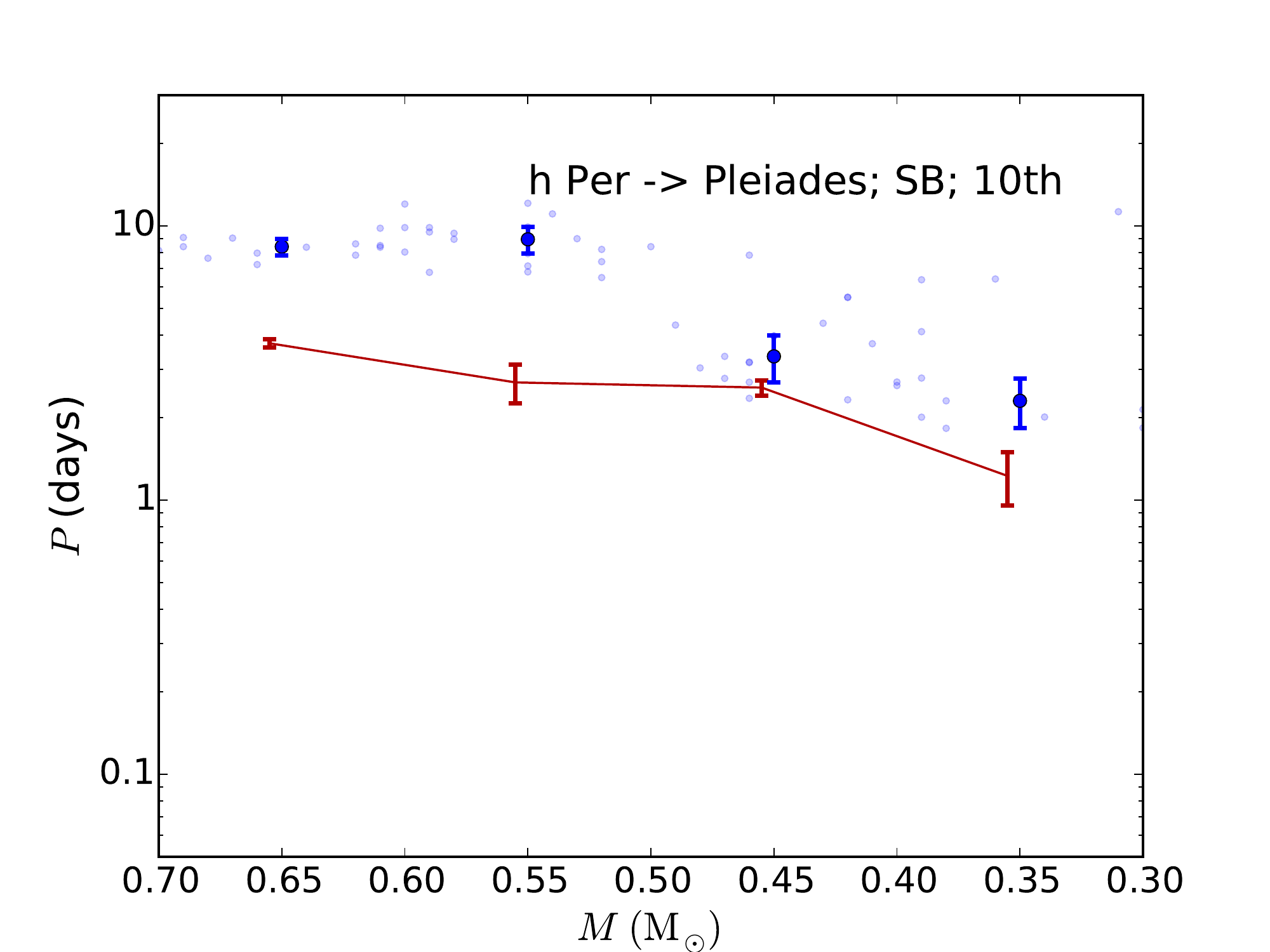}
\includegraphics[scale=0.45]{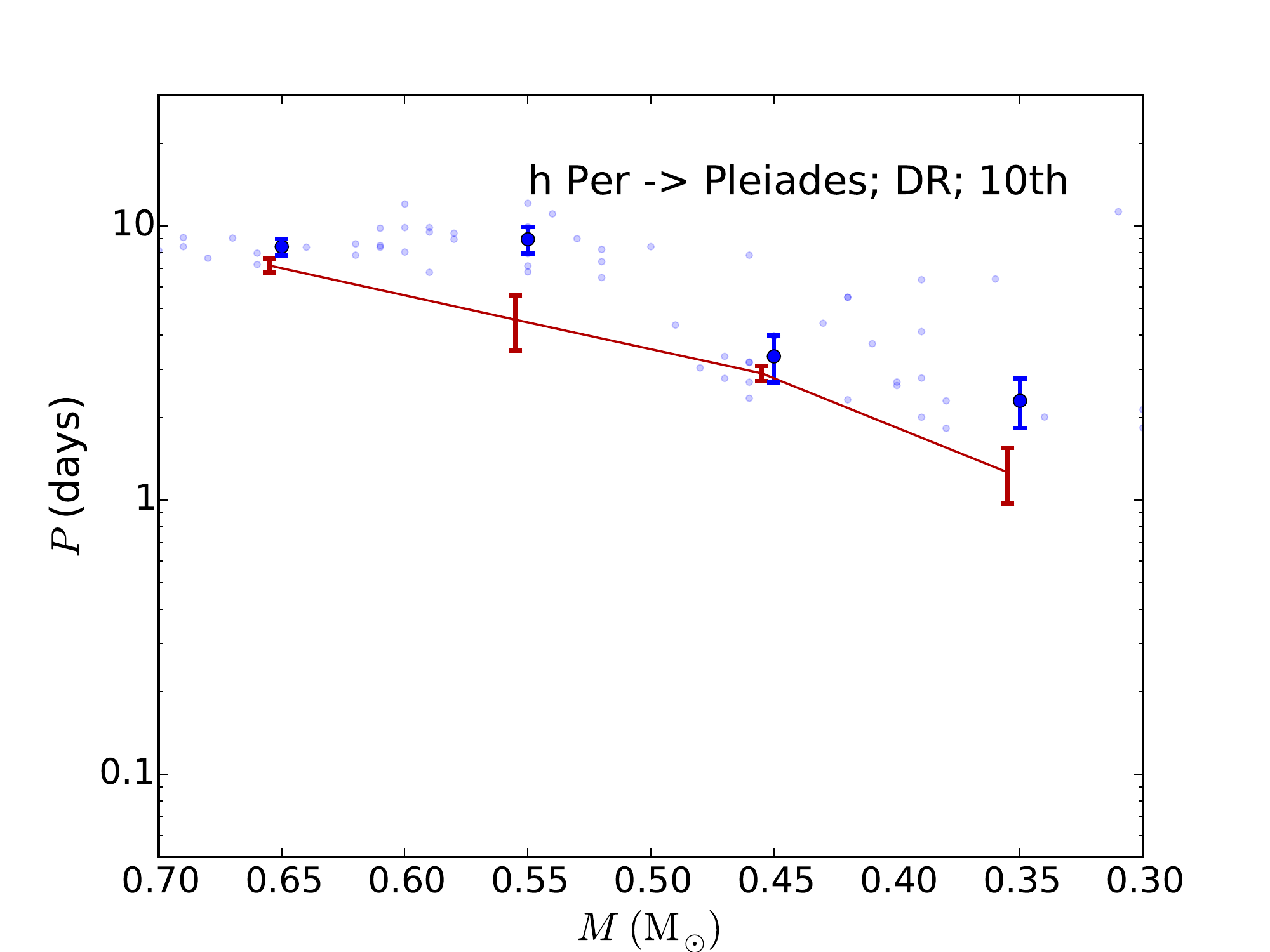}
\includegraphics[scale=0.45]{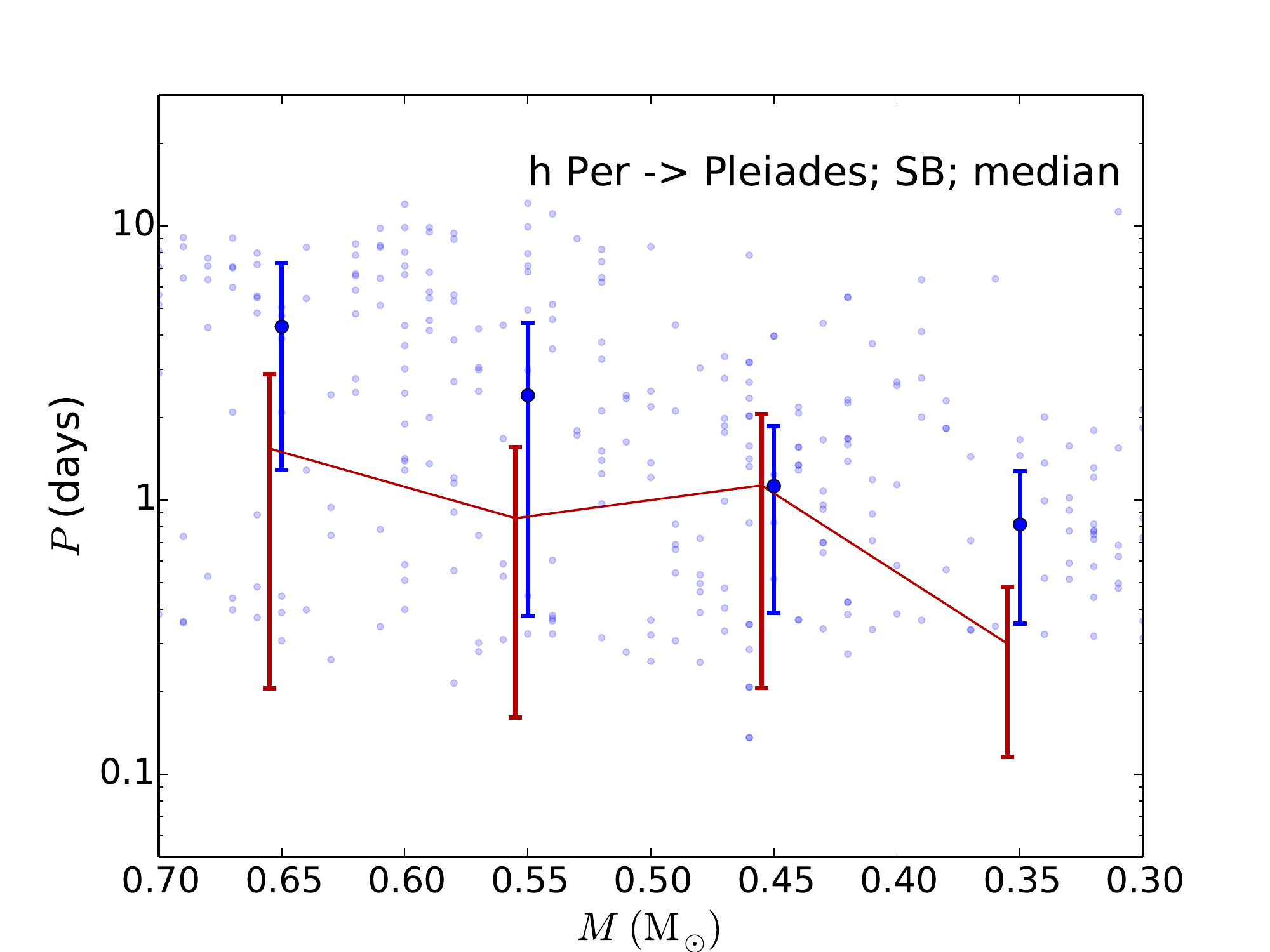}
\includegraphics[scale=0.45]{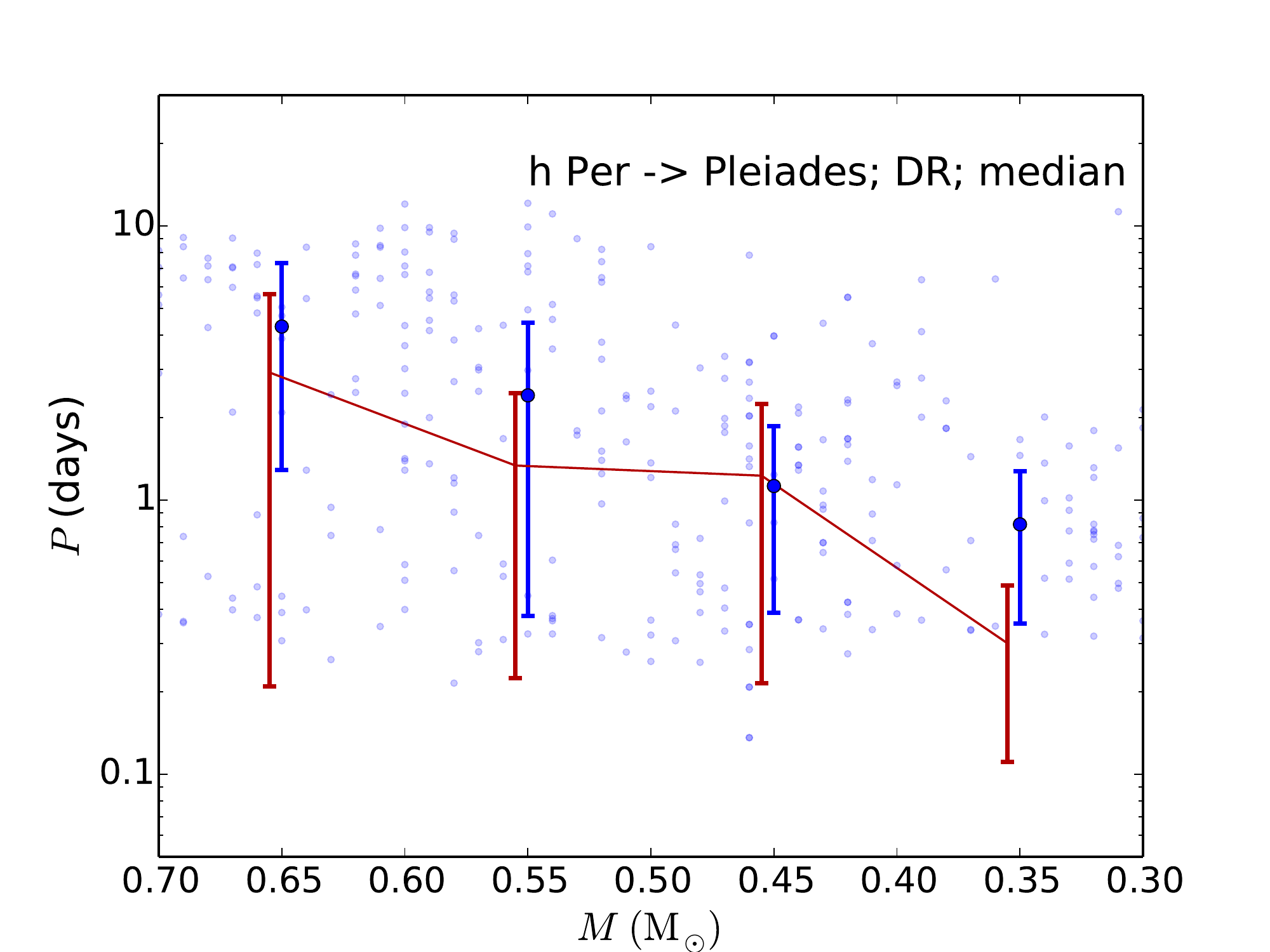}
\includegraphics[scale=0.45]{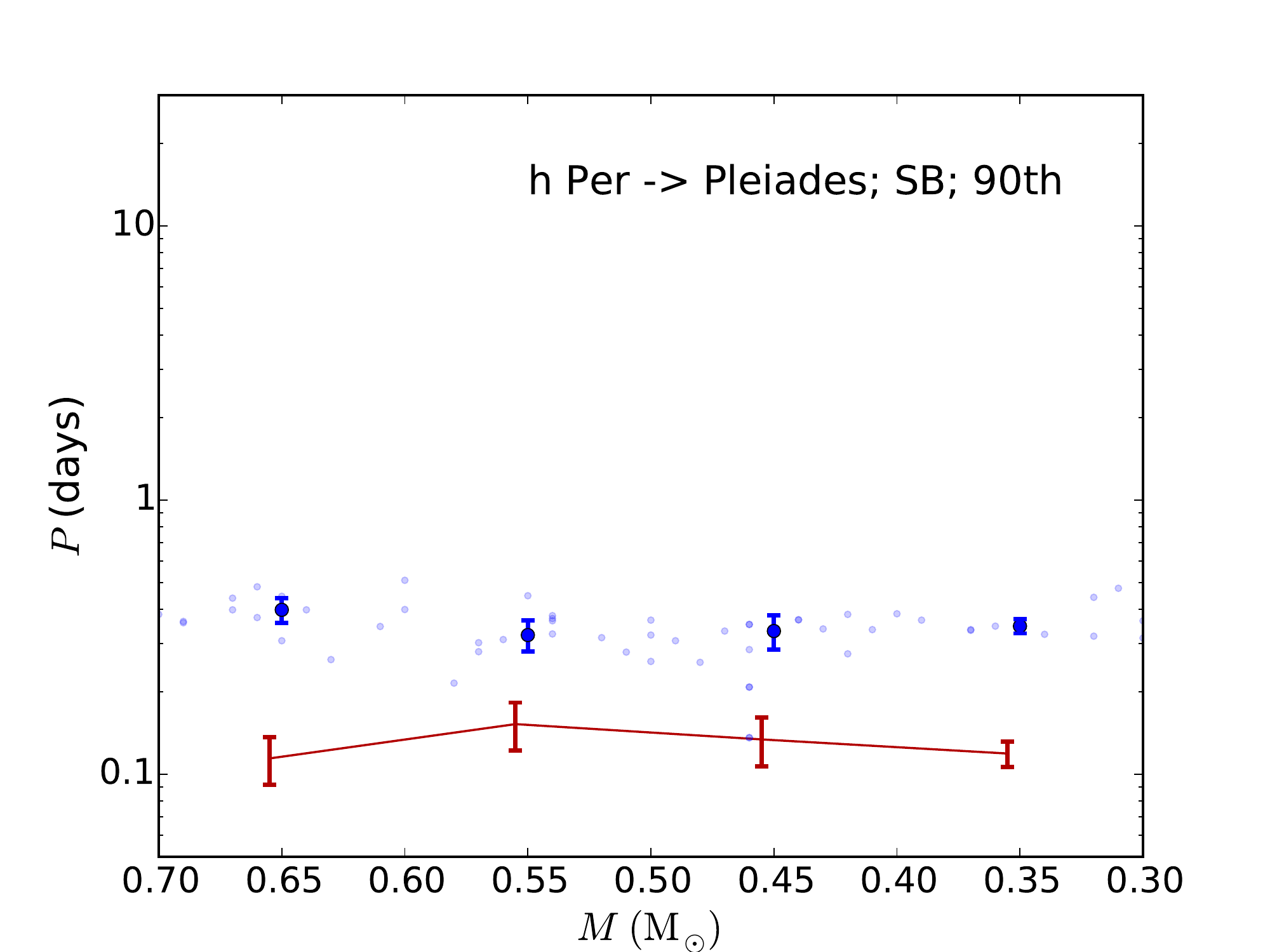}
\includegraphics[scale=0.45]{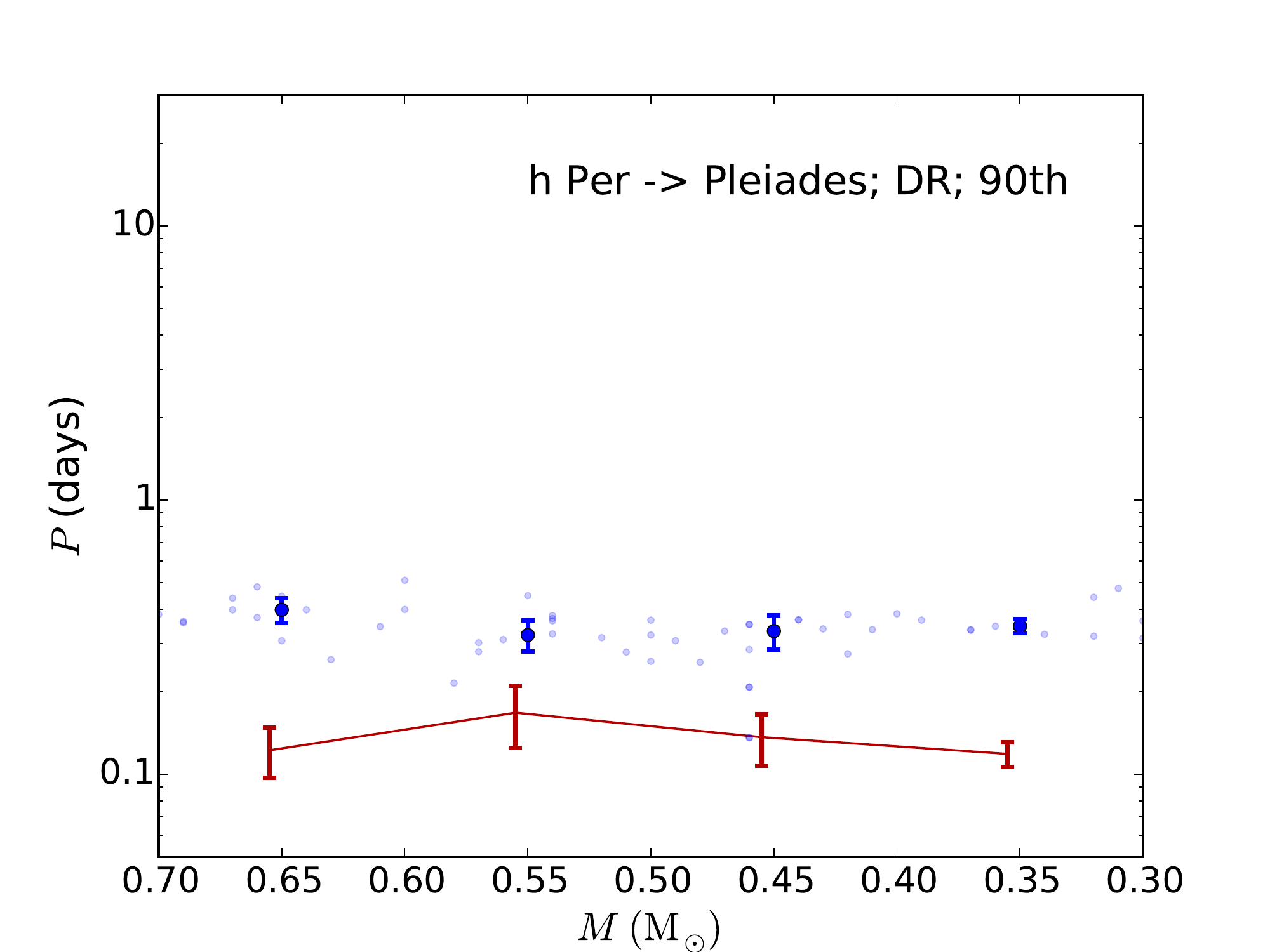}
\caption{Comparison of our h Per simulations (red lines) to the Pleiades data (blue circles) from Hartman et al.~(2010) and Covey et al.~(2016).  Below 0.5~$M_{\odot}$, there is little difference between the solid-body (left column) and the angular momentum transport models (right column).  Our angular momentum transport models match the slow branch extremely well in the mass regime shown.  The top row shows the 10th-percentile periods in each mass bin, the middle row the median periods, and the bottom row shows the 90th-percentile periods.  All the error bars are the 0th-20th percentile MAD, the overall MAD, and 100th-80th percentile MAD, respectively, for the stars in a given mass bin.  The theory and data points are offset from each other by 0.005$M_{\odot}$ for readability's sake.  The 10\% and 50\% simulations, without long-lived disks, are consistent with the data, while the rapidly rotating 90\% bound is not.}
\label{fig:losscomp}
\end{figure*}

\subsubsection{Do Long-lived Disks Exist?} \label{sec:slowrotators}

The maximum lifetime of circumstellar disks is still a matter of research.  Photodissociation/radiation pressure arguments (e.g., Alexander et al.~2006) and some simulations (e.g., Tinker et al.~2002) have argued for disk lifetimes of no more than $\sim10$~Myr, while other authors (see, e.g., Denissenkov et al.~2010) have allowed longer-lived disks (up to 20~Myr) in their simulations in order to fit the data when using solid-body rotation, and as a hedge against observational uncertainties.  Therefore, the question still remains:  do long-lived disks exist?  Our results shown in Figure~\ref{fig:diskfrac} in Section~\ref{sec:bimodality} offer some weak evidence against the idea, as a maximum disk lifetime of 12~Myr almost exactly matched the mean period of the h Per slow rotator peak, and even longer disk lifetimes would presumably keep those stars too slow.  Those results are far from conclusive, however, so we decided to test whether evolving the h Per stars directly without including disks of any kind reproduced our older clusters (these are the same models described in Section~\ref{sec:domains}).

We calibrated our wind model on the old open cluster M37.  However, as noted above, the treatment of the loss law has only a minor impact at the age of the Pleiades.  Therefore, the Pleiades offers a better test of the initial conditions than does M37.  To determine if our models could explain the features in the Pleiades rotation period distribution, we compared them to the actual data from Hartman et al.~(2010) and Covey et al.~(2016).  Figure~\ref{fig:losscomp} shows the 90th percentile, median, and 10th percentile rotators in our models overlaid on the Pleiades data.   The error bars show the MAD of the stellar periods in each mass bin.  For each mass bin, the median periods are consistent with those of the Pleiades within $1\sigma$, although the largest tensions occur in the lowest-mass stars and the higher-mass solid body rotators.  Therefore, our models' deviations from the Pleiades data are not statistically significant by this measure for either the median or lower bound of the distribution (we note, however, that the upper bound of our distribution shows significant deviation from the data; this is discussed further in Sections~\ref{sec:toM37} and \ref{sec:rapidrot}).  This means that long-lived disks are not necessary to explain the observed rotation period distribution, nor is any other angular momentum loss mechanism in addition to stellar winds.  This evidence also further supports the core-envelope decoupling picture in solar-mass stars, since the only way for solid-body rotation to reproduce their periods is to invoke long-lived disks.

\subsubsection{Initial Age and Metallicity} \label{sec:metallicity}

We also investigated the effects of changing the ages of the Pleiades or h Per, as well as the metallicity of our model stars.  Changing the ages changes the relevant timescales in the problem, and affects how long mechanisms such as contraction or stellar winds have to work.  Changing the metallicity changes the opacity of the star and its radius, thus changing its moment of inertia; this both affects the surface rotation period and how it evolves.  Changing the ending age within the error margin on current Pleiades age estimates did not significantly change the results, and neither did changing the distance to the Pleiades, since $M \sim L^{1/4}$ for stars in this mass range.  The HIPARCOS estimate of 118~pc (van Leeuwen~1999) and other estimates of 135~pc (Pinsonneault et al.~1998) produce a $\sim$7\% difference in mass, with the further distance leading to a higher estimated mass.  This difference is not enough to significantly affect our models.  

We also tested whether changing the metallicity within the provided error bars had any effect on our results, since we assume all the stars in our simulations are at solar metallicity.  Changing the [Fe/H] of any of our clusters to 0.05 had no statistically significant effect on the results.  We also ran much lower metallicity simulations of h Per, since the metallicity of that cluster is less well established.  Changing the metallicity from solar to [Fe/H] = -0.5 significantly accelerates the rotational evolution of the stars, but does not resolve the rapid rotator discrepancy at low masses.  Changing the age of h Per does have a measurable effect on our results by shifting the entire distribution to faster (younger h Per) or slower (older h Per) periods, but even going to an age of 20~Myr does not resolve the rapid rotator discrepancy.  Along with the results from Section~\ref{sec:bimodality}, this is evidence for cosmic variance in the stellar rotation period distribution of clusters, as changing the other parameters of the stars involved did not have the required effect.

\subsection{From h Per and the Pleiades to M37} \label{sec:toM37} 

A stringent test of our models is to evolve our stars from h Per all the way to M37, and do so consistently with the cluster data along the way.  This provides a test of the idea that stellar clusters form a single evolutionary sequence, presuming our model physics is correct.  We found that the slow rotators can indeed be fit as a single consistent sequence.  However, we found that the rapid rotators could not be using a single Kawaler wind law.  We argue in Section~\ref{sec:rapidrot} that this discrepancy is potentially the result of cosmic variance.

We started by taking our models and simply extending them from the Pleiades age and going to M37, using our best-fit Rossby-scaled wind law that we developed in Section~\ref{sec:windcalib}.  The results of those simulations are shown in Figure~\ref{fig:M37losscomp}.  The data and simulations match, unsurprisingly, given that we fit to M37 in the first place.  However, the $0.35M_{\odot}$ bins are too fast in both the solid-body and differentially rotating cases.  There are two reasons for this problem with the bin, each of which causes us to caution against reading much into any conclusions we might draw from it.  First, there is simply a lack of data in that bin in M37; second, many of the stars in that bin are fully convective, and $\tau_{\rm cz}$ is ill-defined for such stars.

The above shows that we are able to go from h Per straight to M37, but problems develop when we attempt to match the Pleiades as well.  As shown in Figure~\ref{fig:losscomp}, the 90th percentile fast rotators in our models do not match the ones found in the data; the model stars end up spinning much too fast.  This recalls the point touched on in Section~\ref{sec:windcalib}, where we noted that a single, global wind law fit to h Per and the fast rotators in the Pleiades and M37 was not possible at the $3\sigma$ level.  Figure~\ref{fig:zeropoint} illustrates this.  It shows suites of models run with the best-fit $\omega_{\rm crit}$'s from Section~\ref{sec:windcalib} and solid-body rotation, plotting period vs. age for a number of different starting periods.  The black points show the 90th percentile periods for the Pleiades and M37, while the error bars are the 100th-80th percentile MADs for each bin.  The blue bands show the evolution of the 90th percentile rotators from h Per fit to M37 or the Pleiades; the edges of the bands represent the 100th-80th percentile MADs for each bin.  The problem is not that the fast rotators cannot be matched by some wind law; rather, it is that the fast rotators in the Pleiades and M37 cannot be simultaneously matched by a Kawaler wind law with a single Rossby-scaled saturation threshold.  This problem is most difficult in the lowest mass stars; the periods of these stars are very similar in both h Per and the Pleiades, yet a wind law strong enough to counteract the effect of contraction is much too strong to be valid for any of the other mass bins.

A KS test helps confirm this.  Comparing the solid-body model distribution to the Pleiades data returns a $p$-value of $8\times 10^{-12}$, while doing the same for the transport distribution returns a $p$-value of $10^{-8}$.  Figure~\ref{fig:PleiadesCDF} shows the cumulative distribution of our simulations versus the Pleiades, where red is our models, and black is the data.  The clear discrepancy could be due either to environmental factors, or it could be a defect in the models.  If, when they were much younger, the Pleiades stars overall rotated much more slowly than the stars in h Per do, for example, then the fast rotators in the Pleiades would naturally end up rotating more slowly than our models show.  Given the large bump in the distribution, it could be that there is a much larger fraction of rapid rotators in h Per than there were in the Pleiades when it was younger.  It is also possible that our models do not contain some piece of physics that is relevant to low-mass fast rotators but is not strong elsewhere; if such an effect exists, it would need to somehow slow the surface rotation period of the star, whether that involved losing angular momentum or increasing the moment of inertia of the star.

\subsubsection{Detection Completeness of the Cluster Period Data}

In all of the models discussed above, we have assumed that stars of any rotation period can be detected, or equivalently that the ground-based surveys cited in this paper are mostly complete.  Here we examine this assumption, using tabulated amplitudes of stars in the Hartman et al. (2009) survey of M37 to gauge the effect of censored detection.  In their Figure 17, Hartman et al. show that the amplitudes are correlated with Rossby number, which is the ratio of the rotation period to the convective overturn time, and then derived an algebraic relation between amplitude and Rossby number.  Our analysis of these data shows the amplitudes are approximately (log) normally distributed about the mean relation, and furthermore that the M37 survey had a detection limit that was about 0.007 mag for stars with $M / M_\odot \leq 0.65$ but which rises to about 0.08 mag near $M / M_\odot \leq 0.3$. 

Figure~\ref{fig:completeness} demonstrates that in some circumstances our assumption may be incorrect.  This plot shows the cumulative period distribution for (upper panel) DR models of the M37 period distribution and (lower panel) for SB models; these are shown as thick lines.  The thin lines in each panel display the period distribution after selecting only those stars that would be above the detection limit in amplitude as a function of mass.  The selection was performed by randomly assigning an amplitude based on the Rossby number as a function of stellar mass using the distribution derived for M37.

The SB models are not much affected by censored data, mainly because the model rotation periods are narrowly distributed near the median value ($\sim 10$ days), and such stars have Rossby numbers that imply sufficient amplitudes for detection.  The difference between the two distributions in the lower panel of Figure~\ref{fig:completeness} arises mainly because a small fraction of the stars in the mass range $0.8 \leq (M / M_odot) \leq 1.2$ would have undetectable amplitudes; these are distributed like those actually detected in the Hartman et al. sample.

\begin{figure*}
\includegraphics[scale=0.45]{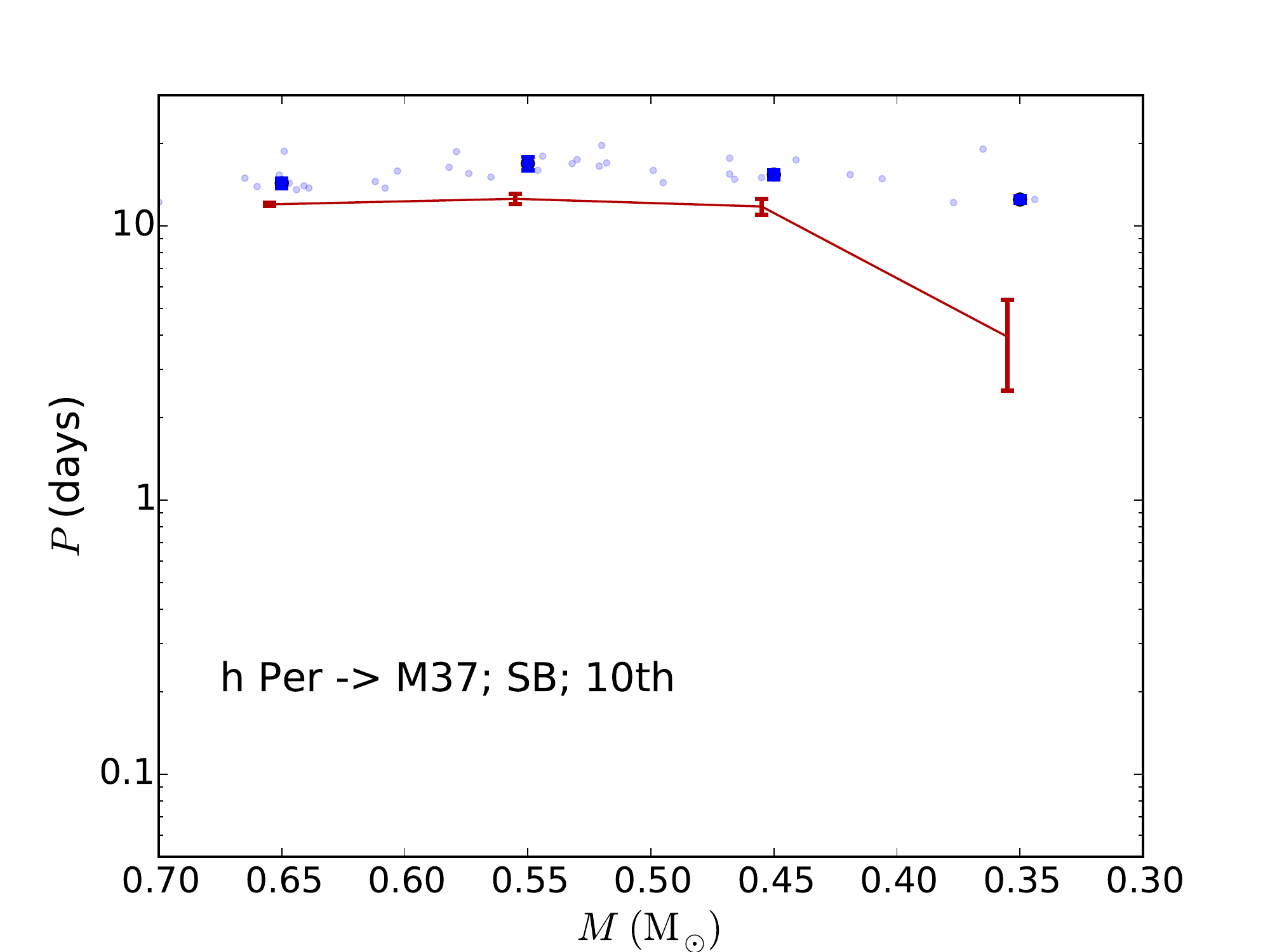}
\includegraphics[scale=0.45]{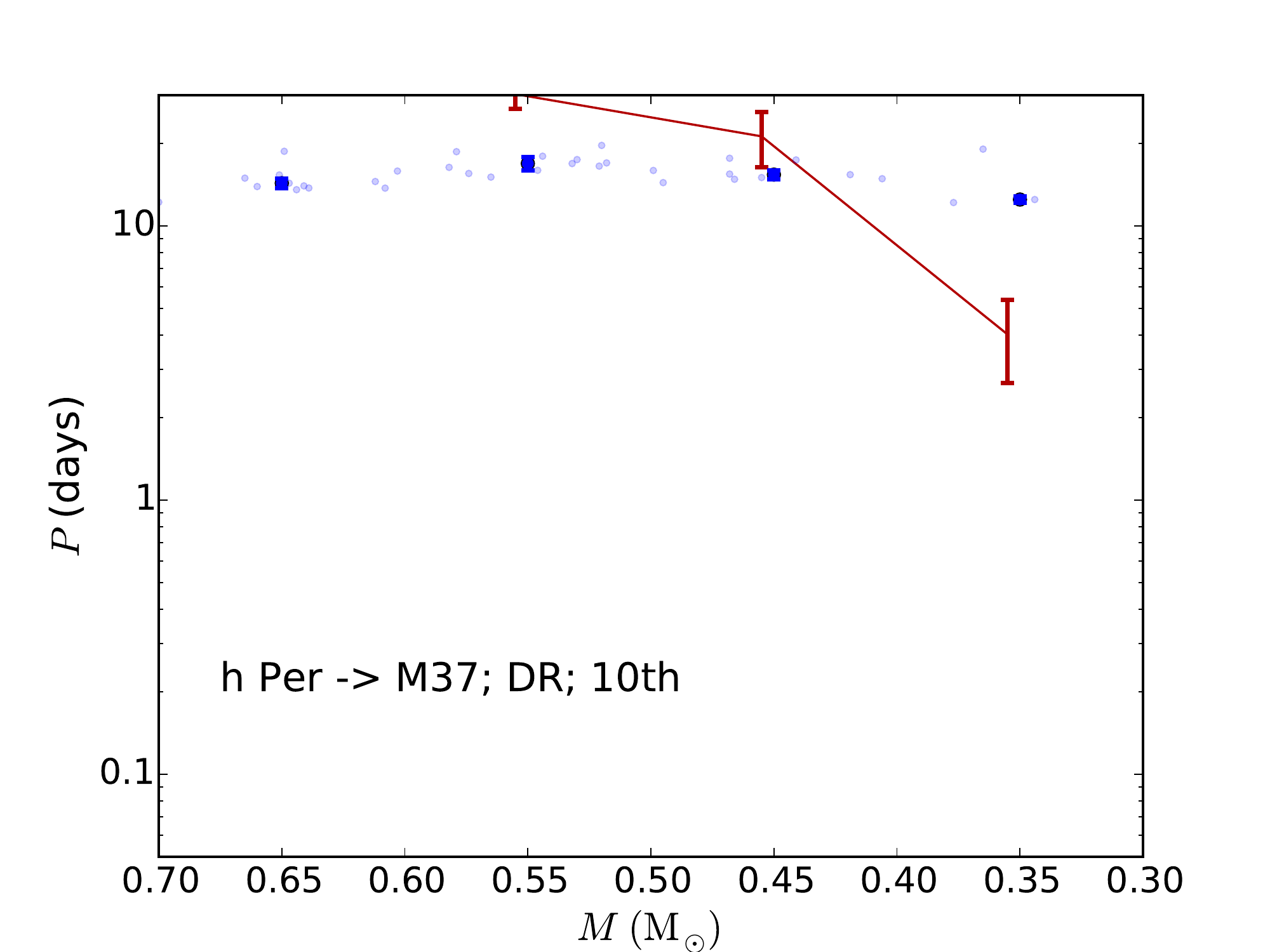}
\includegraphics[scale=0.45]{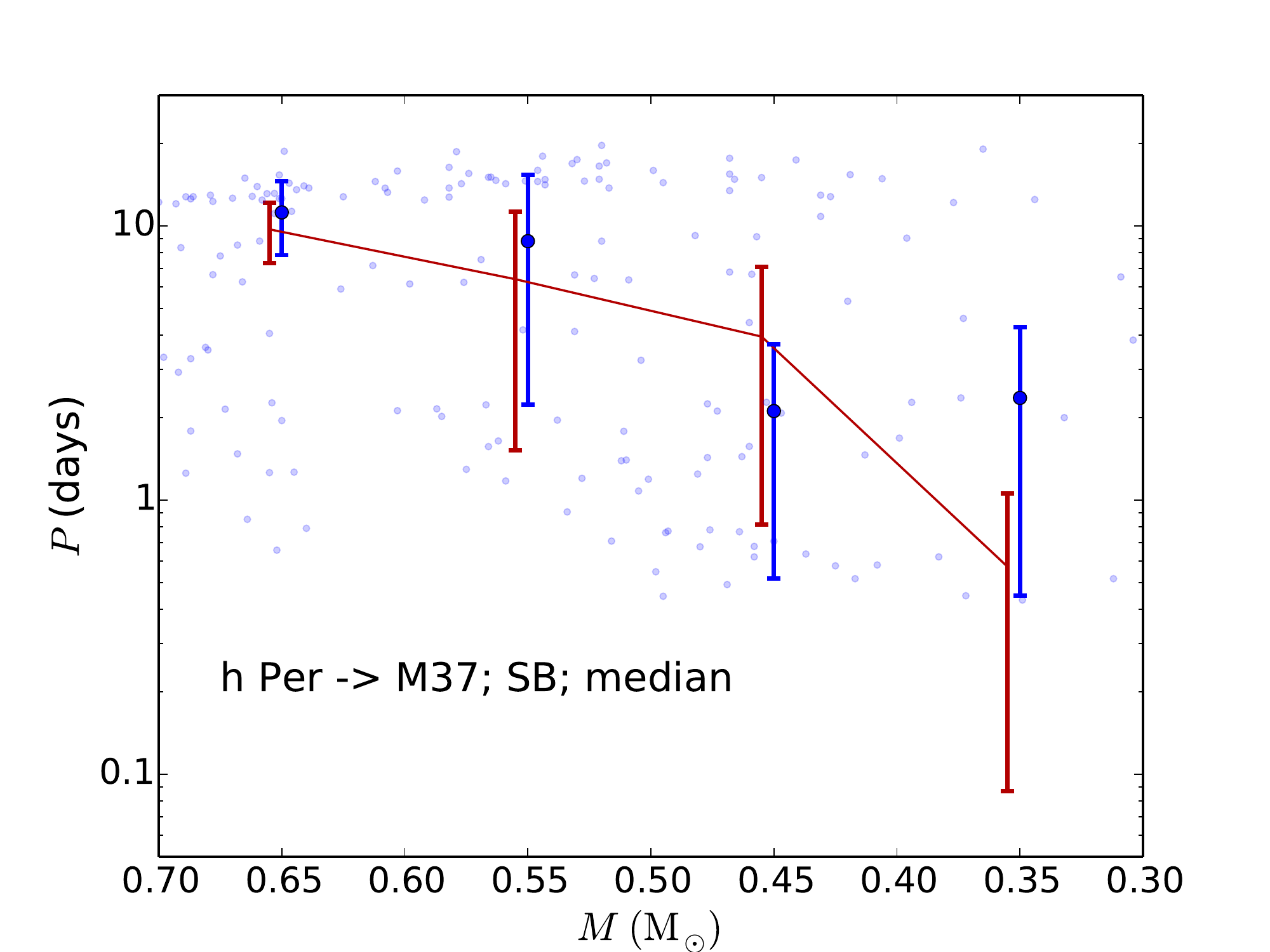}
\includegraphics[scale=0.45]{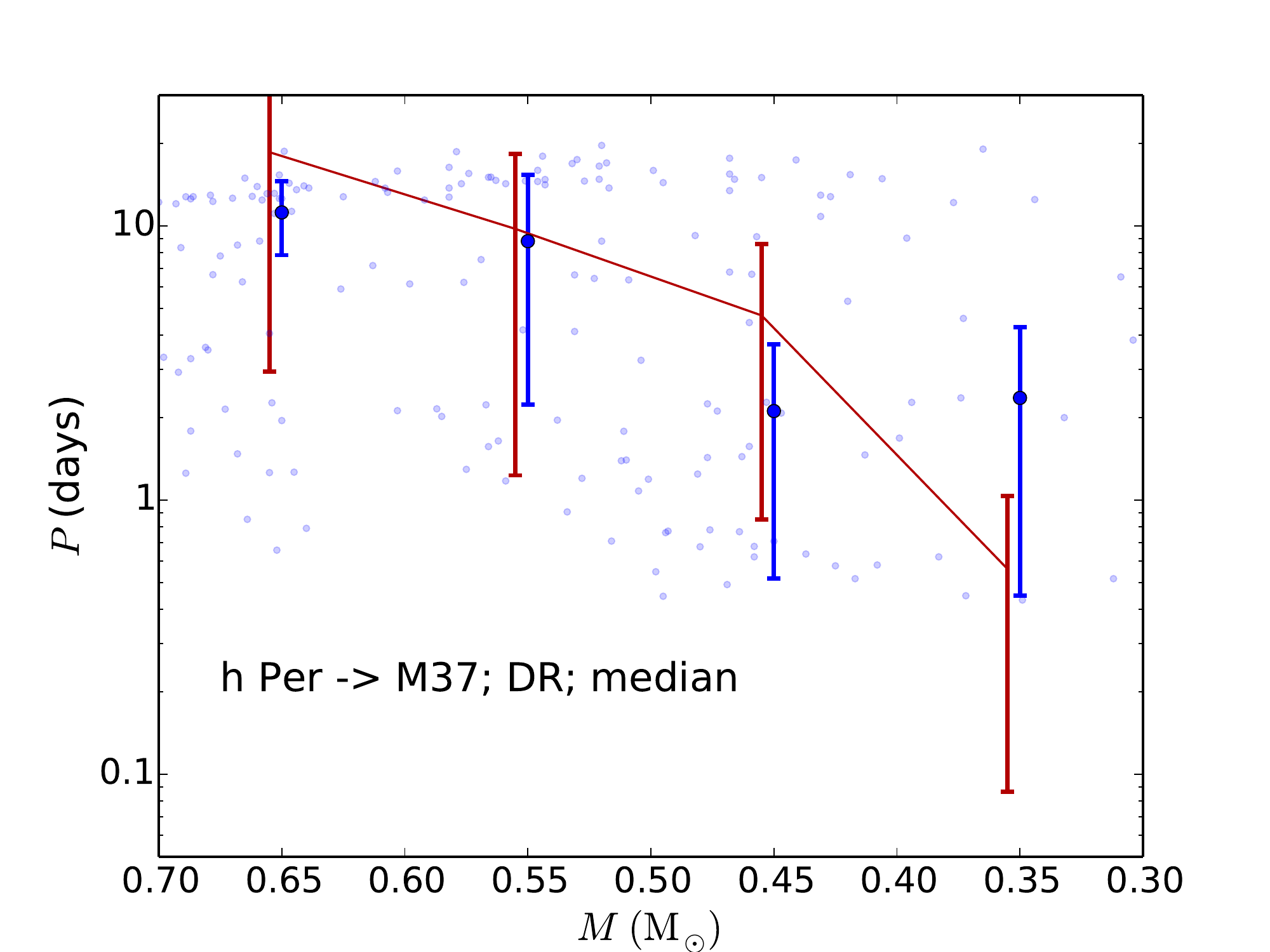}
\includegraphics[scale=0.45]{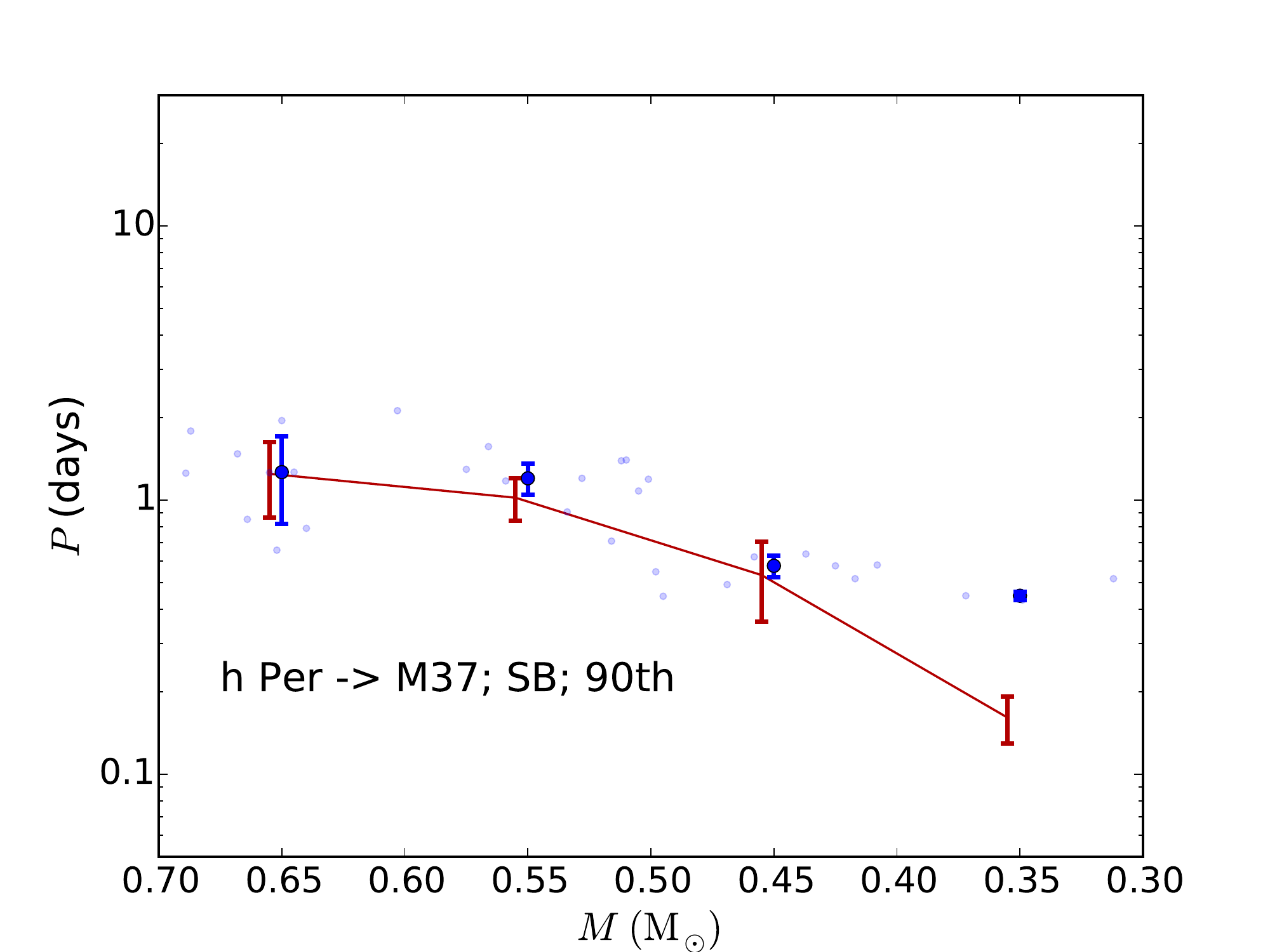}
\includegraphics[scale=0.45]{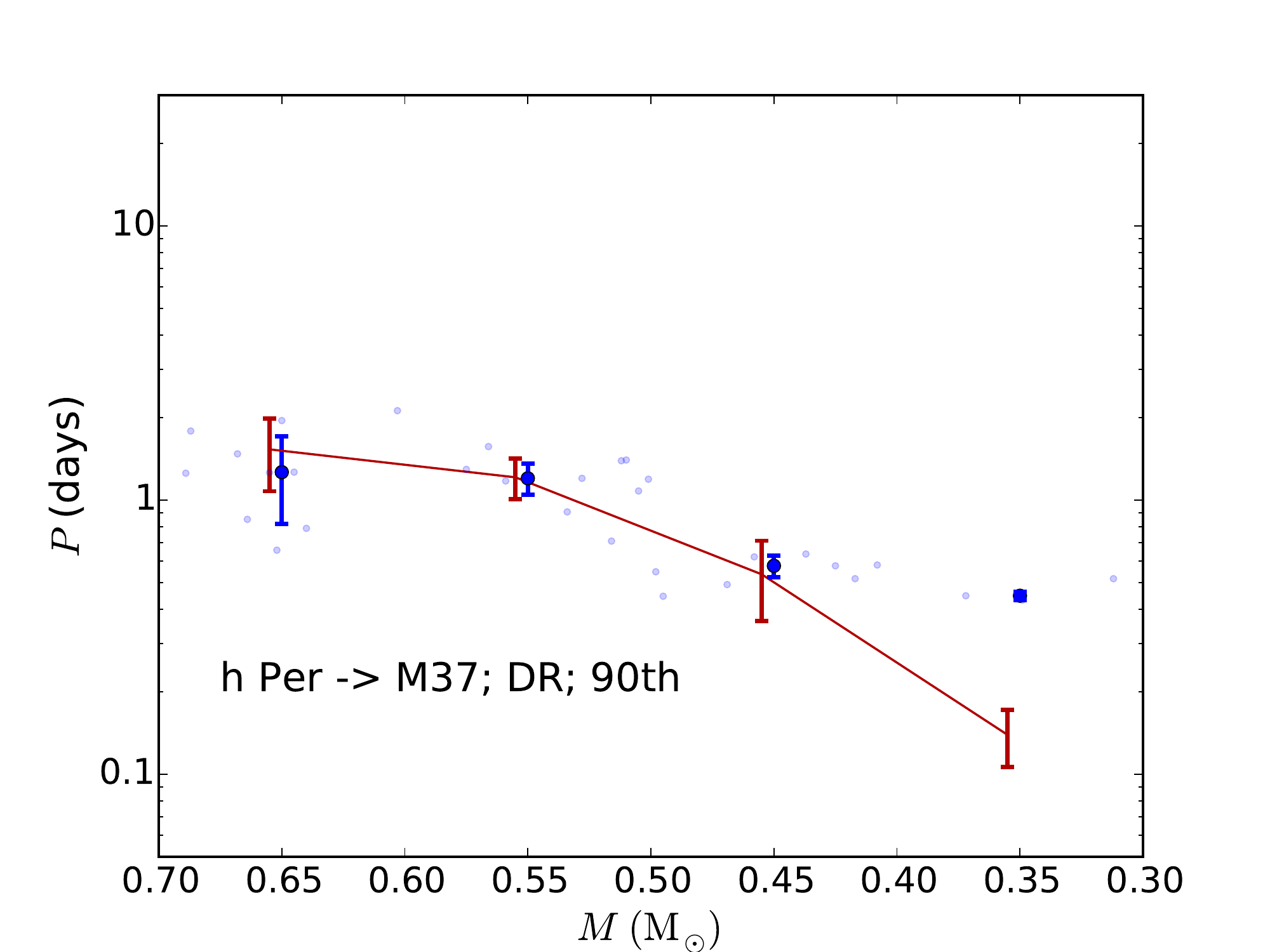}
\caption{Comparison of our h Per simulations (red solid lines) to the M37 data (blue circles) from Hartman et al.~(2009).  The top panels show the 10th-percentile periods from the cluster data and our simulations for each mass bin.  The middle panels show the median periods, while the bottom panels show the 90th-percentile periods.  In all panels, the period data are represented by the unconnected dots, while our simulation results are shown with the lines.  All the error bars are the 0th-20th percentile MAD, the overall MAD, and the 100th-80th percentile MAD, respectively, for the stars in a given mass bin.  The theory and data points are offset from each other by 0.005$M_{\odot}$ for readability's sake.}
\label{fig:M37losscomp}
\end{figure*}

\begin{figure}
\includegraphics[scale=0.45]{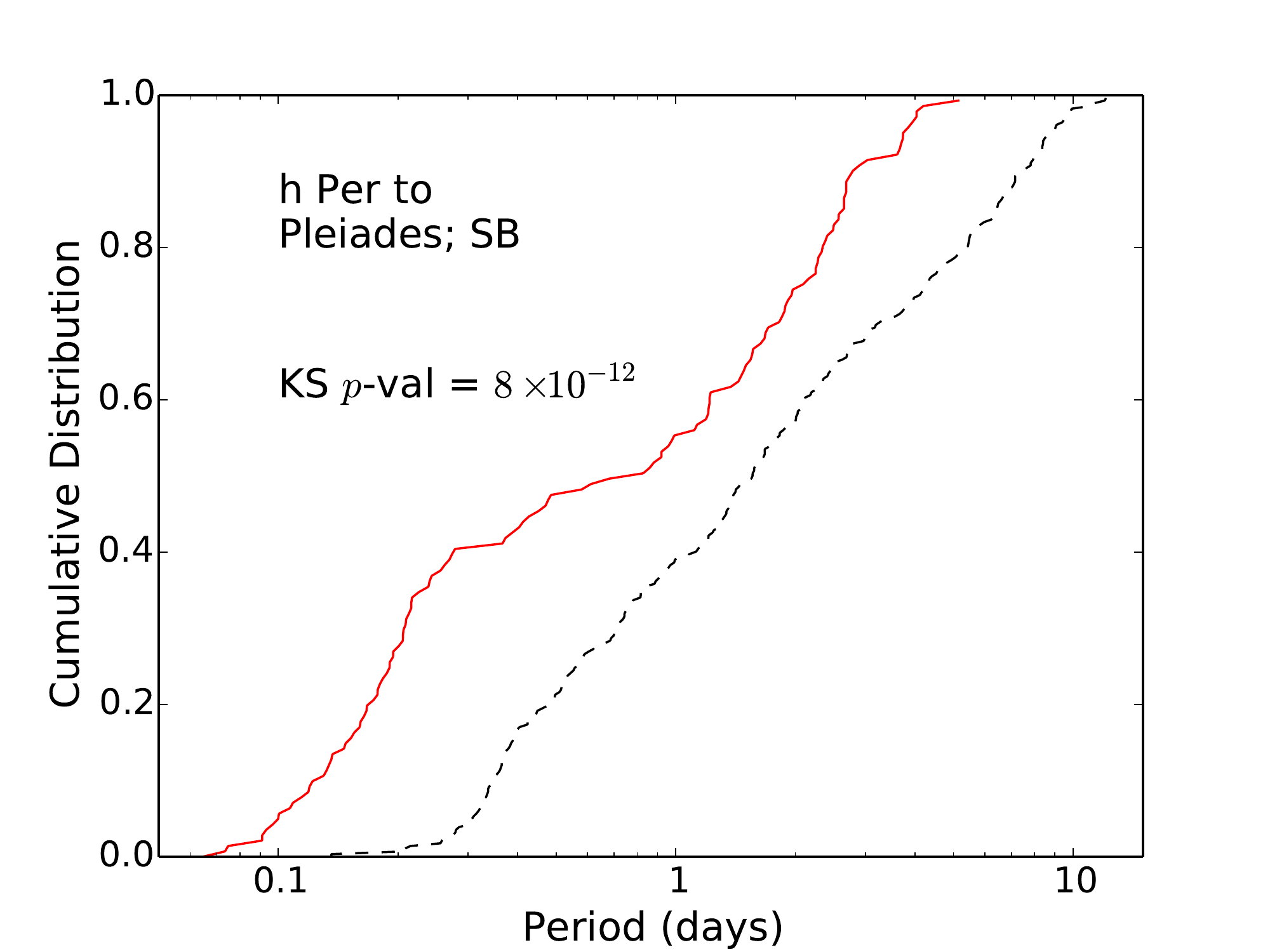}
\includegraphics[scale=0.45]{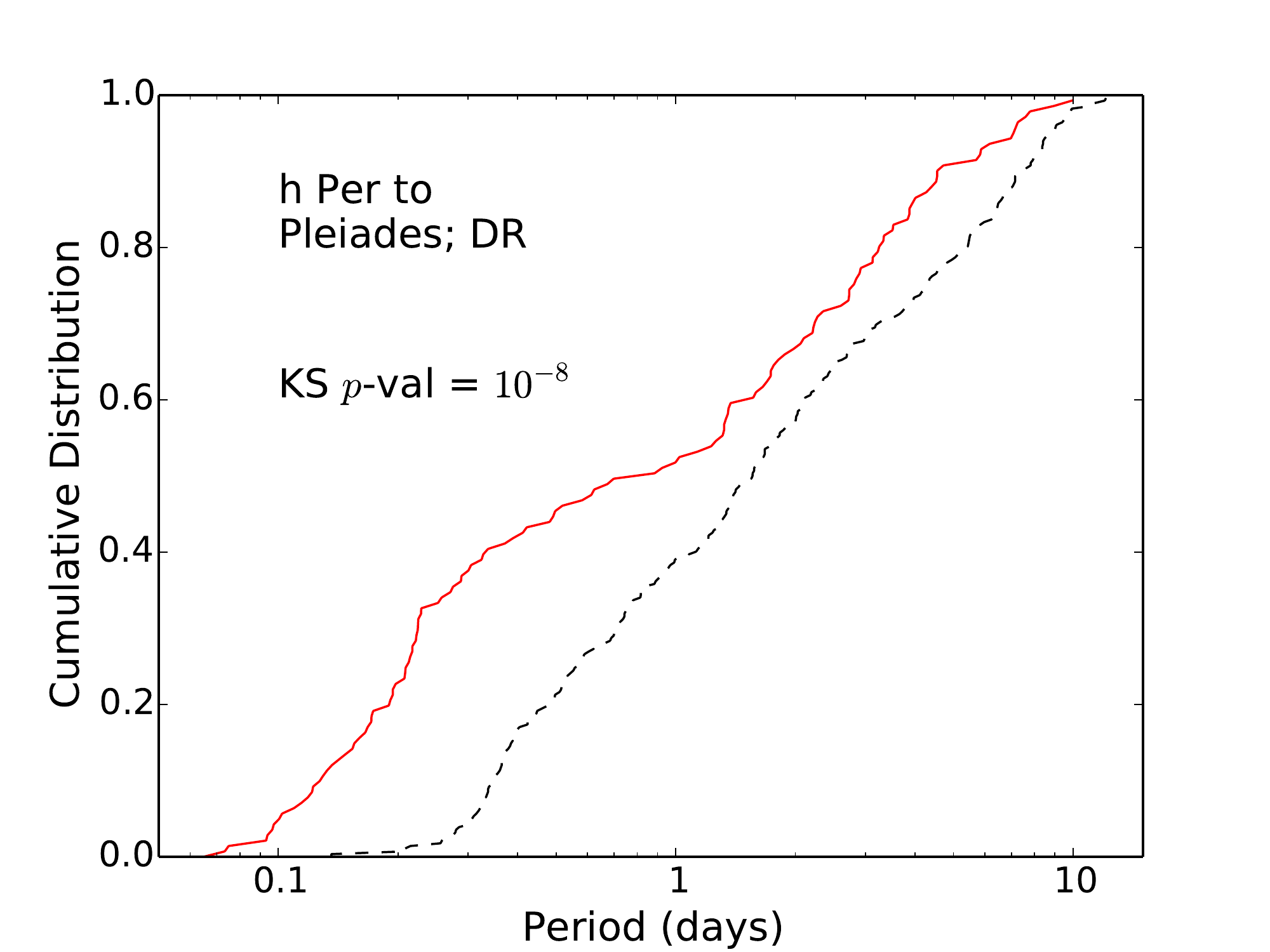}
\caption{Cumulative period distributions of our simulations of $0.3-0.7M_{\odot}$ stars compared to the Pleiades data.  The dashed black lines are the Pleiades data compared with our solid-body models (top panel) and differentially rotating models (bottom panel).}
\label{fig:PleiadesCDF}
\end{figure}

\begin{figure}[b]
\includegraphics[scale=0.5]{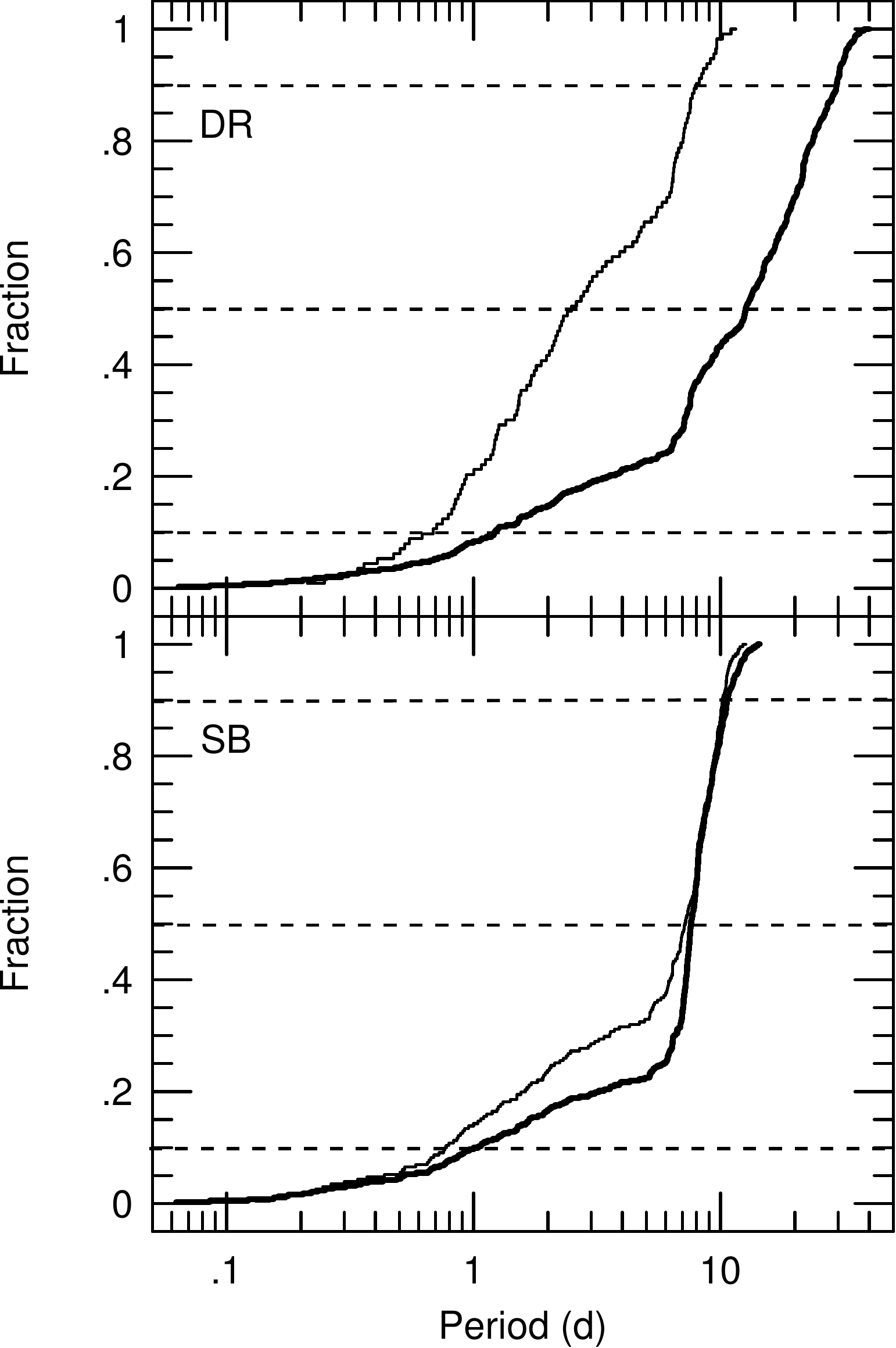}
\caption{Period detection completeness fraction for our model distributions.  The thick lines show the cumulative period distribution for our simulations evolved to M37, while the thin lines show the expected period distribution we would detect based on the limits of current photometry.  Upper panel shows differentially rotating models with angular momentum transport, while the bottom panel shows those with solid-body rotation.  The dashed lines are placed at the 90th-percentile, median, and 10th-percentile periods.  The solid-body distribution is well-detected, while the differentially rotating distribution produces many stars which would have unmeasurably slow periods for current surveys.  We note, however, that at the rapid rotator end of the distribution, virtually all stars have measurable periods, and that this does not affect our conclusions there.}
\label{fig:completeness}
\end{figure}

The situation is very different for the DR models, which predict a wide range of rotation near a solar mass, with periods extending down well below 10 days, and many stars at lower masses with very slow rotation.  Such stars would often have insufficient amplitudes for detection, and as a result the median rotation after amplitude selection is much shorter (2.5 d) than in the uncensored simulation (12.7 d).  A large fraction of the periods in Figure~\ref{fig:M37losscomp} ($70\%$) do not survive the amplitude selection, while the fraction of stars rejected in the DR models is much lower ($40\%$).

This means that the seeming disagreement between our higher-mass differentially rotating models and M37 may not be quite so large, since any stars much slower than those seen would not have detectable rotation periods.  However, this result does not affect the rapid rotators at all.  As seen in Figure~\ref{fig:completeness}, almost all stars with periods faster than the 90th-percentile are detected.

\subsubsection{Resolving the Rapid Rotator Discrepancy} \label{sec:rapidrot}

We are left with a conundrum:  what is it about the fast rotators that is resulting in these large discrepancies between the model and the data?  It is of course quite possible that our angular momentum evolution models are simply wrong.  However, our models describe the slow rotators well, and they perform well in other age and mass domains.  Testing alternate basic physical prescriptions and new alternative angular momentum loss laws such as those used by Matt et al.~(2015) could be fruitful, but such models are tuned to reproduce the same empirical constraints that we do, and are thus likely to have the same problem.  It could be that the Rossby/mass scaling of the loss law is invalid.  However, we found that changing the scaling, even fitting bin-by-bin following the approach of Sills et al.~(2000) (which found that the Rossby scaling did not work for low-mass stars such as modeled in this paper), did not resolve the discrepancy, and was not statistically different from the Rossby-scaled model in the bins where we had adequate data.

We have been assuming that the initial conditions of stellar evolution are universal - that is, star clusters of a given metallicity all form a single evolutionary sequence.  This approach is common throughout the literature, and is supported by the results of authors such as Tinker et al.~(2002), who found that it was possible to evolve the ONC forward to the Pleiades without trouble using the data available at the time.  If h Per, the Pleiades, and M37 do not lie on the same sequence because their initial conditions were all different due to cosmic variance, however, then this could cause the discrepancy we see.

\begin{figure*}
\includegraphics[scale=0.45]{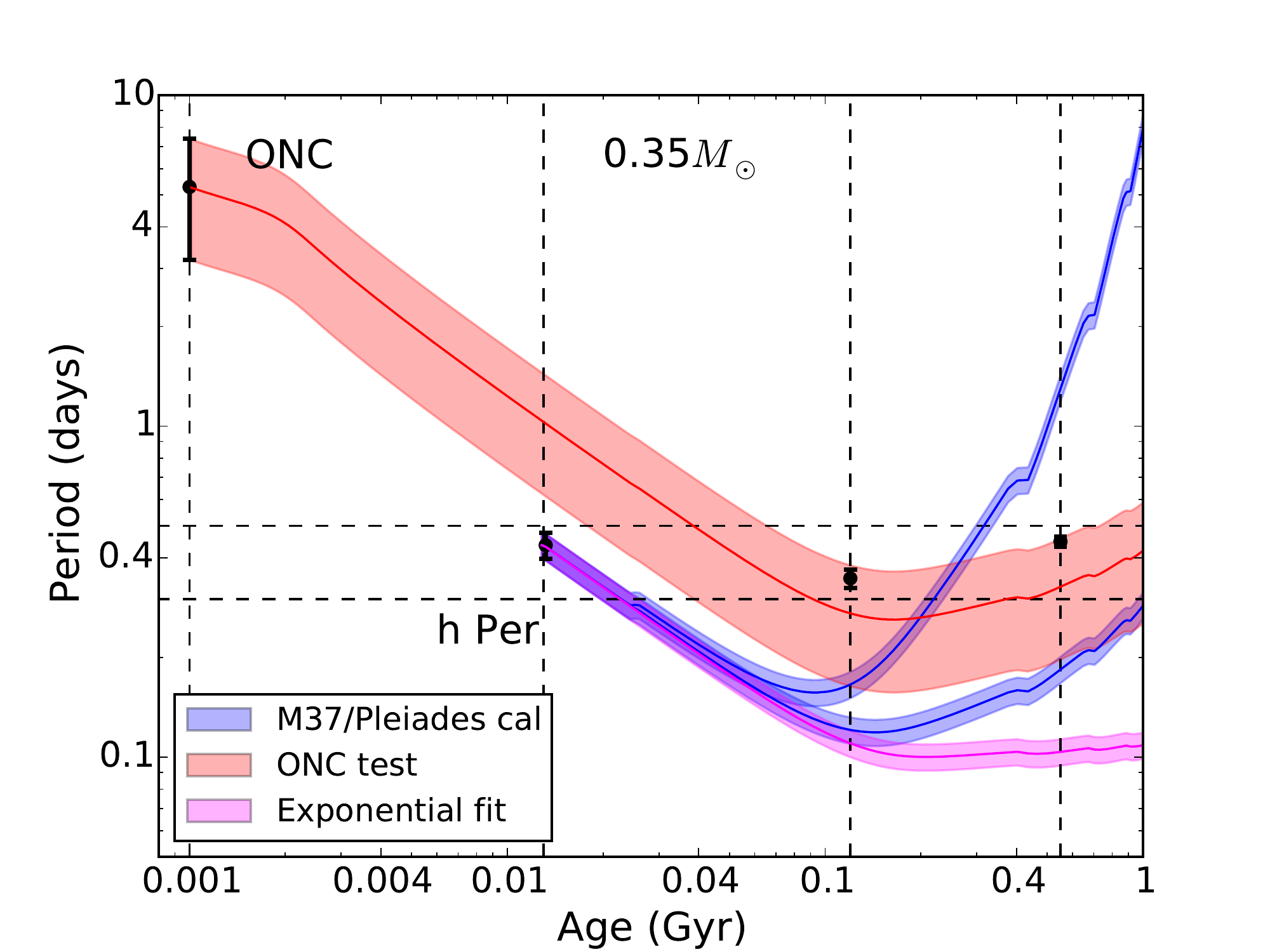}
\includegraphics[scale=0.45]{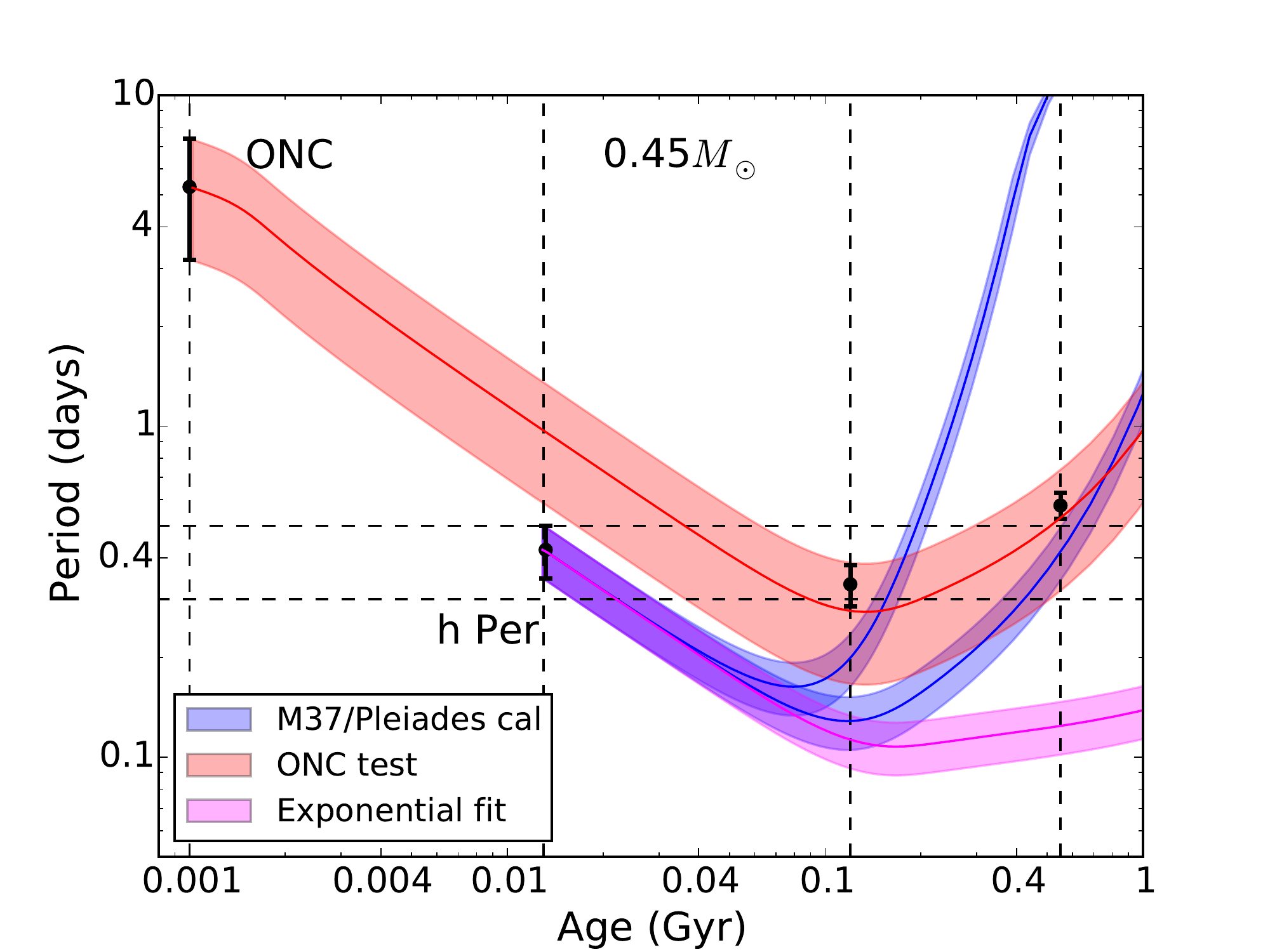}
\includegraphics[scale=0.45]{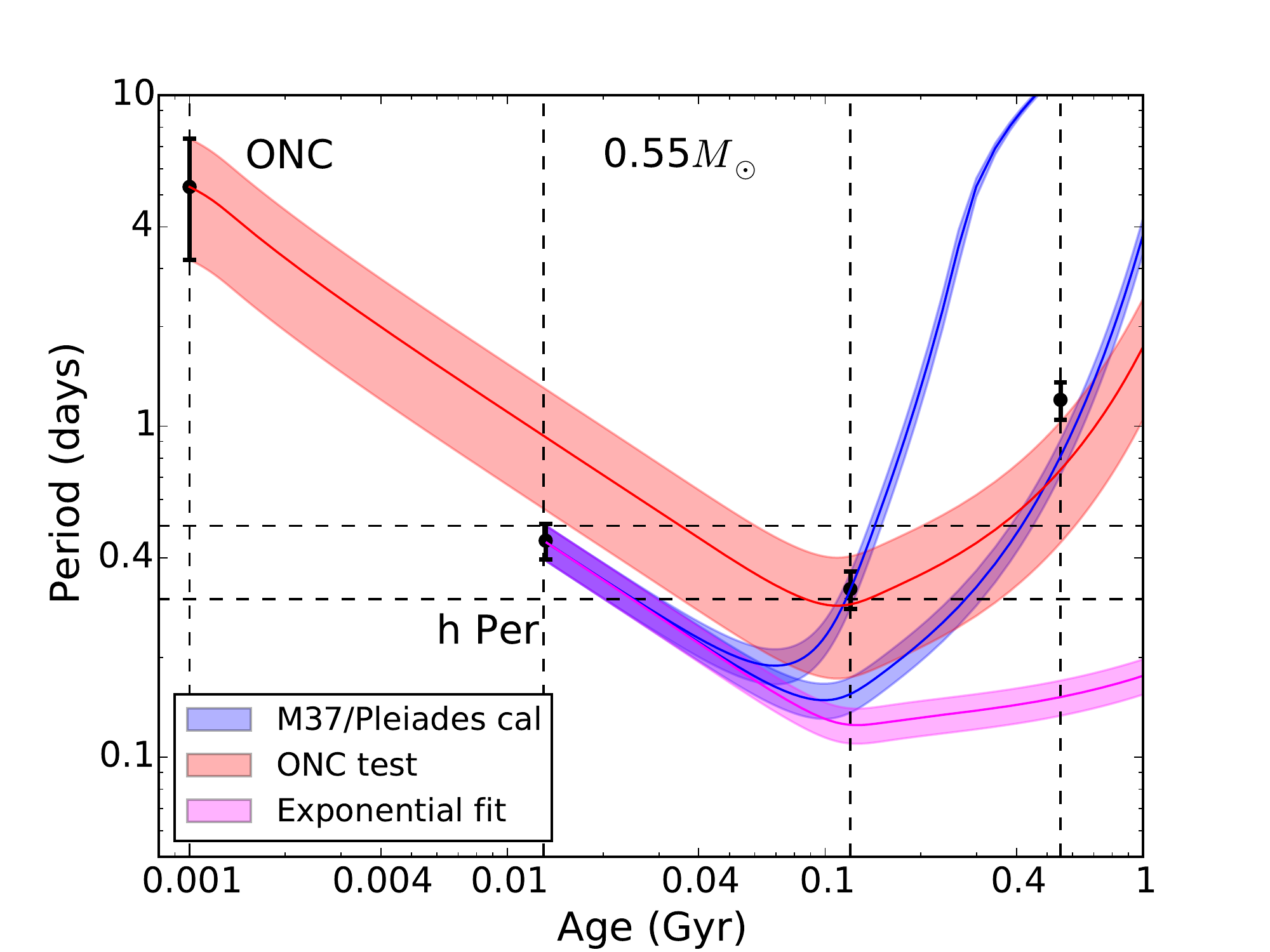}
\includegraphics[scale=0.45]{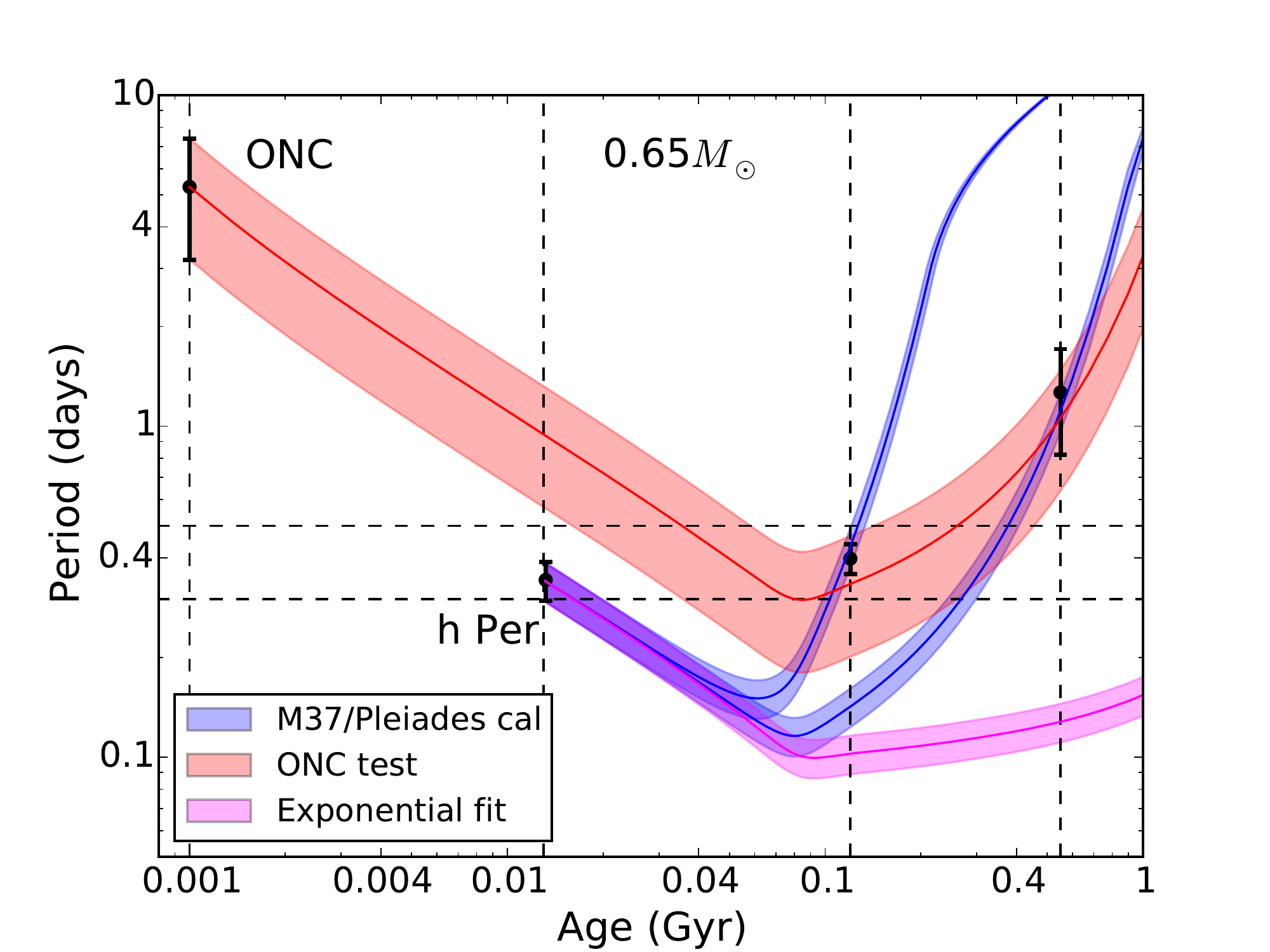}
\caption{Comparison of different wind law fits.  The red track uses the mean period of the non-disked stars in the ONC; the width of the band is set by the MAD of their periods.  The blue tracks are evolved from the h Per data, calibrated to M37 or the Pleiades; their widths are set by the MAD of the 100th-80th percentile periods for h Per, with the centerline being the 90th-percentile period for that mass bin in h Per.  The magenta track is the simple exponential fit between the Pleiades and M37; its width is the same as that of the blue tracks.  The black points are the actual cluster datapoints.  The ONC data point is offset so it can be seen.  The much better agreement between the ONC evolution and the Pleiades and M37 data argues that h Per may not be representative of clusters like the Pleiades or M37 early in their lives, and that environment may play a large role in the formation and evolution of rapidly rotating stars.}
\label{fig:zeropoint}
\end{figure*}

We therefore attempted to find a model using the standard Kawaler wind law which matched both the Pleiades and M37.  We decided to use the statistics for the ONC, using the non-disked stars as our starting point.  This model was Rossby-scaled.  We found that starting from the mean period of the ONC non-disked stars, and using a saturation threshold of $8.2\omega_{\odot}$ came very close to matching the 90th percentile points of both the Pleiades and M37.  Figure~\ref{fig:zeropoint} shows the period evolution of this model for each mass bin.  For the red band, we used the same fractional error as we did for the $0.65M_{\odot}$ bin in the other fits.  In no case is the model more than $1.5\sigma$ away from the data points, even in the $0.35M_{\odot}$ bin where we feel our M37 data is insufficient to draw rigorous conclusions from. This clearly shows that the Rossby scaling and Kawaler wind law work for this model.  This is clear evidence that cosmic variance may play a strong factor in the formation and evolution of rapid rotators in clusters.

There is also a more general way to test the data.  In our mass range, the Pleiades and M37 rotation periods have self-similar distributions, as verified by a KS test; applying a simple scale factor in each mass bin statistically matches the clusters.  Evolution of the form $\omega = \omega_0 e^{-kt}$ implies self-similar distributions and supports a saturated angular momentum loss law (Tinker et al.~2002).  Therefore, finding an exact wind law which matches the Pleiades and M37 should be as simple as solving for the decay constant.  If h Per, the Pleiades, and M37 form one cluster sequence, then, inputting this constant into our models should put the model distributions through each cluster point nicely.  However, as the magenta line in Figure~\ref{fig:zeropoint} shows, starting from h Per, using this decay constant fails to match either cluster, much less both.

\section{Discussion} \label{sec:discussion}

In this paper, by using a newly available extensive dataset of pre-MS stellar periods, we have obtained three overarching main results:  first, that a standard Kawaler wind law is sufficient to explain the angular momentum evolution of slowly rotating stars after the first $\sim$13~Myr; second, that a standard Kawaler wind law cannot simulatenously match the fast rotators in the Pleiades and M37 when starting from h Per; and third, that disk-locking can produce a bimodality in the rotation period distribution of young low-mass stars similar to that observed in h Per, so long as the disk-locking timescale is allowed to extend up to more than $\sim$10~Myr, in contradiction of previous studies such as Tinker et al.~(2002).

In previous studies of pre-MS rotation, if disks were included in the simulations, the disk-locking timescale was treated as a free parameter which could extend to 20~Myr or more (see, e.g., Denissenkov et al.~2010), even though observational studies of disks have found no evidence of them lasting longer than about $\sim$10~Myr (see, e.g., Hillenbrand~1997; Bell et al.~2013).  Until the release of the Moraux et al.~(2013) h Per dataset, there was a distinct lack of data from $\sim$10~Myr-old systems with which to test for upper limits to the disk-locking timescale.  Using this new dataset, we found that we did not need to invoke any sort of long-lived disks to match the slowly rotating population in h Per to that in the Pleiades and M37; a standard Rossby-scaled Kawaler wind law provided sufficient angular momentum loss.  This result establishes an upper limit to the disk-locking timescale of $\sim$13~Myr, the age of h Per; this is in line with the aforementioned observational studies.  Therefore, long-lived gaseous disks do not solve the core-envelope decoupling problem for solar-type stars.

The discrepancy between the fast rotators in the Pleiades and M37 in our models opens up a puzzling conundrum.  It is worth noting that a similar discrepancy appears in Gallet \& Bouiver~(2015), where their best-fit line is many $\sigma$ away from the fitting points, but they do not comment on this.  We found that the two populations could not be matched by a Rossby-scaled Kawaler wind law using a single saturation threshold when starting from h Per.  This is in contrast to previous studies of solar-mass stars, for example Gallet \& Bouvier~(2013), which has no such difficulty in matching clusters of different ages.  However, as shown in Section~\ref{sec:domains}, core-envelope decoupling is important in solar-mass stars, and provides an extra theoretical degree of freedom over the lower-mass stars we tested.  In fact, Gallet \& Bouvier~(2013) require long core-envelope decoupling timescales in rapid rotators in order to fit the data; this approach is contradicted by Pinsonneault et al.~(1990), which found that solar-mass rapid rotators must be strongly coupled.  Combined with our result, that would suggest that our understanding of the physics of rapid rotators, magnetically driven winds, or both is incomplete.

If, however, our understanding of the physics of rapidly rotating stars is complete enough that we should expect a single wind law to easily match both the Pleiades and M37, then it is possible that rapid rotators in h Per are anomalously fast.  The fact that our ONC simulations from Section~\ref{sec:bimodality} failed to replicate the fast-rotator peak seen in the h Per data despite a starting age of 1~Myr and no significant angular momentum loss for the non-disked stars, combined with the fact that slowing the initial period of our fast rotators allowed us to easily fit to the Pleiades and M37 without problems, further suggests that this could be the case.  If so, this result could indicate that environmental effects are important for understanding the rotation period distribution of young low-mass stars.  H Per is a very dense cluster, much denser than the ONC or the Pleiades (Currie et al.~2010), and it has been suggested that massive stars in such dense clusters may significantly affect the evolution of disks around nearby lower-mass stars (St{\"o}rzer \& Hollenbach 1998), potentially evaporating or significantly attenuating them within 1~Myr.  This would obviously eliminate the possibility for disk-locking to slow the rotation of such stars and generate a more rapidly rotating population.

\begin{acknowledgements}

We acknowledge Kevin Covey and collaborators for providing us with the Pleiades rotation period data for use with this work before the data was published, and also thank Kevin for his many thoughtful and insightful comments as we were writing this paper.  We thank Luisa Rebull for providing us with her dataset and mass estimates for the ONC stars, as well as her comments on this paper.  Marc Pinsonneault and Don Terndrup acknowledge support from NSF grant 1411685.

\end{acknowledgements}

\end{document}